\documentclass{jfm_arxiv}
\usepackage{graphicx}
\usepackage{epstopdf, epsfig}
\usepackage{amsmath,amsfonts}

                           \def\u{\mathbf{u}}
                           \def\v{\mathbf{v}}

                            \def\g{\mathbf{g}}
                            
                            \def\z{\mathbf{z}}
                            \def\s{\mathbf{s}}
                             
                             \def\p{\partial}
                            \def\S{{\bf{S}}}
                            
                            \def\F{\mathbf{F}}

                           \def\be{\begin{equation}}
                           \def\ee{\end{equation}}
                           \def\bea{\begin{eqnarray}}
                           \def\eea{\end{eqnarray}}
                           \def\bex{\begin{example}}
                           \def\eex{\end{example}}
\newtheorem{theorem}{Theorem}

\newtheorem{example}[theorem]{Example}

\usepackage{color}
\usepackage{subfigmat}

\shortauthor{Bryan Glaz, Igor Mezi{\'c}, Maria Fonoberova, Sophie Loire}

\title{Quasi-Periodic Intermittency in Oscillating Cylinder Flow}

\author{Bryan Glaz\aff{1}
  \corresp{\email{bryan.j.glaz.civ@mail.mil}},
   Igor Mezi{\'c} \aff{2},
  Maria Fonoberova\aff{3},   \and Sophie Loire \aff{3}}

\affiliation{\aff{1}Vehicle Technology Directorate, U.S. Army Research Laboratory,
Aberdeen Proving Ground, MD, USA
\aff{2}Department of Mechanical Engineering, University of
California Santa Barbara, Santa Barbara, CA, USA,
\aff{3} Aimdyn, Inc., Santa Barbara, CA, USA}

\begin{document}

\maketitle

\begin{abstract}
Fluid dynamics induced by periodically forced flow around a cylinder is analyzed computationally for the case when the forcing frequency is much lower than the von K{\'a}rm{\'a}n vortex shedding frequency corresponding to the constant flow velocity condition. By using the Koopman Mode Decomposition approach, we find a new normal form equation that extends the classical Hopf bifurcation normal form by a time-dependent term for Reynolds numbers close to the Hopf bifurcation value. The normal form describes the dynamics of an observable and features a forcing (control) term that multiplies the state, and is thus a parametric -  i.e. not an additive - forcing effect. We find that the dynamics of the flow in this regime are characterized by alternating instances of quiescent and strong oscillatory behavior, and that this pattern persists indefinitely. Furthermore, the spectrum of the associated Koopman operator is shown to possess quasi-periodic features. We establish the theoretical underpinnings of this phenomenon -- that we name Quasi-Periodic Intermittency -- using the new normal form model and show that the dynamics are caused by the tendency of the flow to oscillate between the unstable fixed point and the stable limit cycle of the unforced flow. The quasi-periodic intermittency phenomena is also characterized by positive Finite-Time Lyapunov Exponents that, over a long period of time, asymptotically approach zero. 
\end{abstract}


\section{Introduction}

Hopf bifurcations and the ensuing dynamics are ubiquitous throughout physics. In fluid dynamics, the interest in Hopf bifurcations is motivated by both practical and theoretical considerations. From a practical perspective, Hopf bifurcations are inherent features of many fluid dynamic systems with engineering significance, such as vortex-induced vibrations, wake dynamics, and instabilities \citep{forced_cyl_1, sreenivasan_cyl, HuerreandMonkewitz:1990}. Theoretical investigations are of interest due to the impact on a variety of nonlinear dynamical systems phenomena, including theories of  the onset of turbulence  \citep{landau_book, ruelle_takens_chaos}. In fluid dynamics, the near wake dynamics of a stationary cylinder subjected to sufficiently large steady streamwise flow Reynolds number (the bifurcation parameter) is a well established physical realization of a Hopf bifurcation \citep{sreenivasan_cyl}.

In order to understand the development of Hopf bifurcation dynamics when subjected to external influences, it is informative to study systems with applied forcing. The forcing could represent a controller input, or it could result from coupling between the fluid dynamics and another dynamical system such as in fluid-structure interaction problems. Correspondingly, flows over oscillating cylinders have been the subject of several numerical and experimental studies \citep{forced_cyl_1, osc_cyl_jfm2001, viv_cyl, leontini_cyl, forced_cyl_2}. Typically, these studies have been motivated by the fluid-structure interaction vortex-induced vibration (VIV) problem. Although a variety of oscillation (forcing) amplitudes have been investigated, much of the streamwise oscillating cylinder studies have focused  on VIV relevant frequency conditions in which a prescribed cylinder oscillation frequency, $f$, corresponds to a low order harmonic of the stationary cylinder natural shedding frequency, $f_0$. The $1 \le f/f_0 \le 2$ regime has been thoroughly investigated \citep{leontini_cyl, forced_cyl_2}.  Synchronization between the vortex shedding dynamics and $f$, as well as quasi-periodicity  for non-synchronized conditions, are well established phenomena that can occur in this regime.

However, the simple harmonically forced cylinder system is not limited to synchronization or quasi-periodic dynamics. Two-dimensional simulations by \citet{cyl_chaos} have indicated the presence of chaos at certain amplitudes when $f/f_0 = 1$. \citet{cyl_chaos} attributed aperiodic behavior observed in the lift force as a manifestation of chaos caused by competition between the spatial structures of the natural shedding mode (spatially anti-symmetric) and the forced mode (spatially symmetric). These results from \citet{cyl_chaos} demonstrate the rich and potentially complex dynamical states that can be achieved from simple harmonic forcing of the canonical cylinder flow problem. It is interesting to note that the explanation offered for aperiodicity in the observable functional (i.e. lift force) was based on the underlying Navier-Stokes system states (i.e. velocity mode shapes), rather than through a direct mathematical expression of the observable dynamics that exhibited chaotic symptoms. In this study, a dynamics of observables perspective is taken in which an equation directly describing the evolution of the functional of interest (lift force) is used to provide theoretical understanding for the observed phenomena. 

Although lock-on \& synchronization regimes of systems forced near resonances are well understood, the underlying chaotic or quasi-periodic dynamics when forcing far away from a natural frequency are not. This behavior can be critical for systems of practical significance, such as oscillating airfoils undergoing dynamic stall \citep{mccroskey1982} where $f$ can be at least one order of magnitude slower than the separated flow shedding frequency of the stationary airfoil. Therefore, study of canonical systems such as the oscillating cylinder for $f << f_0$ can lend insight into unexplored periodically forced bifurcation parameter regimes that may advance fundamental understanding of more complex practical situations.

As discussed, periodically forced Hopf bifurcation flows such as oscillating cylinders can exhibit a variety of dynamical phenomena, ranging from lock-on/synchronization to chaos. Typically, a spectral perspective corresponding to a local measurement (i.e. velocity at a particular location in the wake) or an integrated measurement (i.e. lift force) has been utilized to characterize the dynamics. The spectral perspective is ideal for cylinder wake shedding dynamics given that much of Hopf bifurcation theory deals with parametric changes in eigenvalues in the vicinity of a critical value, e.g. \citet{wiggins}.  Therefore, a spectral approach seems appropriate for extension to forced bifurcation flows in which we seek to track the development of dynamically relevant post-bifurcation features, such as the shedding mode, as forcing is applied. However, typically the transition from the nonlinear, infinite-dimensional Navier-Stokes description to a reduced order set of equations considered in Hopf bifurcation normal form theory is difficult. Recently, modal decomposition methods for the analysis of both finite and infinite-dimensional nonlinear systems based on spectral properties of evolution operators have emerged \citep{mezic:2005, mezic_annual_reviews, rowleyetal:2009, applied_koop}.  This framework has the remarkable capability of capturing the full nonlinear dynamics by merging applied ergodic theory with operator theory to identify key spectral properties of an evolution operator attributed to Koopman \citep{Koopman:1931, mezic:2005}. These operators are defined for any nonlinear system, and modal decompositions of nonlinear systems based on spectral analysis of the Koopman operator can be derived from measured data without the need for linearization \citep{rowleyetal:2009}. Moreover, spectral decompositions derived from the Koopman operator lead to spatial modes valid in all of phase space, not just locally, and the modes correspond to specific frequencies and growth/decay rates \citep{mezic_annual_reviews, Koopman:1931}. Thus, these so-called ``Koopman modes" provide a qualitative fully nonlinear analog to the familiar notion of global modes (see the discussion in \citet{HuerreandMonkewitz:1990} on linearized global modes). While recent progress illuminating the spectral properties of the Koopman operators show considerable promise, there remain a number of fundamental research issues. For instance, the important effects of periodic excitation of bifurcation parameters have not been studied.   

Various approaches for computing Koopman modes are summarized in \citep{mezic_annual_reviews}. The interpretation of the Dynamic Mode Decomposition (DMD) method \citep{schmid:2010} as an approximate approach for calculating Koopman modes was established in \citep{rowleyetal:2009}. Koopman and DMD analysis have recently been applied to post-bifurcation wake dynamics of flow past a stationary cylinder \citep{chenetal:2012, bagheri:2013, bagheri_noise, Wynn_OMD}.  \citet{chenetal:2012} and \citet{Wynn_OMD} introduced improved numerical methods to generalize DMD and obtain more accurate approximations of the Koopman modes, while \citet{bagheri:2013} established the linkages between the complex-valued Landau equation (i.e. first order normal form) describing Hopf bifurcations and the Koopman operator. The effects of noise on the Koopman decompositions of stationary cylinder flow were later treated in  \citet{bagheri_noise}. However, Koopman-theoretic analysis of periodically forced bifurcation systems have not been pursued - neither within the context of fluid dynamics, nor outside of it -  and we do that here. By considering a separation in scale between the forcing and natural frequencies, we find a physical realization of a novel dynamics regime that we dub the Quasi-Periodic Intermittency due to the interchange of quiescent and oscillatory regimes of flow. This regime is interesting since it combines the properties of quasi-periodicity with intermittency, and presents us with a novel type of an attractor in a simple form that reflects more complex attractors responsible for a variety of  transition regimes from laminar to turbulent flows. Methodologically, we develop a procedure for deriving normal forms in nonlinear field theories, such as the Navier-Stokes equations, by utilizing the Koopman Mode Decomposition. 


In this paper, we utilize Koopman decompositions of cylinder flow-fields due to prescribed streamwise velocity oscillations superimposed on a steady flow component to gain understanding  of dynamics for the separation-of-scale regime $f<<f_0$. The cylinder flow problem, two-dimensional computational fluid dynamics (CFD) implementation, and the connections to Hopf bifurcation dynamics are briefly described in Section \ref{cyl_hopf}.  A summary of Koopman-operator based theory and numerical implementation is provided in Section \ref{koop_methods}. In Section \ref{cmnf_section}, the effect of forcing is derived from a dynamics-of-observables perspective where it is shown that oscillating the streamwise Reynolds number is equivalent to exciting the Hopf bifurcation parameter. A center-manifold reduction using Koopman modes is described next, followed by further simplification of nonlinearities using normal form theory. Finally, results are presented in Section \ref{results} which establish the accuracy of the reduced-order normal form mathematical models, and provide theoretically grounded evidence of a new dynamical systems phenomena that occurs in the $f<<f_0$ regime.

\section{Cylinder Flow}
\label{cyl_hopf}

Two-dimensional CFD simulations of a streamwise oscillating circular cylinder were considered. For the stationary cylinder, the system can be expressed in the general dynamical systems form given by 
\be
\dot \z=\F(\z),
\label{DSGen}
\ee
where $\z$ is the vector of streamwise and transverse velocity components in the discretized domain, and $\F$ are the discretized incompressible Navier-Stokes equations. When the oncoming streamwise flow is oscillated, the forced Navier-Stokes equations can be written in the format of (\ref{excited_NS}), 
\be
\label{excited_NS}
\dot{\z} = \F(\z) + B\u (t) . 
\ee
The linear operator $B$ allows us to take into account the fact that forcing (i.e. prescribed oscillation of the cylinder) can enter via boundary conditions, like in our problem, and thus the forcing does not necessarily affect all the states in structurally the same way. Similar to the implementation described in \citet{cyl_chaos}, at the domain inlet boundary, the elements of $B\u$ are $2 q \sin (\omega_f t)$ for the streamwise velocity components, and 0 for the transverse components. Furthermore,  $\F(\z) = 0$ is at the inlet boundary. At all other locations in the domain away from the inlet boundary, $B = 0$. The amplitude of the oscillation is controlled by $q$, and the prescribed frequency is $\omega_f = 2 \pi f$.

As in \citet{forced_cyl_2}, the CFD equations are solved in a frame of reference attached to the moving cylinder. Therefore, the resultant oncoming streamwise flow velocity $u_\infty$ over the cylinder is
\be
u_\infty(t) = u_0 - \frac{2q}{\omega_f} \cos \left(\omega_f t \right),
\ee
where $u_0$ is the steady component of the freestream. Since sinusoidal cylinder motions are prescribed, the oscillatory Reynolds number $Re$ is
\be
Re(t) = Re_0 - Re_q \sin \left(\omega_f t \right),
\ee
where $Re_0 = u_0 D/\nu$, $Re_q = 2q D/(\omega_f \nu)$, $\nu$ is the kinematic viscosity, and $D$ is the cylinder diameter. 
The commercial software package, CFD++ was used for all CFD solutions of (\ref{DSGen}) and (\ref{excited_NS}). The CFD++ framework is a finite-volume based solver and is second order accurate in space. Implicit, dual time-stepping was employed.

The connections between Hopf bifurcation dynamics and flow past a stationary cylinder are well known \citep{sreenivasan_cyl}. In such systems, a critical Hopf bifurcation parameter $\mu$ exists such that:

\begin{itemize}

\item For $\mu < 0$: perturbations to the system decay.
\item For $\mu = 0$: the system is neutrally stable and initial perturbations induce oscillations that neither decay nor grow.
\item For $\mu > 0$: initial perturbation amplitudes grow in a transient manner until reaching a saturation value; the system is characterized by limit cycle oscillations for long time. 

\end{itemize}

The Hopf bifurcation limit cycle frequency corresponds to the von K{\'a}rm{\'a}n wake shedding mode and $\mu$ is proportional to $\left( Re_0 - Re_c \right)$, where $Re_c$ is the critical value at which vortex shedding initiates \citep{sreenivasan_cyl}. Without loss of generality, we consider the critical bifurcation value to be $\mu=\mu_0 = 0$. These dynamics can be expressed by the classical Landau equation (i.e. first-order normal form equation),
 \be
\label{nf_2}
\dot{\eta} = \lambda_1 (\mu) \eta + \beta(\mu) \eta^2 \bar{\eta} ,
\ee
where $\eta$ represents a time-dependent state governed by the bifurcation dynamics (e.g. velocity in the cylinder wake, integrated lift force, etc.), $\lambda_1 (\mu)$ is complex valued such that $\mathrm{\Im} [\lambda_1 (\mu)]$ corresponds to the von K{\'a}rm{\'a}n wake shedding frequency, $\mathrm{\Re} [\lambda_1 (\mu)]$ is the associated growth/decay rate, and $\beta(\mu)$ is a complex valued parameter that effects the limit cycle saturation amplitude and frequency. Identification of the parameters in (\ref{nf_2}) as functions of $Re_0$ for stationary cylinder wake shedding was established by \citet{sreenivasan_cyl}.  In the following sections, we extend the standard Hopf bifurcation normal form (\ref{nf_2}) to account for the effects of forcing, $B\u(t)$.


\section{Koopman Mode Decomposition}
\label{koop_methods}

A summary of Koopman modal analysis is provided in this section. The modal analysis stems from a decomposition of an infinite linear dimensional operator that captures the dynamics of an arbitrary observable functional, even if the observable is a nonlinear function of a dynamical system state. By utilizing such an approach, the dynamics of some measurable quantity can be described directly and without linearization of the underlying dynamical system. Additional details can be found in \citet{mezic_annual_reviews}. 

\subsection{Continuous time}
For the general dynamical system (\ref{DSGen}) defined on a state-space $A$ (i.e. $\z\in A$), where $\z$ is a $\mathbb{R}^{\mathrm{N}}$ vector and $\F$ is a possibly nonlinear vector-valued function of the same dimension as its argument $\z$. We denote by $\S^t(\z_0)$ the position at time $t$ of a trajectory corresponding to (\ref{DSGen}) that starts at initial condition $\z_0$. 

An arbitrary vector-valued observable from $A$ to $\mathbb{R}^p$ is denoted by $\g$. The value of $\g$ at time $t$ starting from the system trajectory initial condition $\z_0$ is
 \begin{equation}
 \g(t,\z_0)=\g(\S^t(\z_0)).
 \end{equation}
Note that the space of observables $\g$ is a vector space.  The family of operators $U^t,$ acting on the space of observables parameterized by time $t$ is defined by
 \be
 U^t\g(\z_0)=\g(\S^t(\z_0)).
 \label{Koopdef}
 \ee
Thus, for a fixed time $\tau$, $U^\tau$ maps the vector-valued observable $\g(\z_0)$ to $\g(\tau,\z_0)$. With some abuse of
language we will call the family of operators $U^t$ the Koopman operator of the continuous-time system given by Eqn.~(\ref{DSGen}).
This operator was defined for the first time in \citet{Koopman:1931}, for Hamiltonian systems. In operator theory, such operators, when defined for general dynamical systems,  are often called composition operators, since $U^t$ acts on observables by composing them with the mapping $\S^t$ \citep{sm:1993}.

The operator $U^t$ is linear as can be easily seen from its definition by Eqn.~(\ref{Koopdef}), and thus it makes sense to consider its spectral properties in the context of analyzing Eqn.~(\ref{DSGen}).
In this direction, we will be looking for special observables $\phi(\z):A\rightarrow\mathbb{C}$ on the state space that have the evolution in time given by
 \begin{equation*}
U^t\phi(\z_0)=\phi(\S^t(\z_0))=\exp(\lambda t)\phi(\z_0).
 \end{equation*}
 Such observables (functions) $\phi$ are the eigenfunctions of $U^t$, and the associated complex numbers $\lambda$ are the eigenvalues of $U^t$.

For quasi-periodic attractors, the observables can be expanded onto a basis system spanned by the Koopman eigenfunctions as \citep{mezic:2005,mezic_annual_reviews}
\be
\label{proj}
\g(t,\z_0)  = \sum_{j=1}^{\infty} \phi_j (\z) \s_j = \sum_{j=1}^{\infty} U^{t} \phi_j (\z_0) \s_j = \sum_{j=1}^{\infty} e^{\lambda_j t} \v_j(\z_0) +e^{\bar{\lambda}_j t} \bar{\v}_j(\z_0),
\ee
where $\s_j$ are the Koopman modes and represent the projections of the observables onto the eigenfunctions, $\v(\z_0) = \phi_j (\z_0) \s_j$, and $\bar{\lambda}_j$, $\bar{\v}_j$ are the complex conjugates of $\lambda_j$ and $\v_j$ respectively.

\subsection{Discrete time and computation of Koopman modes}

Due to the discrete nature of numerical simulations or experimental data, the dynamical system can be described by a discrete sequence of state values or observables:
$$\mathbf{z}_{k+1}=\F (\mathbf{z}_k)$$
and a discrete sequence of $U^{k \Delta t};~k = 0;\cdots ;N$ is obtained. It is easy to show that the discrete version of (\ref{proj}) is \citep{mezic:2005,mezic_annual_reviews}
\be
\g(\z_k)  = \sum_{j=1}^{\infty} \phi_j (\z_k) \s_j = \sum_{j=1}^{\infty} U^{k} \phi_j (\z_0) \s_j = \sum_{j=1}^{\infty} \lambda^k_j \v_j(\z_0) +\bar{\lambda}^k_j  \bar{\v}_j(\z_0).
\ee

Koopman modes can in principle be computed directly based on snapshots of the flow, using the Generalized Laplace Analysis \citep{mezic_annual_reviews}. The Koopman Modes can also be approximated by using an Arnoldi-like algorithm \citep{schmid:2008, rowleyetal:2009} which computes eigenvalues based on the so-called companion matrix.

Given a sequence of equispaced in time snapshots from numerical simulations or physical experiments, with $\Delta t$ being the time interval between snapshots, a data matrix is formed, columns of which represent the individual data samples $u_j \in \mathbb{R}^n,~j=0;\cdots;m$ with $j$ representing time $j \Delta t$.  The companion matrix is then defined as:
\begin{equation}\label{eq.comp}
C=\left(
    \begin{array}{ccccc}
      0 & 0 & \cdots & 0 & c_0 \\
      1 & 0 &   & 0 & c_1 \\
      0 & 1 &   & 0 & c_2 \\
      \vdots &  & \ddots &  & \vdots \\
      0 & 0 & \cdots & 1 & c_{m-1} \\
    \end{array}
  \right)
\end{equation}
where $c_i,~i=0,\cdots,m-1$ are such that:
$$
u_m = \sum_{j=0}^{m-1} c_i u_i +r
$$
and $r$ is the residual vector.

The spectrum of the Koopman operator restricted to the subspace spanned by $u_j$ is equal to the spectrum of the infinite-dimensional companion
matrix and the associated Koopman modes are given by $Ka$ (provided that $a$ does not belong to the null space of K), where
$K = [u_0;~ u_1 \cdots;~ u_{m-1}]$ is the column matrix (vector-valued) of observables snapshots at times $0;\Delta t; \cdots; (m-1)\Delta t$ and $a$ is an eigenvector of the shift operator restricted to Krylov subspace spanned by $u_i$ which the companion matrix is an approximation of.
The approximate Koopman eigenvalues and eigenvectors obtained by the Arnoldi's algorithm are sometimes called Ritz eigenvalues and eigenvectors.

The standard Arnoldi-type algorithm to calculate the Ritz eigenvalues $\beta_j$ and eigenfunctions $v_j$ is as follows:
\begin{enumerate}
  \item Define $K = [u_0, u_1 ,\cdots ,u_{m-1}]$.

  \item Find constants $c_j$ such that:
  $$
    r = u_m - \sum_{j=0}^{m-1} c_j u_j = u_m - Kc,~r \bot span\{u_0, . . . , u_{m-1}\}.
    $$
    This can be done by defining $c=K^{+} u_m$ where $K^+$ is the pseudo inverse of $K$.

  \item Define the companion matrix $C$ by Eqn.~(\ref{eq.comp}) and find its eigenvalues and
eigenvectors:
$$
C = T^{-1} \alpha T, \alpha = diag(\alpha_1, \cdots , \alpha_m),
$$
where eigenvectors are columns of $T^{-1}$.
Note that the Vandermonde matrix, $\tilde T$:
$$
\tilde T=\left(
    \begin{array}{ccccc}
      1 & \alpha_1 & \alpha_1^2 &\cdots & \alpha_1^{m-1} \\
      1 & \alpha_2 & \alpha_2^2 &\cdots & \alpha_2^{m-1}  \\
      \vdots & \vdots & \vdots & \ddots & \vdots \\
      1 & \alpha_m & \alpha_m^2 &\cdots & \alpha_m^{m-1}  \\
    \end{array}
  \right)
$$
diagonalizes the companion matrix $C$ , as long as the eigenvalues
$\alpha_1, \cdots , \alpha_m$ are distinct.

  \item Define $v_j$ to be the columns of $V=K\tilde T^{-1}$.
\end{enumerate}

Then, the Arnoldi-type Koopman Mode Decomposition gives:
\begin{equation*}
\forall k=[0,1,\cdots, m],~u_k = \sum_{j=1}^{m} \alpha_j^k V(:,j).
\end{equation*}



To fairly compare the contribution of all the Koopman modes, the eigenvectors of the companion matrix $C$ can be normalized to be unitary i.e the column of $T^{-1}$ are normalized to have norm 1.
Let $N_j=\|T^{-1}(:,j)\|$ 
and $V_N(:,j)= K T^{-1}(:,j) /N_j$, then:
\begin{equation}
\forall k=[0,1,\cdots, m],~u_k = \sum_{j=1}^{m} \alpha_j^k V_N(:,j)N_j.
\end{equation}

\section{Dynamics of Observables with Applied Forcing}
\label{cmnf_section}

The evolution of an observable function of interest subject to applied forcing is developed in this section. The full-order representations are derived first, which explicitly show how forcing effects appear in the dynamics-of-observables equations. In order to obtain a lower order representation that can be used to understand phenomena observed in the forced bifurcation systems, projections onto the unforced system Koopman shedding mode basis are described. This basis corresponds to the center manifold of the unforced system. Finally, the nonlinearity in the reduced-order equations are simplified further for $f << f_0$ to obtain remarkably simple mathematical model equations that explain the salient features of the seemingly complex dynamical states.

\subsection{Full Order Equations}

Considering the observables evolved by the Koopman operator as in (\ref{Koopdef}), we denote 
$$
 \g (t, \z_0)=\g (S^{t}  \z_0).
 $$
Note that $ \g (t, \z_0)$ depends on the initial condition $\z_0$ so we can think of it as a function in the Lagrangian frame - but in the state space, not in the physical space! We have
\bea
\frac{\p \g (t, \z_0)}{\p t}&=& \frac{\p \g (S^{t}  \z_0)}{\p t}
= \nabla \g (S^{t}  \z_0)) \cdot \frac{\p S^{t} \z_0}{\p t} \nonumber \\
&=& \dot{\z} \cdot \nabla \g (t, \z_0) 
= \F(\z) \cdot \nabla \g (t, \z_0) 
\label{Koopevo}
\eea
where (\ref{DSGen}) has been substituted for $\dot{\z}$ and the gradient operator corresponds to the state space. Equation (\ref{Koopevo}) describes the evolution of observables starting from a smooth initial condition given by $\g (0, \z_0)$. In operator form,  (\ref{Koopevo}) can be written as
\be
\label{koop_sl}
\frac{\p \g }{\p t} = \left( \F \cdot \nabla \right) \g = L \g,
\ee
where $L$ is the linear infinite dimensional operator - the generator of Koopman operator evolution - that fully characterizes the evolution of $\g$, which can be a nonlinear function of $\z$.  From (\ref{koop_sl}), the Koopman operator is given by
\be
\label{koop_sl_2}
U^t = e^{Lt}.
\ee
The linkages between the Landau equation (\ref{nf_2}) and the Koopman operator (\ref{koop_sl_2}) for post-critical flow past a stationary cylinder were established in \citet{bagheri:2013}.

Now, an applied forcing $B\u (t),$ where $B$ is a linear operator, and $\u$ is a vector in the same space as $\z$, can be added to the Navier-Stokes equations, as in (\ref{excited_NS}). Substituting the forced system (\ref{excited_NS}) into (\ref{Koopevo}) leads to 
\be
\label{Koopevo_excited}
\frac{\p \g (t, \z_0)}{\p t} = (\F(\z) + B\u (t)) \cdot \nabla \g (t, \z_0) = L \g + \left( B\u \cdot \nabla \right) \g.
\ee
Although the external forcing in the underlying system (\ref{excited_NS}) appears as an additive term, it is clear from (\ref{Koopevo_excited}) that forcing appears as a bi-linear term in $\g$ and $\u$ when considering the time evolution of a general observable. Similarly, it is clear from (\ref{Koopevo_excited}) that additive forcing can only occur when considering dynamics-of-observables in the special case when $\g$ is a linear function of $\z$. Therefore, the original additive forcing will lead to a parametric type excitation in the observable evolution equation for the general case when $\g$ is a nonlinear function of $\z$. Note that the forcing $\u$ could also be thought of as a control input; thus (\ref{Koopevo_excited}) also demonstrates that control studies involving dynamics-of-observables models based on additive forcing in the underlying system must involve a term multiplicative in control $\u$, which in the simplest case reduces to a bi-linear term.

\subsection{Koopman Modes and the Unforced System Reduced-Order Model}


In the vicinity of a non-hyperbolic fixed point in state space,  the long-time behavior can be accurately represented by the dynamics on the lower dimensional center manifold \citep{wiggins}. Therefore, we truncate the projection in (\ref{proj}) by considering a subspace spanned by the Koopman eigenfunction and eigenvalue corresponding to $\mathrm{\Re} [\lambda_j (\mu_0)] = 0$ and $\mathrm{\Im} [\lambda_j (\mu_0)] = \omega_0 (\mu_0) \ne 0$. For flow over a stationary cylinder, this is the well known von K{\'a}rm{\'a}n wake shedding (bifurcation) mode. If the Koopman eigensolutions are ordered such that $j=1$ is the bifurcation mode, then (\ref{proj}) is approximated by the truncated expansion
\be
\label{rop}
\g(t,\z_0)  \approx \gamma_1(t) \v_1(\z_0) + \bar{\gamma_1}(t) \bar{\v}_1(\z_0) = \mathbf{V}  \pmb {\Gamma},
\ee
where $\mathbf{V} = [\v_1, \bar{\v}_1] \in \mathbb C^{p \times 2}$, and $\pmb{\Gamma} = [\gamma_1, \bar{\gamma}_1]^{T} \in \mathbb C^{2 \times 1}$ contains the the complex valued time-dependent coefficients $\gamma_1$ and $\bar{\gamma}_1$ which replace the exponential terms when considering reduced-order approximations \citep{susukiandmezic:2012}. Substituting (\ref{rop}) into (\ref{Koopevo}) and noting that $\z = \g^{-1}(\g(\z))$ leads to the reduced-order model, 
\be
\label{rom_unforced_1}
\dot{\pmb{\Gamma}} =  \left(\mathbf{V}^{\dag} \mathbf{V} \right)^{-1} \mathbf{V}^{\dag} \left[\F \circ \g^{-1} \left(\g \left(\mathbf{V}  \pmb {\Gamma} \right )\right) \cdot \nabla \left(\mathbf{V} \pmb{\Gamma} \right) \right] .
\ee
This two-dimensional reduced-order model can be re-written as 
\be
\label{rom_unforced}
\left[ \begin{array}{c} \dot{{\gamma_1}} \\ \dot{\bar{\gamma_1}} \end{array} \right] =\left[ \begin{array}{cc} \lambda_1 (\mu) & 0 \\ 0 & \bar{\lambda}_1 (\mu) \end{array} \right]  \left[ \begin{array}{c} \gamma_1 \\ \bar{\gamma_1} \end{array} \right]  + \hat{\F} (\gamma_1, \bar{\gamma_1}, \mu) 
\ee
where the right-hand side of (\ref{rom_unforced_1}) has been expressed as a linear term plus a nonlinear term, i.e.
\be
 \left(\mathbf{V}^{\dag} \mathbf{V} \right)^{-1} \mathbf{V}^{\dag} \left[\F \circ \g^{-1} \left(\g \left(\mathbf{V}  \pmb {\Gamma} \right )\right) \cdot \nabla \left(\mathbf{V} \pmb{\Gamma} \right) \right] =  \left[ \begin{array}{cc} \lambda_1 (\mu) & 0 \\ 0 & \bar{\lambda}_1 (\mu) \end{array} \right]  \left[ \begin{array}{c} \gamma_1 \\ \bar{\gamma_1} \end{array} \right]  + \hat{\F} (\gamma_1, \bar{\gamma_1}, \mu) .
\ee
Equation (\ref{rom_unforced}) can be thought of as the projection of (\ref{Koopevo}) onto the center manifold spanned by the basis corresponding to the primary Koopman shedding (bifurcation) mode at $\mu = \mu_0$. Since $\gamma_1$ and $\bar{\gamma}_1$ in (\ref{rom_unforced})  correspond to complex conjugate states, it is sufficient to solve one of the complex valued differential equations,
\be
\label{rom_unforced_g1}
\dot{\gamma}_1 = \lambda_1 (\mu) \gamma_1 + \hat{F}_1 (\gamma_1, \bar{\gamma}_1, \mu) ,
\ee
where $\hat{F}_1$ is the first element of the vector $\hat{\mathbf{F}}$. It is important to note that, (\ref{rom_unforced}) - (\ref{rom_unforced_g1}) are expressed as functions of the bifurcation parameter $\mu$, even though the reduced order projections are based on the Koopman modes at $\mu = \mu_0$. This is valid for values of $\mu$ in the vicinity of $\mu_0$ for dynamics projected onto the center manifold; i.e. the Koopman modes at $\mu = \mu_0$ are a valid projection basis for all values of $\mu$ in the vicinity of $\mu_0$ since the orbit of the full order system (\ref{koop_sl}) near $\mu_0$ is determined by the solution restricted to the center manifold \citep{wiggins}. Furthermore, it should be noted that for $\mu=\mu_0$, $\hat{F}_1=0$ and (\ref{rom_unforced_g1}) is linear.

\subsection{Koopman Modes and the Forced System Reduced-Order Model}

The appropriate basis for model order reduction of the forced Hopf bifurcation system is considered next. We make the argument that the unforced Koopman modes at the Hopf bifurcation value of the Reynolds number (i.e. $\mu=\mu_0$) can be used as the projection bases, even for the forced  system. From (\ref{Koopevo_excited}), the dynamics of the forced system are described by the unforced linear operator $L$ plus an additional time-variant forcing term that is bi-linear in $\u$ and $\g$. We know that the unforced Koopman eigenvalues and eigenfunctions satisfy
\be
\label{L_eig}
L \phi_i(\z) = \lambda_i \phi_i(\z), i =0, 1, 2, \ldots, \infty.
\ee
As a result, if we select a basis spanned by unforced system Koopman eigenfunctions, as in (\ref{proj}), then the full action of $L$ on $\g$ in (\ref{Koopevo_excited}) will be captured. If the forcing term is properly projected onto the new basis, then $\dot{\g}$ will be represented without any approximation. Therefore, the modes computed from the unforced system can be used as an appropriate basis for model order reduction of the forced system. As we will show to be the case in Section~\ref{osc_cyl_results}, this implies that the spatial modal structures of the unforced bifurcation system persist into the forced case, though the corresponding time dependence $\gamma_1(t)$ will differ between the two cases. However, just as in the case for the unforced reduced-order model, truncation to a finite number of modes will introduce error. Therefore, one would still need to be careful to include modes that may not be significant in unforced conditions, but become relevant when the forcing is applied. Since we are primarily interested in the shedding dynamics under forcing, we restrict our projection to just the primary bifurcation mode (and its complex conjugate), just as in the unforced case.

Proceeding with the reduced-order model development for the forced system, the expansion (\ref{rop}) is substituted into (\ref{Koopevo_excited}), which leads to
\be
\label{rom_forced}
\left[ \begin{array}{c} \dot{{\gamma_1}} \\ \dot{\bar{\gamma_1}} \end{array} \right] =\left[ \begin{array}{cc} \lambda_1 (\mu) & 0 \\ 0 & \bar{\lambda}_1 (\mu) \end{array} \right]  \left[ \begin{array}{c} \gamma_1 \\ \bar{\gamma_1} \end{array} \right]  + \hat{\F} (\gamma_1, \bar{\gamma_1}, \mu) + \left(\mathbf{V}^{\dag} \mathbf{V} \right)^{-1} \mathbf{V}^{\dag} \left [ \mathbf{u} \cdot \nabla (\mathbf{V} \pmb{\Gamma}) \right ].
\ee
In this study, we assume simple harmonic time-dependance for $\u$ with frequency $\omega_f = 2 \pi f$. When the elements of $\mathbf{u}$ corresponding to the streamwise velocity components equal $2 q \sin (\omega_f t)$, and all other elements equal zero, then  
\be
\label{rom_forced_1}
\dot{\gamma}_1 = \lambda_1 (\mu) \gamma_1 + \hat{F}_1 (\gamma_1, \bar{\gamma}, \mu) + 2 q \sin(\omega_f t) \left[ \gamma_1 a  + \bar{\gamma}_1 \bar{a} \right ] ,
\ee 
where $a(\z_0)$ corresponds to the first element of the vector $\left (\mathbf{V}^{\dag} \mathbf{V} \right)^{-1} \mathbf{V}^{\dag} \sum_{i=1}^{N} \partial{\v_1} / \partial{z_i}$, and $\bar{a}(\z_0)$ is its complex conjugate. Furthermore, we replace the explicit time dependence in the forcing term by an additional state, $\zeta = e^{i \omega_f t}$ such that (\ref{rom_forced_1}) becomes 
\be
\label{rom_final}
\dot{\gamma}_1 = \lambda_1 (\mu) \gamma_1 + \hat{F}_1 (\gamma_1, \bar{\gamma}_1, \mu) - i q (\zeta - \bar{\zeta}) \left( \gamma_1 a  + \bar{\gamma}_1 \bar{a} \right )  .
\ee
Note the bi-linear terms involving $\zeta$ and $\gamma_1$ originate from $\u \cdot \nabla \g$ in (\ref{Koopevo_excited}).

\subsection{Normal Form Theory}
\label{nf_section}

Further simplification of the reduced-order equation (\ref{rom_final}) is possible by eliminating nonlinear terms that are not necessary to retain in the ordinary differential equation. Instead, the eliminated terms can be added after solving the simplified differential equation to obtain the solution for $\gamma_1 (t)$. To do this, we apply normal form theory \citep{gh, wiggins,nayfeh_normal_forms}. As we will show, normal form theory leads to explicit necessary conditions for the forcing to induce spectral broadening, along with the conditions for which chaos can be ruled out \emph{a priori} as a first order phenomena. To derive the normal form of (\ref{rom_final}), we assume the following near identity transformations \citep{nayfeh_normal_forms}
\begin{equation}
\label{zeta}
\gamma_1 =   \eta + \epsilon^1 f_1 (\eta, \bar{\eta}, \zeta, \bar{\zeta}, \mu) + O(\epsilon ^2) + \ldots,
\end{equation}
and
\begin{equation}
\label{eta_dot}
\dot{\eta} = i \omega_0 (\mu) \eta + \epsilon^1 h_1 (\eta, \bar{\eta}, \zeta, \bar{\zeta}, \mu) + O(\epsilon ^2) + \ldots
\end{equation}
The small parameter $\epsilon$ is a book-keeping parameter used to track the order of the normal form expansion. In normal form theory, the transformation function $f_1$ is chosen to eliminate all non-resonant terms after substitution of (\ref{zeta}) into (\ref{rom_final}), while resonant terms proportional to $e^{i \omega_0 t}$ and near-resonant terms proportional to $e^{i (\omega_f+\omega_0 \approx \omega_0) t}$ are retained in $h_1$ \citep{gh, nayfeh_normal_forms}. The solution for $\gamma_1$ is obtained by solving the simplified ordinary differential equation $(\ref{eta_dot})$ for $\eta(t)$, and then substituting the solution into (\ref{zeta}).  In this process, transformation functions are selected to eliminate terms of progressively higher order; i.e. $f_1$ is first chosen to eliminate $\epsilon^1$ terms, then $f_2$ is chosen to eliminate remaining non-resonant terms proportional to $\epsilon^2$, etc.  In classical unforced Hopf bifurcation problems, the leading order terms proportional to $\epsilon^1$ are typically sufficient to accurately approximate the salient features of the dynamics, i.e. transient growth followed by nonlinear saturation into a limit cycle attractor \citep{wiggins}. 

Substituting (\ref{zeta}) and (\ref{eta_dot}) into (\ref{rom_final}) and retaining terms proportional to $\epsilon^1$ gives the homological equation which need to be solved for $f_1$ and $h_1$  \citep{nayfeh_normal_forms}, 
\begin{equation}
\label{homological}
h_1 + \mathcal{L} (f_1) =  \sigma(\mu) \eta + \hat{F_1}(\eta, \bar{\eta}, \mu) - i q (\zeta - \bar{\zeta}) \left( \eta a  + \bar{\eta} \bar{a} \right ) 
\end{equation}
in which $\sigma = \Re[\lambda_1]$, and $\mathcal{L}$ is the Lie operator
\begin{equation}
\label{Lie}
\mathcal{L}(f_1) = i \omega_0 \left (\eta \frac{\partial f_1}{\partial \eta} - \bar{\eta} \frac{\partial f_1}{\partial \bar{\eta}} - f_1 \right) + i \omega_f \left (\zeta \frac{\partial f_1}{\partial \zeta} - \bar{\zeta} \frac{\partial f_1}{\partial \bar{\zeta}} \right).
\end{equation}
Note that we have used the chain rule for $\dot{h}_1$ and $\dot{\zeta} = i \omega_f \zeta$ to obtain the Lie operator.

To solve (\ref{homological}), a polynomial expansion for the nonlinear function $f_1$ can be assumed. For example, 
\begin{equation}
\label{f_1}
f_1 = \Lambda_1 \eta^2 + \Lambda_2 \bar{\eta}^2 + \Lambda_3 \eta \bar{\eta} + \Lambda_4 \eta \zeta + \Lambda_5 \eta \bar{\zeta}+ \Lambda_6 \eta^3 + \Lambda_7 \bar{\eta}^3 + \Lambda_{8} \bar{\eta}^2 \eta + \Lambda_{9} \eta^2 \bar{\eta} + \ldots.
\end{equation}
The complex coefficients, $\Lambda_i$, are selected to eliminate corresponding polynomial terms on the right hand side of (\ref{homological}). However, resonant terms proportional to $e^{i \omega_0 t}$, such as the term corresponding to $\eta^2 \bar{\eta}$, cannot be eliminated from the right hand side of (\ref{homological}) no matter what coefficient selection is made because they identically vanish in (\ref{Lie}); e.g. $\mathcal{L}(\Lambda_9 \eta^2 \bar{\eta}) = 0$ for any choice of $\Lambda_9$. In this manner, resonant terms in the right hand side of (\ref{homological}) cannot be eliminated by $f_1$ and must be retained in $h_1$. 

One approach to solving (\ref{homological}) would be to only eliminate terms that do not involve the forcing, i.e. assume $f_1$ is of the form $f_1 (\eta, \bar{\eta}, \mu$) instead of the more general $f_1 (\eta, \bar{\eta}, \zeta, \bar{\zeta}, \mu$). We refer to this approximation as a \emph{semi normal form} since potentially non-resonant terms due to coupling between $\zeta$ and $\eta$ are retained. When considering leading order approximations, the semi-normal form is
\be
\label{semi_nf}
\dot{\eta} = \lambda_1 (\mu) \eta + \beta(\mu) \eta^2 \bar{\eta} - i q (\zeta - \bar{\zeta}) \left( \eta a  + \bar{\eta} \bar{a} \right ).
\ee
One arrives at the cubic term $\eta^2 \bar{\eta}$ by recognizing that this leading order resonant term is universal to Hopf bifurcation dynamics when $q=0$ \citep{wiggins}. Other non-resonant nonlinearities, such as terms proportional to $\eta^2$ and $\eta \bar{\eta}^2$ for example, are retained in $f_1$.  When $q=0$, (\ref{semi_nf}) is typically referred to as the Landau equation and accurately describes dynamics associated with the von K{\'a}rm{\'a}n wake shedding mode \citep{sreenivasan_cyl, HuerreandMonkewitz:1990}. Therefore, the Landau equation describing Hopf bifurcation dynamics corresponds to a leading order normal form expansion. It is important to note that the semi-normal form (\ref{semi_nf}) is valid for \emph{any} forcing frequency since no assumptions on $\omega_f$ were made. Although the cubic structure of (\ref{semi_nf}) is universal to any Hopf bifurcation problem, explicit expressions for $\beta$ as a function of $\mu$ must be estimated empirically when explicit expressions for $\hat{F}_1$ are not available.

Assumptions on $\omega_f$ are necessary in order to further simplify (\ref{semi_nf}) by considering $f_1 (\eta, \bar{\eta}, \zeta, \bar{\zeta}, \mu$). We refer to this approach as the \emph{full normal form}. The full normal form reduction requires the identification of near resonant terms associated with $\omega_f$. These terms must be retained in $h_1$ because they are proportional to $e^{i (\omega_f+\omega_0 \approx \omega_0) t}$, and thus indistinguishable from resonant terms. Clearly, identification of such terms is specific to the value of $\omega_f$ relative to $\omega_0$. For example, the $\zeta \eta$ term appearing in the homological equation (\ref{homological}) will be proportional to $e^{i({\omega_0 + \omega_f})t}$ to a leading order. Such a term is only near-resonant when $\omega_f << \omega_0$. Similar arguments would apply to the $\eta \bar{\zeta} $ term. Equivalently, terms that must be retained in the normal form ordinary differential equation can be identified by recognizing transformation function coefficients $\Lambda_i$ that are singular, or near-singular. \citep{nayfeh_normal_forms, Tsarouhas_1}. If there is no separation in scale between $\omega_f$ and $\omega_0$, then an appropriate coefficient in (\ref{f_1}) could be selected to eliminate the $\eta \zeta$ term from the right hand side of (\ref{homological}). Substituting (\ref{f_1}) into (\ref{homological}) shows that the coefficient is
\be
\label{lambda_singular}
\Lambda_4 = -\frac{q a}{\omega_f} = -\frac{q a}{\omega_0(1-\alpha)}.
\ee
where $\alpha = 1 - f/f_0$.  If $f <<  f_0$, then $\alpha$ is $\mathcal{O}(1)$ and (\ref{lambda_singular}) is near-singular, which would lead one to retain the term proportional to $\eta \zeta$ in $h_1$. Therefore, when $f << f_0$, the full normal form approximation is
\be
\label{nf_1}
\dot{\eta} = \lambda_1 (\mu) \eta + \beta(\mu) \eta^2 \bar{\eta} - i Q (\zeta - \bar{\zeta}) \eta.
\ee
where $Q = Q_R + i Q_I = q a$. For other frequency scenarios, $\zeta \eta$ and $\bar{\zeta} \eta$ would be non-resonant terms and thus would appear in $f_1$ rather than in the differential equation (\ref{eta_dot}). These scenarios would lead to the full normal form differential equation (\ref{nf_2}) while the effects of the forcing (up to a leading order) would be captured as additive nonlinearities in (\ref{zeta}). 

Although the full normal forms, (\ref{nf_1}) and (\ref{nf_2}), are less general than the semi-normal form since (\ref{semi_nf}) is valid for any $f$, the full normal forms provide theoretical insight into when the forcing can induce broadly distributed spectral content. When there is no separation in scale between $f$ and $f_0$, the spectrum corresponding to the first order approximation of $\gamma_1(t)$ (and thus $\g(t, \z_0)$) would be narrowly concentrated about $f$, $f_0$, and $f_0 \pm f$. This is because the spectral content resulting from the solution of (\ref{nf_2}) would correspond to the limit cycle frequency and combinations of $f$ and $f_0$ that arise from the transformation function $f_1$.  Furthermore, the leading order normal form differential equation (\ref{nf_2}) is two-dimensional, and thus could never be chaotic. Therefore, the full normal form approximation shows that $f << f_0$ is a necessary (though not sufficient) condition for the simple harmonic forcing to induce broadly distributed spectra as leading order phenomena in the observable since their occurrence could not be ruled out \emph{a priori} from (\ref{nf_1}) when $f << f_0$. 


Replacing the complex variable $\eta$ by the polar coordinate representation $\eta = r(t) e^{i \theta(t)}$ and $-i (\zeta - \bar{\zeta}) = 2 \sin (\omega_f t)$ in (\ref{nf_1}) gives
\be
\label{rdot}
\dot{r} = \sigma(\mu) r + \beta_R (\mu) r^3 + 2 r Q_R \sin(\omega_f t)
\end{equation}
and
\begin{equation}
\label{theta_dot}
\dot{\theta} = \omega_0(\mu) + \beta_I r^2 +  Q_I \sin(\omega_f t)
\end{equation}
where $\beta_R$ and $\beta_I$ are the real and imaginary parts of $\beta$. It can be shown that this normal form system is analogous to a nonlinear mechanical oscillator with a parametrically excited damper corresponding to $2 Q_R \sin(\omega_f t)$, and a parametrically driven spring stiffness associated with $Q_I\sin(\omega_f t)$. Furthermore, for the range of parameters corresponding to cylinder flow considered in this study, $\omega_0 + \beta_I r^2 >> Q_I $. Thus, the forcing primarily affects the dynamics through the parametrically excited damping. Note that if $\g$ is linear in $\z$, then the forcing would not appear as a radial excitation in (\ref{rdot}). 

Interestingly, a similar set of equations as (\ref{rdot}) and (\ref{theta_dot}) were studied by  \citet{LS_young_2003, LS_young_2008} in which it was shown that \emph{shear induced chaos} could be induced. However, a fundamental difference between the shear induced chaos attractor and (\ref{rdot}) and (\ref{theta_dot}) is that the equation governing $\dot{r}$ in \citep{LS_young_2008} contained functional dependance on $\theta$ in the bi-linear forcing term. Clearly, this is not the case for (\ref{rdot}) and leads to different behavior compared to the shear induced chaos attractor. As we will show, (\ref{rdot}) and (\ref{theta_dot}) do not lead to chaos, and in fact describe a fundamentally different phenomena.

\section{Results}
\label{results}

\subsection{Koopman Decompositions}

\subsubsection{Stationary Cylinder}

For the CFD simulations, the cylinder was 0.002 m in diameter and the time step was 1e-4 s. The simulation was run for 60000 time steps, which was sufficient to allow the near-wake dynamics to settle into the well known post-bifurcation limit cycle behavior. The oncoming freestream was $Re_0 = 53$. A low freestream Reynolds number was selected so that turbulence could be neglected. The CFD data corresponding to the first 525 time steps were discarded in order to eliminate numerical transients. The remaining data record was used to conduct the Koopman modal analysis. Note that the CFD cylinder simulation exhibited the von K{\'a}rm{\'a}n vortex instability at $Re_c = 46$, which is in good agreement with experimental results presented in the literature \citep{sreenivasan_cyl}. In addition, the calculated vortex shedding frequency at $Re_0 = 53$ was 26 Hz, which is consistent with empirical relations from the literature. For instance, the Roshko number from our simulations is $f_0 D^2/\nu = 7$ at $Re_0 = 53$, which agrees with the empirical relationship from \citet{sreenivasan_cyl} in which $f_0 D^2/\nu = 5.46 + 0.21 \left(Re_0  - Re_c  \right) = 7$.

\begin{figure*}
 \begin{subfigmatrix}{1}
  \subfigure[Using normalization]{\includegraphics{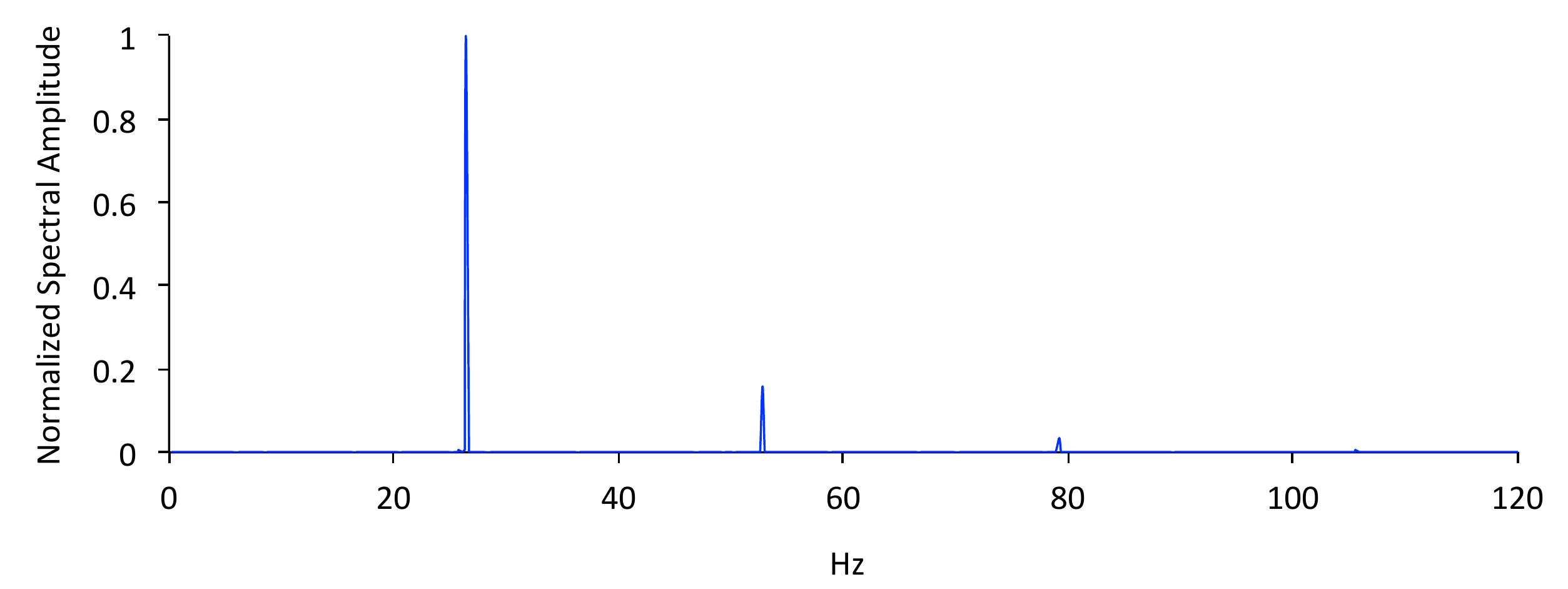}}
 \subfigure[Direct data]{\includegraphics{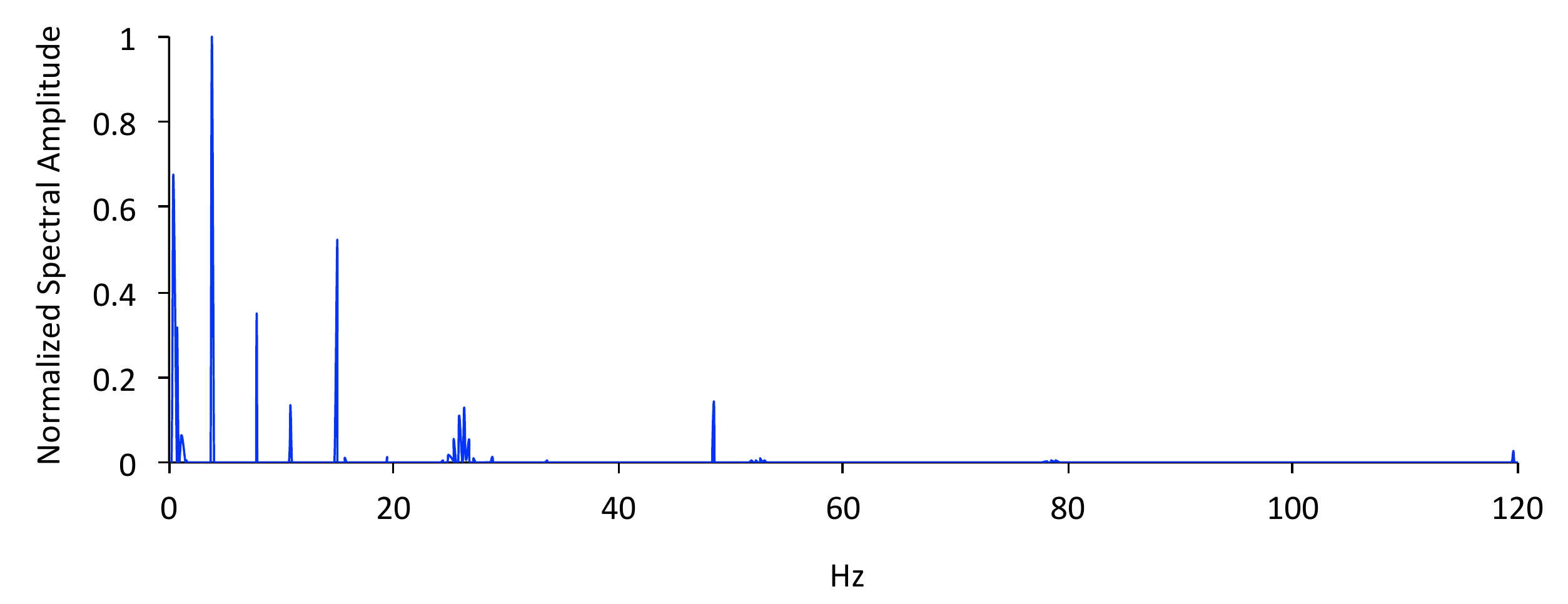}}
  \end{subfigmatrix}
 \caption{Imaginary components of Koopman eigenvalues corresponding to streamwise velocity field or flow over a stationary cylinder; the mean is removed for clarity and spectral amplitudes are normalized such that the maximum value is 1.}
 \label{cyl_stationary}
\end{figure*}

\begin{figure*}
 \begin{subfigmatrix}{2}
   \subfigure[Using normalization]{\includegraphics{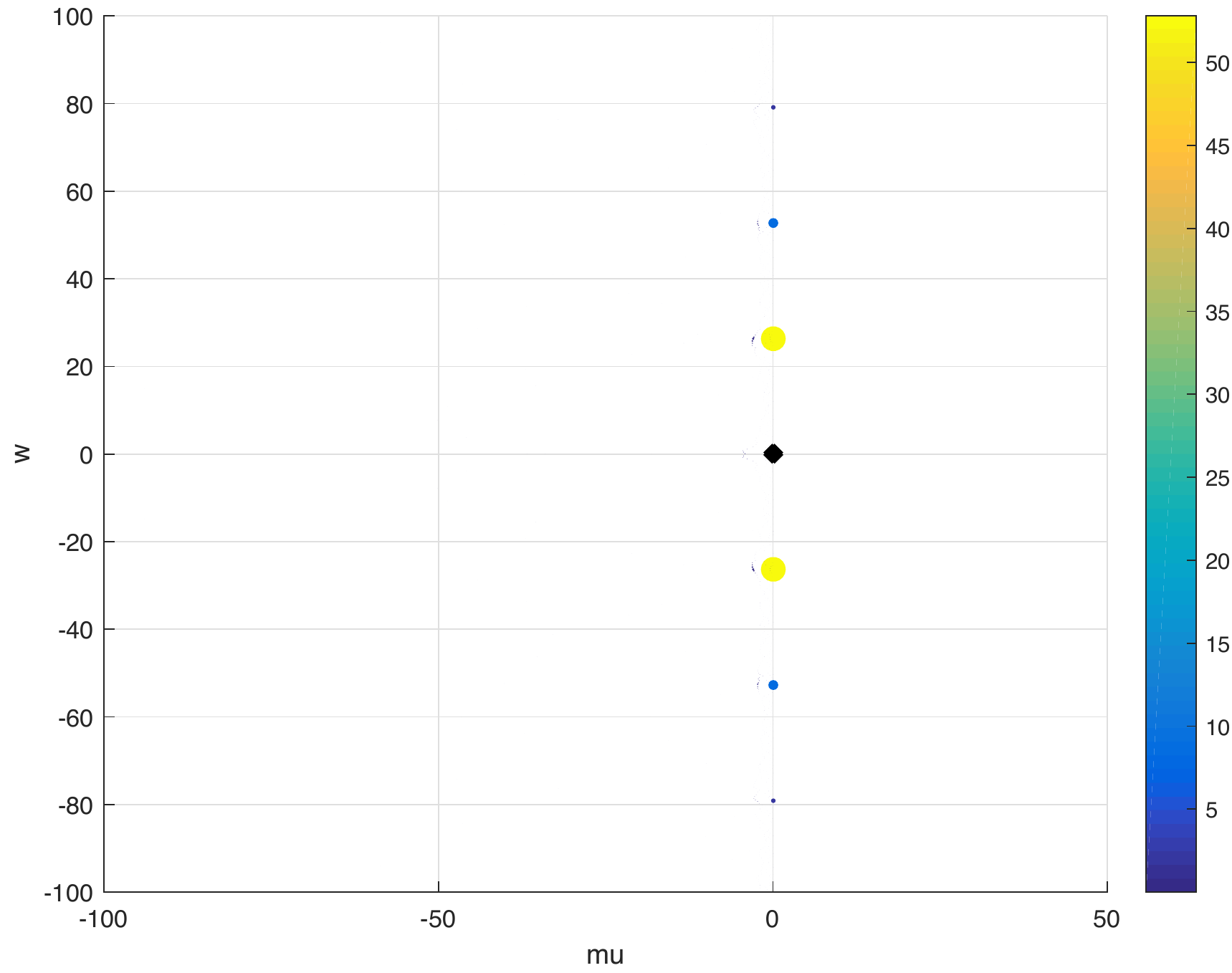}}
  \subfigure[Direct data] {\includegraphics{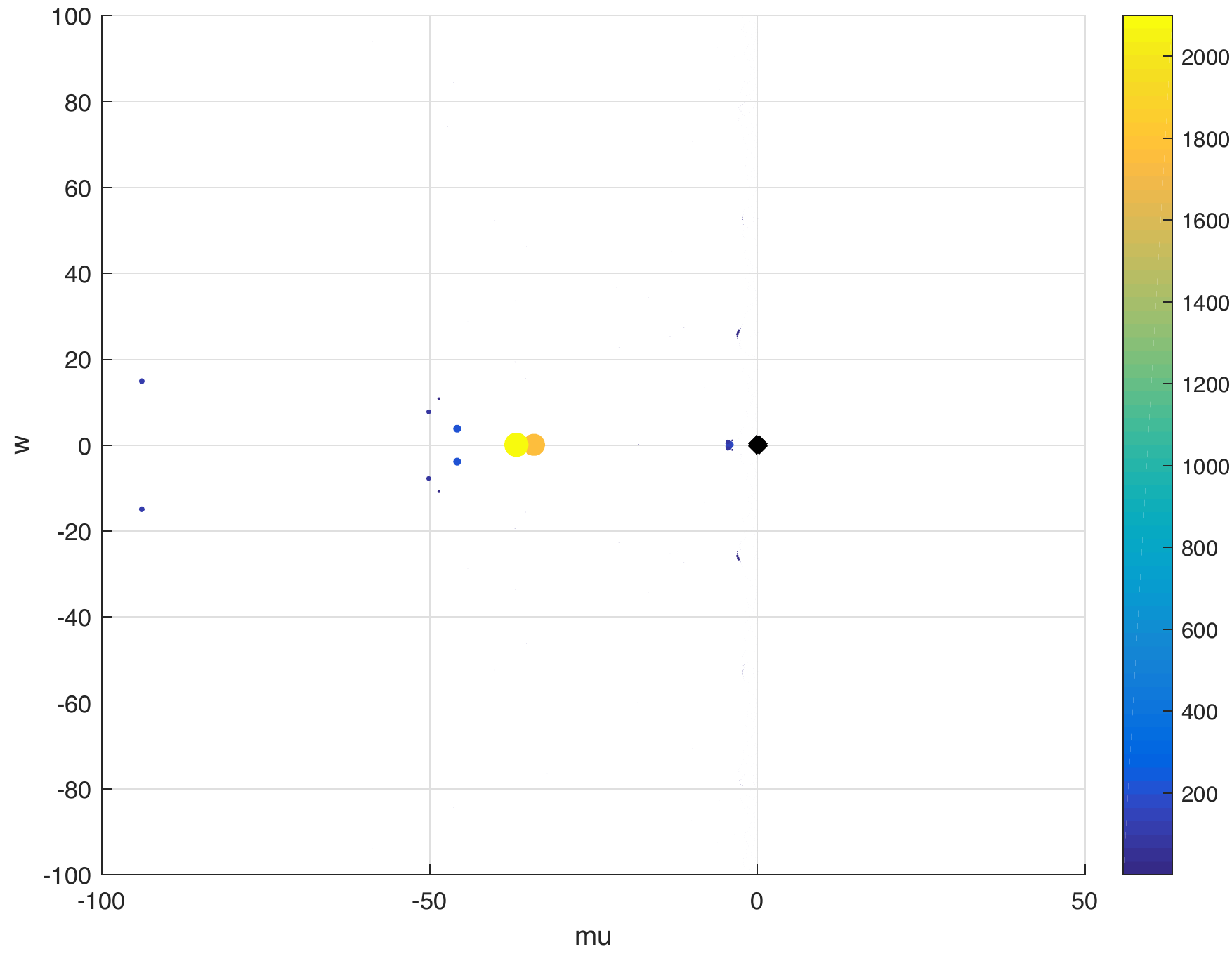}}
  \end{subfigmatrix}
\caption{Koopman eigenvalues in the complex plane for the streamwise velocity component; $\Re [\lambda]$ on the horizontal axis and $\Im [\lambda]$ on the vertical axis (Hz).}
\label{koop_eigenvalues_cyl}       
\end{figure*}

The generalized harmonic analysis -- approximated by finite-time Fast Fourier Transforms -- gives projections that enable computation of Koopman modes. In addition, DMD yields such projections as well, so the two approaches are compared along with the effects of normalization. Koopman decomposition spectra calculated using normalization, and directly from the CFD data are compared in Figs. \ref{cyl_stationary}(a) and \ref{cyl_stationary}(b) respectively. The dominant mode at the von K{\'a}rm{\'a}n shedding frequency of 26 Hz is captured when using normalization. We have also verified that the normalization recovers a spectrum closer to that predicted by Fourier analysis at local probe locations in the wake. However, the spectrum computed from the direct data is somewhat contaminated by low frequency content and predicts maximum spectral content at 4 Hz. The Koopman eigenvalues computed from the direct data set and using normalization are shown in Fig.~\ref{koop_eigenvalues_cyl}. The eigenvalues are colored and sized by the magnitude of each mode; i.e. larger magnitude modes appear as larger circles. The dominant eigenvalues computed using  normalization are on the imaginary axis and consist of the base frequency and harmonics. In contrast, the eigenvalues computed from the direct data set show significant contributions from non-periodic modes close to 0 Hz. Even if all Koopman modes are on the attractor, i.e $\Re [\lambda] = 0$, the numerical computation will give values slightly different than 0, as shown in Fig.~\ref{koop_eigenvalues_cyl}. This is amplified for the Koopman modes based on the direct data because the normalized Koopman modal analysis diminishes the importance of the non-periodic modes. The results in Fig. \ref{cyl_stationary} and Fig. \ref{koop_eigenvalues_cyl} indicate that Koopman modal decompositions based on normalization is better suited for analyzing dynamics on a the attractor, while utilizing the direct data set may yield better results for transient dynamics. Normalization is used throughout the remainder of the paper since long-time dynamics are of interest. 

\begin{figure*}
 \begin{subfigmatrix}{2}
  \subfigure[0 Hz]{\includegraphics{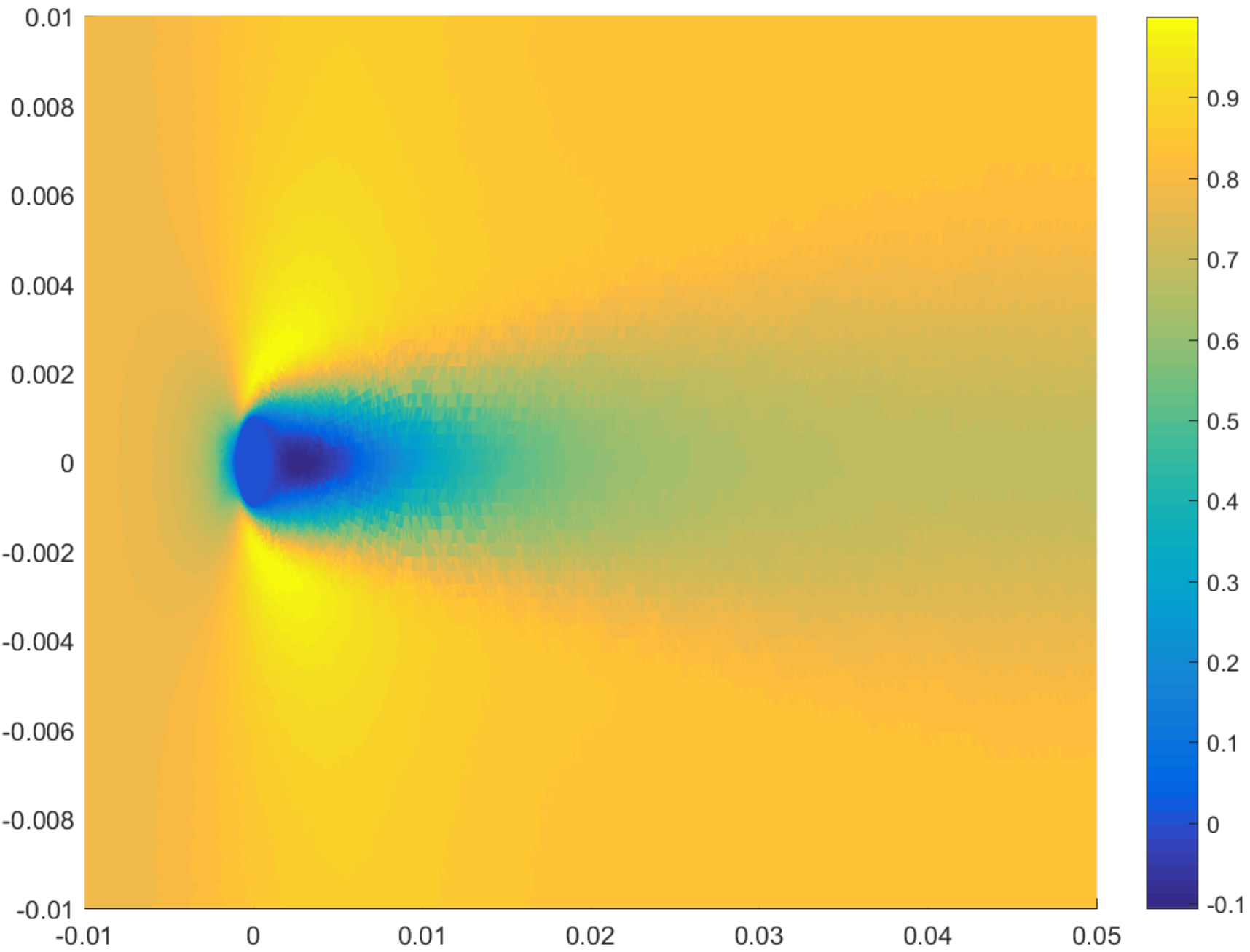}}
 \subfigure[26 Hz]{\includegraphics{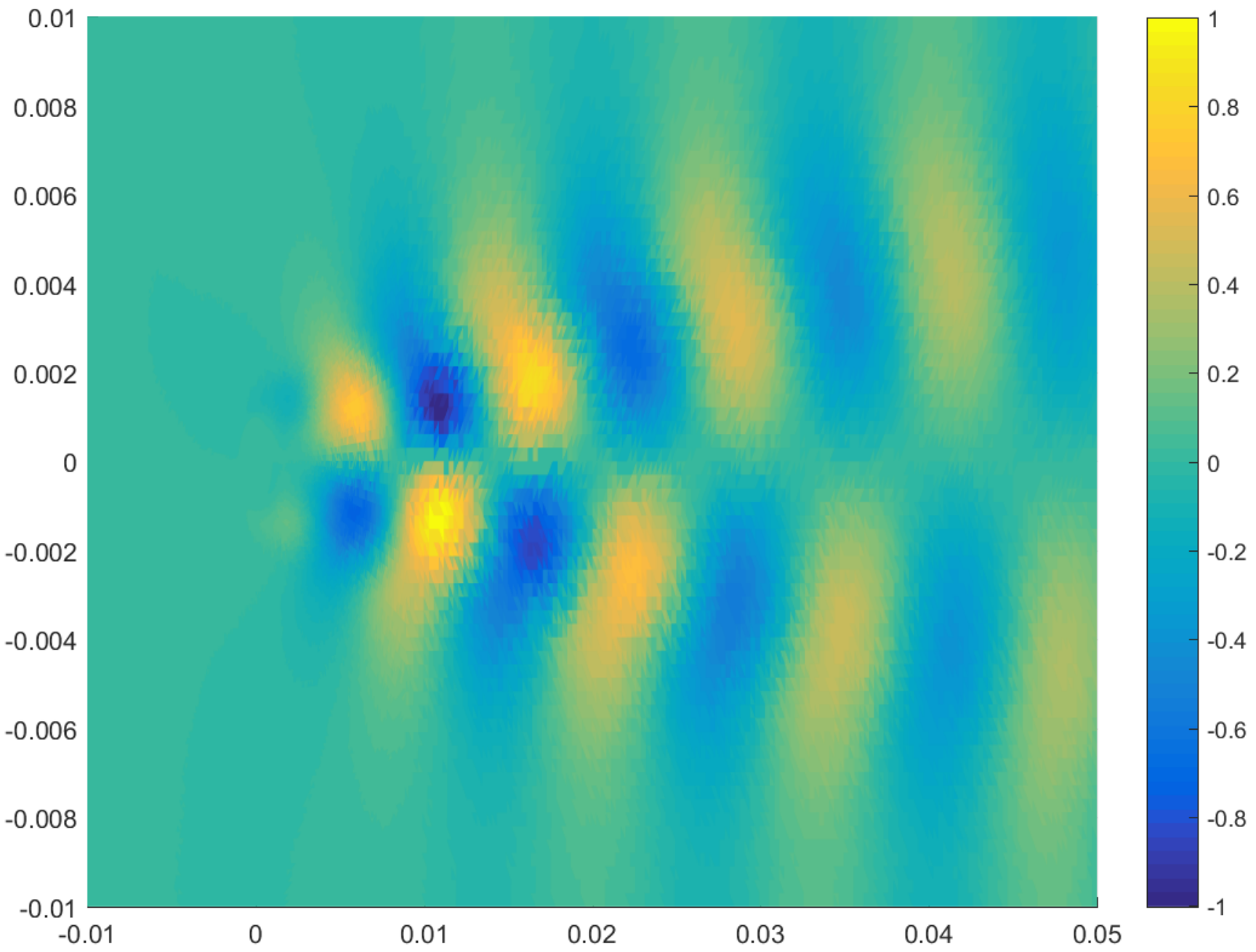}}
 \subfigure[53 Hz]{\includegraphics{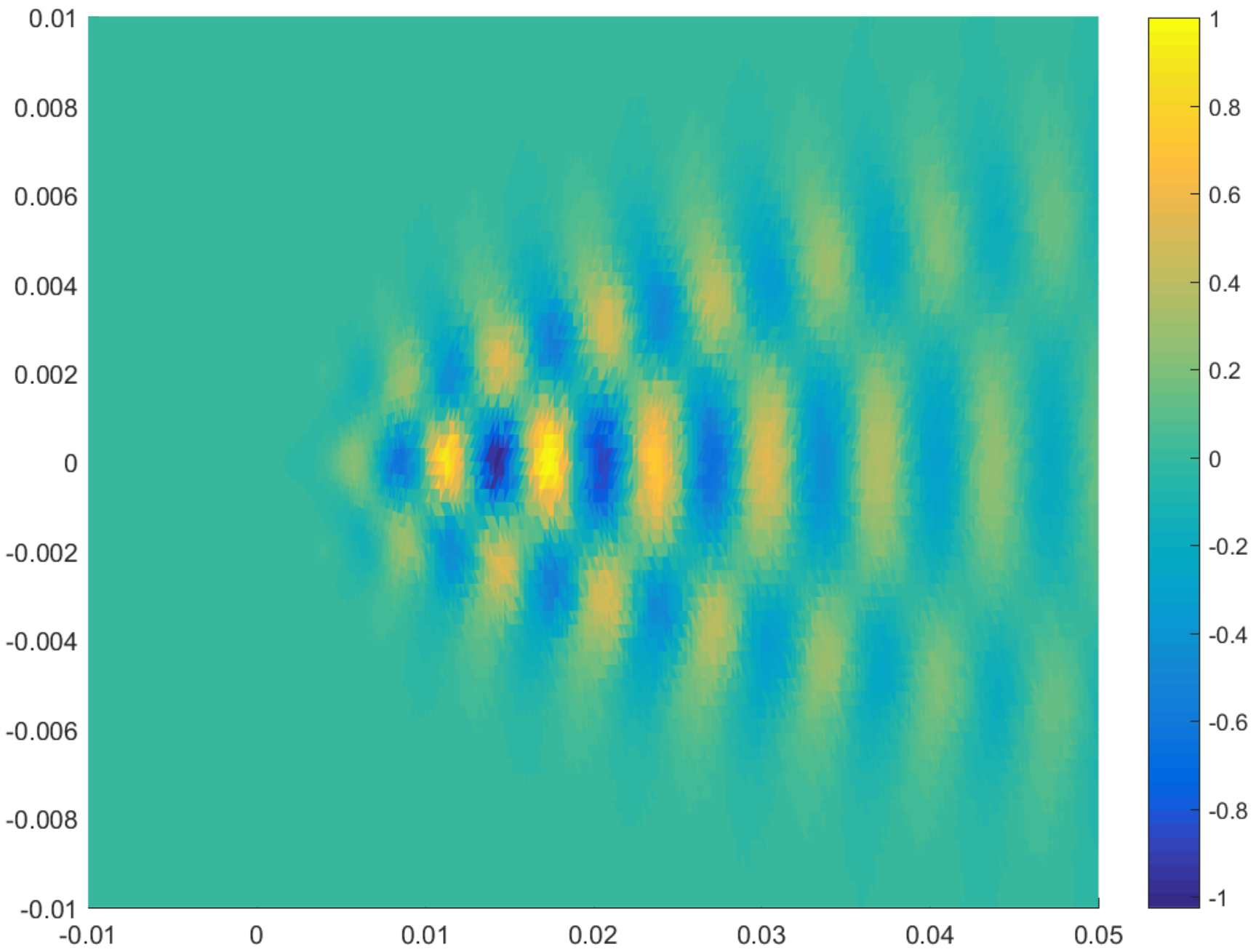}}
 \subfigure[79 Hz]{\includegraphics{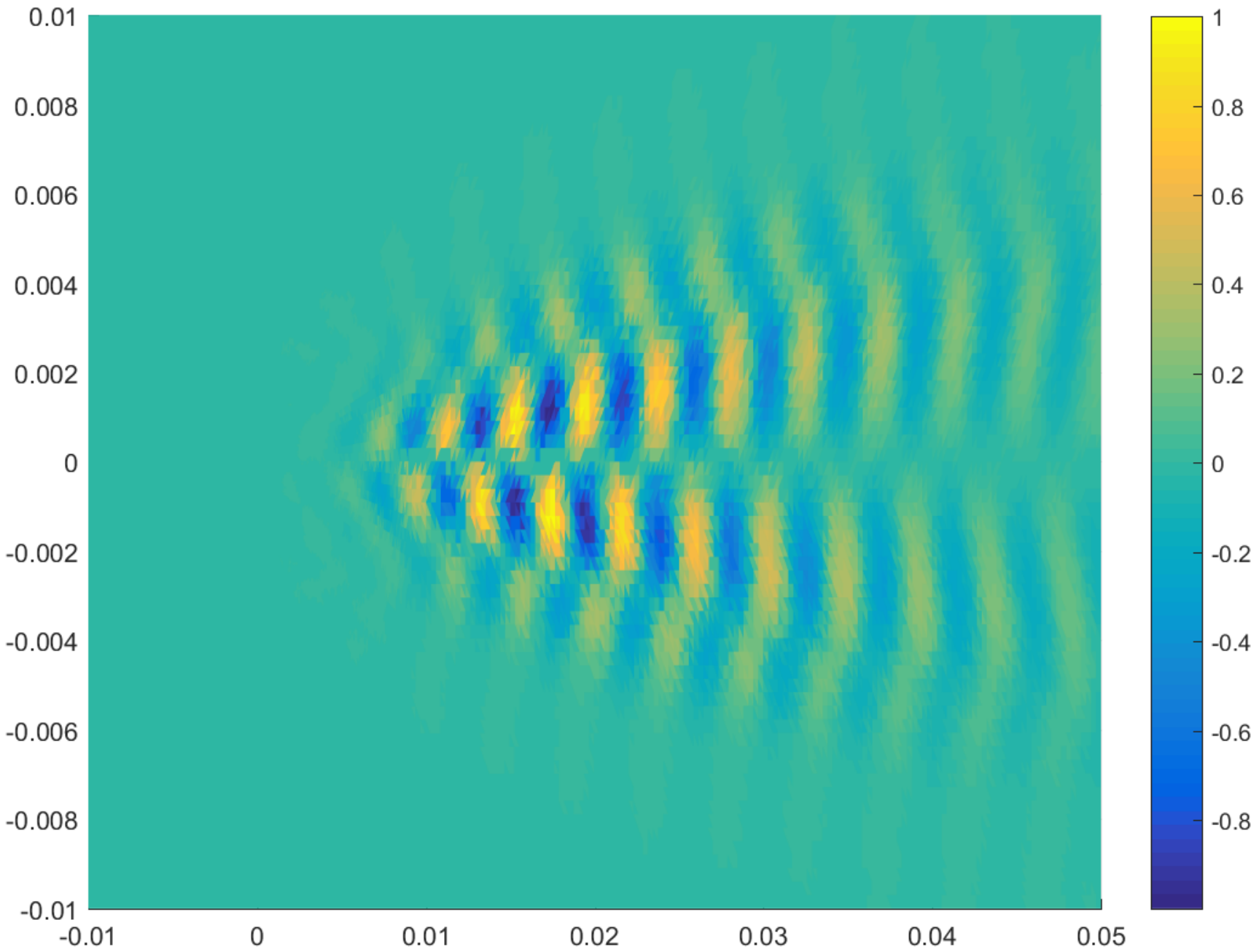}}
  \end{subfigmatrix}
 \caption{Real parts of Koopman modes corresponding to streamwise velocity for the stationary cylinder. Axes correspond to spatial coordinates in meters; the 0.002 m diameter cylinder is centered at (0,0) and the flow moves from left to right. All mode shapes are normalized such that the maximum value is 1.}
 \label{cyl_koop_modes}
\end{figure*}

Streamwise velocity Koopman modes corresponding to the mean (0 Hz), the von K{\'a}rm{\'a}n shedding frequency (26 Hz), and higher harmonics of the shedding frequency (53 and 79 Hz) are shown in Fig.~\ref{cyl_koop_modes}. The modes were  normalized. The 26 Hz Koopman mode clearly reproduces the anti-symmetric character associated with von K{\'a}rm{\'a}n wake vortex shedding, while the higher harmonic shedding modes exhibit finer scale spatial structure. Note that only the real parts of the Koopman modes are shown since the imaginary parts exhibit similar spatial structure. The spatial structures shown in Fig.~\ref{cyl_koop_modes} match with previously reported Koopman decompositions of the stationary cylinder wake \citep{chenetal:2012, bagheri:2013}.

\subsubsection{Oscillating Cylinder}
\label{osc_cyl_results}

The streamwise velocity oscillations were prescribed at $f = 0.5$ Hz for $Re_0 = 53$, which is two orders of magnitude slower than the shedding frequency for the stationary cylinder. As a result of the separation of scales associated with $f << f_0$, we focus on dynamical states other than lock-on/synchronization phenomena that occur when $f \sim f_0$. Oscillation amplitudes from $Re_q = 1$ to $Re_q = 20$ were considered in order to compare large amplitude cases which oscillate through $Re_c$, and smaller amplitude cases which are above $Re_c$ throughout the oscillation. The oscillating cylinder cases were computed for a total time record of twelve forcing periods. 

\begin{figure}
 \begin{subfigmatrix}{1}
  \subfigure[$Re_q = 1$]{\includegraphics[width = 1 \textwidth]{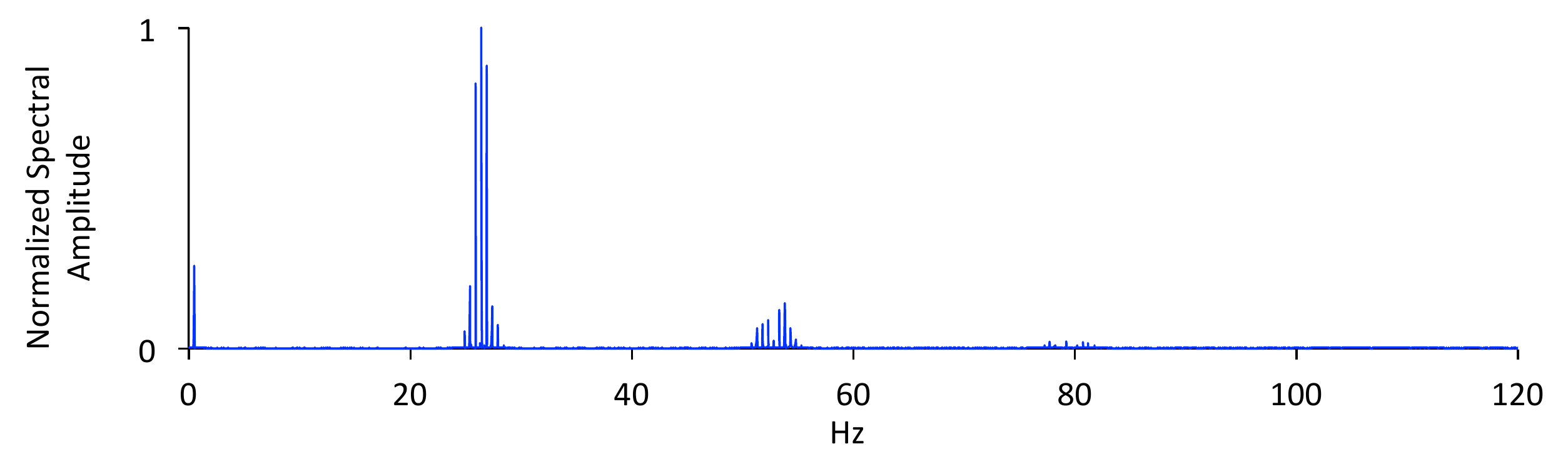}}
 \subfigure[$Re_q = 10$]{\includegraphics[width = 1 \textwidth]{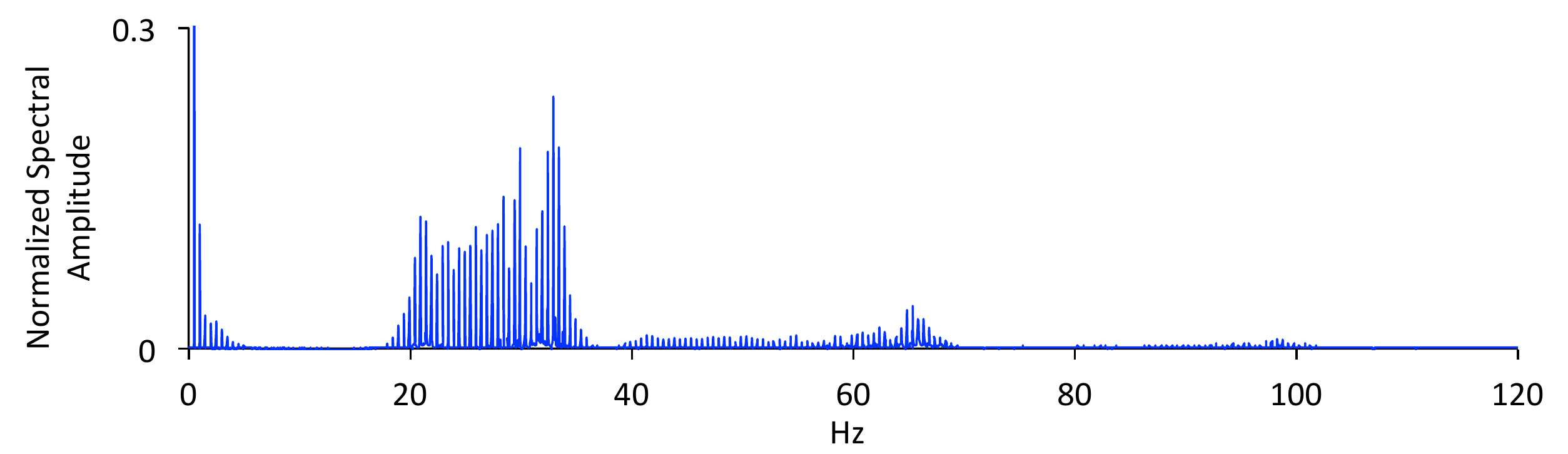}}
 \subfigure[$Re_q = 20$]{\includegraphics[width = 1 \textwidth]{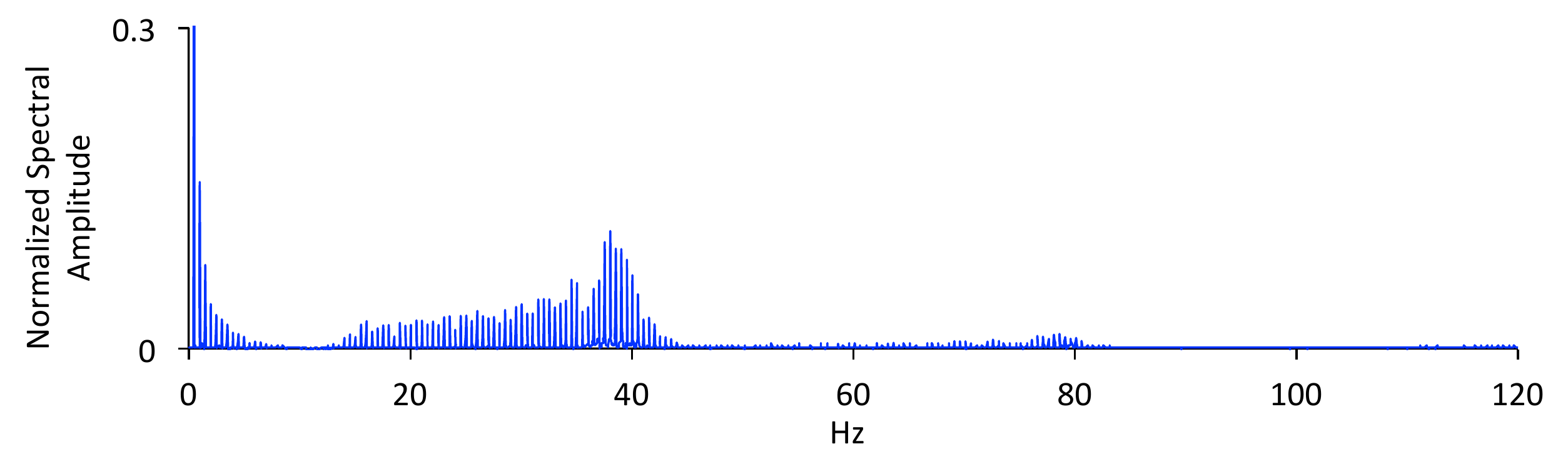}}
  \end{subfigmatrix}
 \caption{Imaginary components of Koopman eigenvalues for streamwise velocity components corresponding to the oscillating cylinder. The mean is removed for clarity and spectral amplitudes are scaled by the amplitude of the prescribed $f = 0.5$ Hz mode; vertical axes are adjusted in order to emphasize the wake shedding spectral content.}
 \label{cyl_osc_spectra}
\end{figure}

The spectra corresponding to various oscillation amplitudes are provided in Fig.~\ref{cyl_osc_spectra}. Progressive broadening of the wake spectral content with increasing amplitude is apparent from Fig. \ref{cyl_osc_spectra}. At low amplitude $Re_q = 1$ oscillations, the spectrum becomes quasi-periodic with narrow bands of discrete spectral content centered about the natural shedding frequency $f_0$ and its harmonics. In Fig. \ref{cyl_osc_spectra}(a), the two largest spikes surrounding $f_0$ and its $n^{\mathrm{th}}$ harmonic are due to interactions with the applied forcing and correspond to $n f_0 \pm f$. This follows from the algebraic properties of the Koopman eigenfunctions discussed in \citet{mezic_annual_reviews} where it was shown that if $\phi_1$ is an eigenfunction of the Koopman operator associated with $\lambda_1$, and $\phi_2$ is an eigenfunction associated with $\lambda_2$, then $\lambda = \lambda_1+ \lambda_2$ is also an eigenvalue and is associated with eigenfunction $\phi = \phi_1 \phi_2$. Thus if we have an eigenvalue $\lambda_f$ associated with the forcing frequency $f$, and $\lambda_0$ associated with $f_0$, then $\lambda_0 + n \lambda_f$ is an eigenvalue with frequency $f_0 +  n f$. A similar argument applies to $2 f_0 + f$, $3 f_0 +f$, etc.  As the oscillation amplitude is increased to $Re_q=20$, the spectral broadening increases while still remaining discrete since the spacing between each frequency spike is $\Delta f = f = 0.5$ Hz.  

\begin{figure*}
 \begin{subfigmatrix}{3}
 \subfigure[$f_0 - f$]{\includegraphics{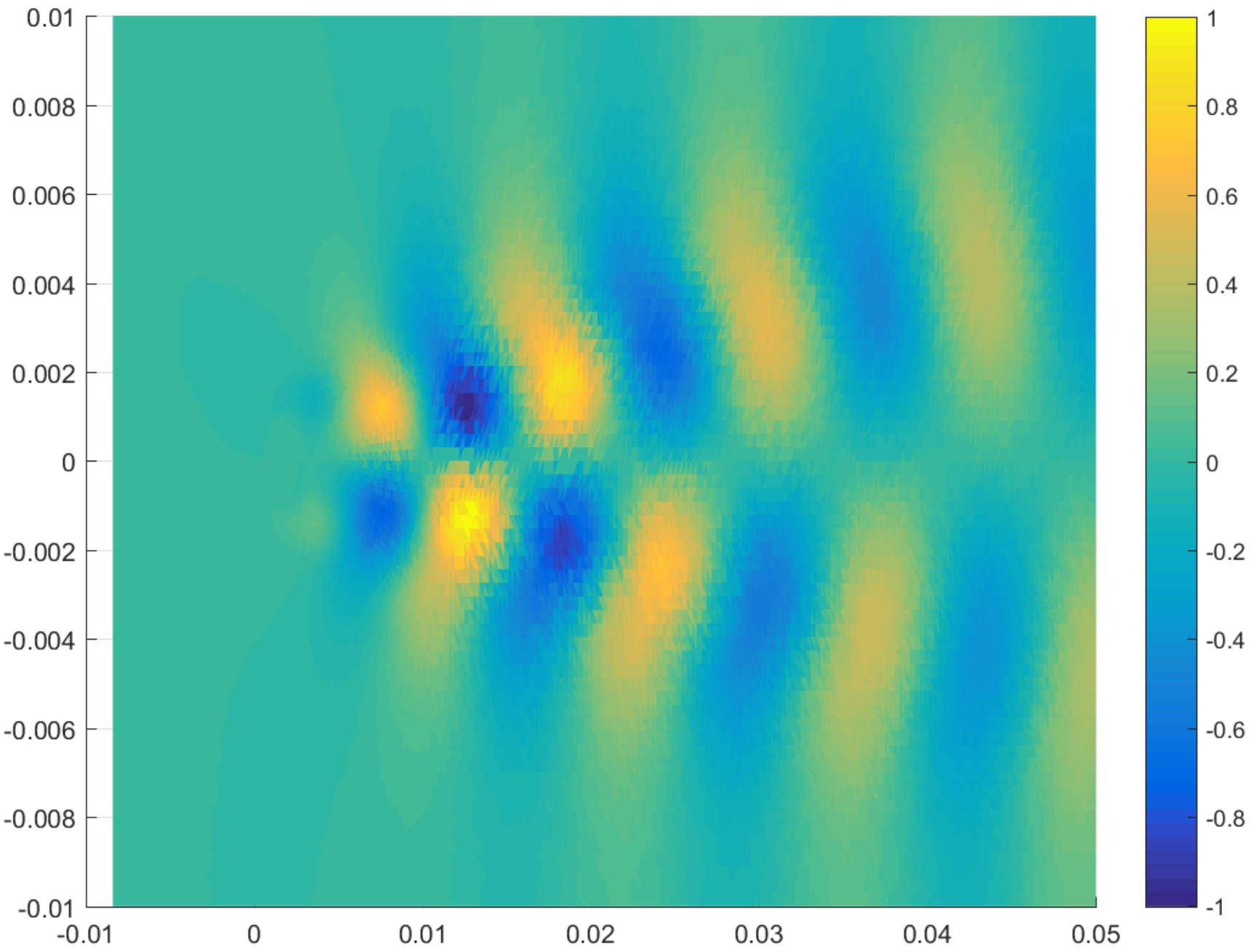}}
 \subfigure[$f_0$]{\includegraphics{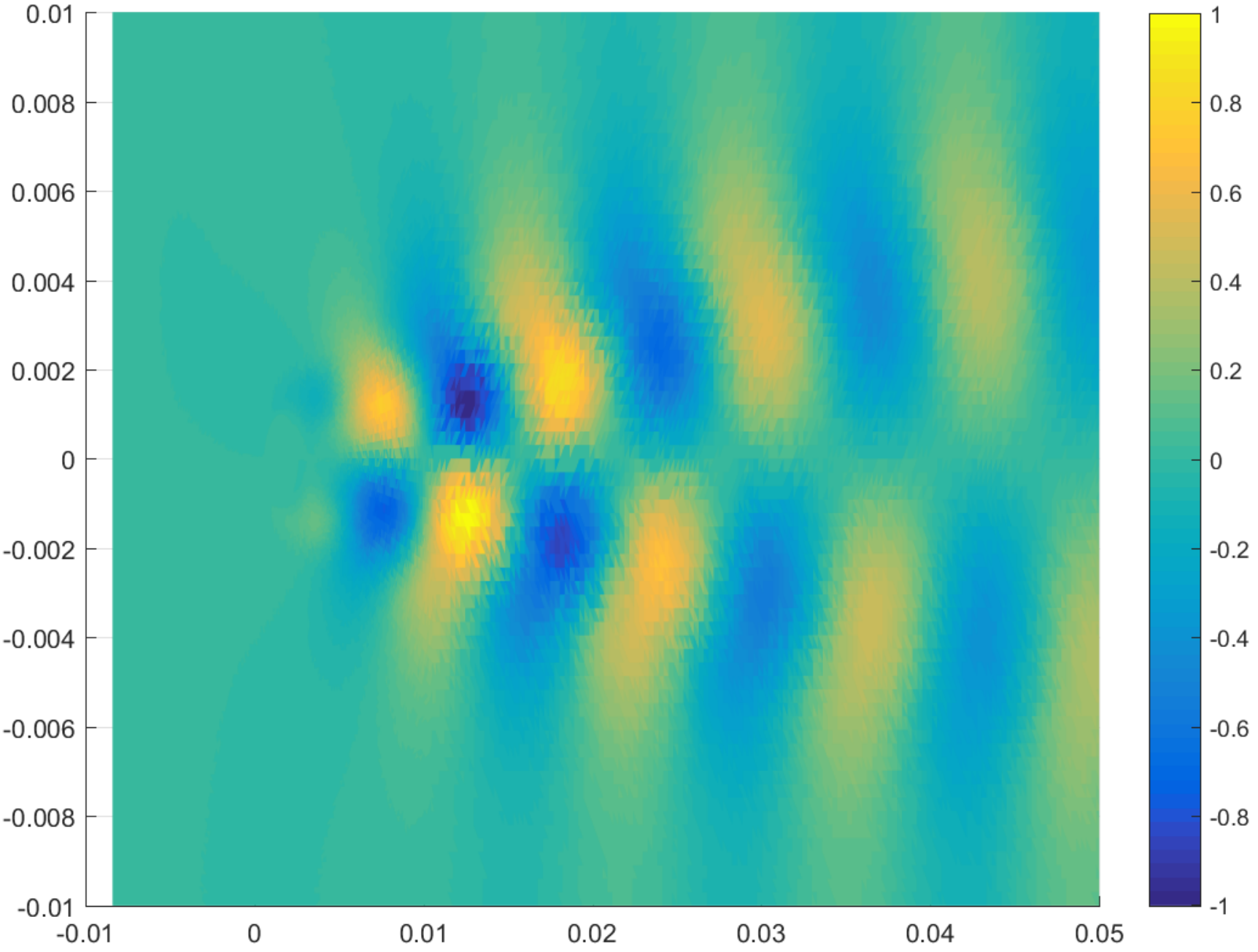}}
 \subfigure[$f_0 + f$]{\includegraphics{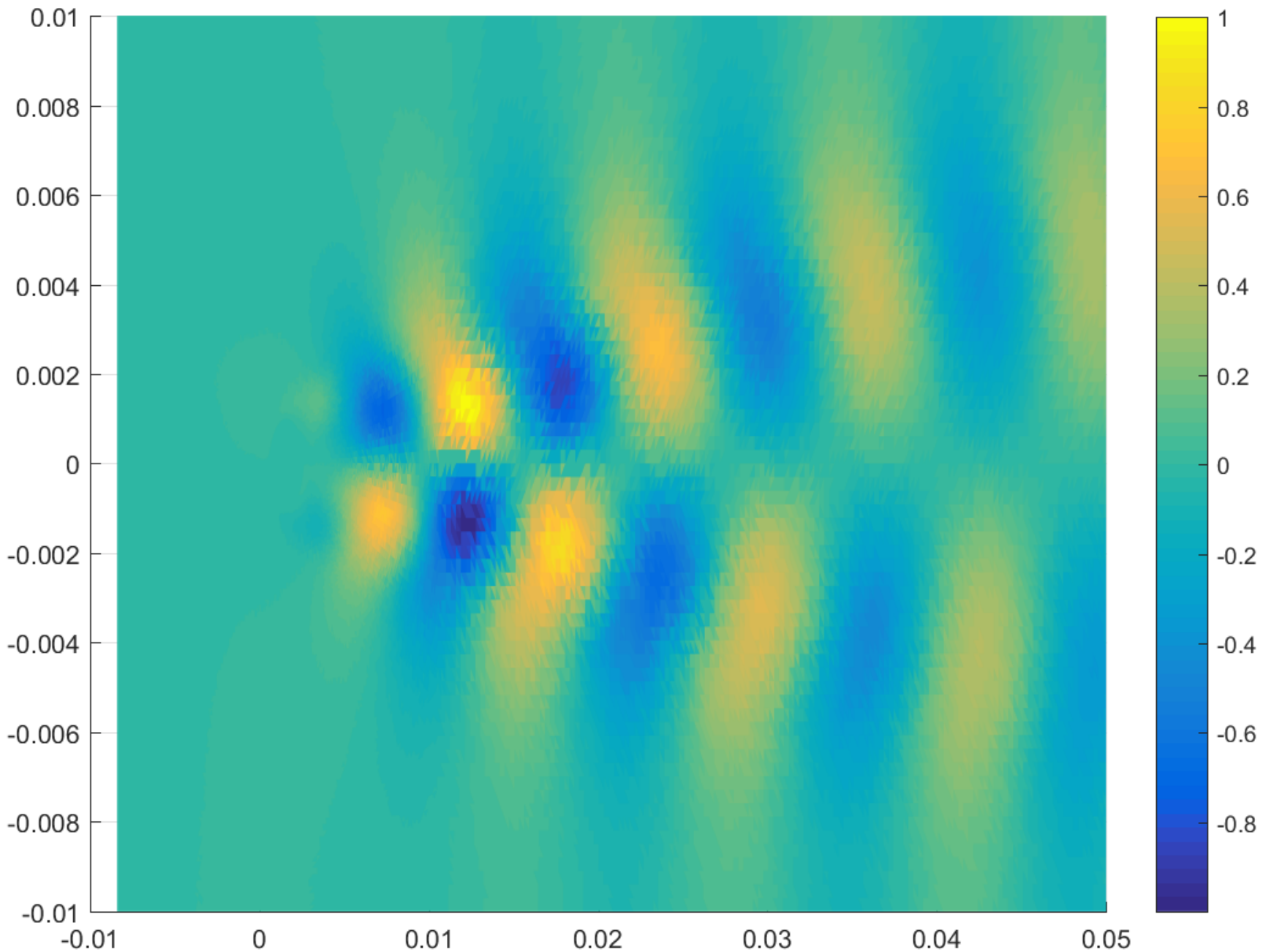}}
  \subfigure[$2 f_0 - f$]{\includegraphics{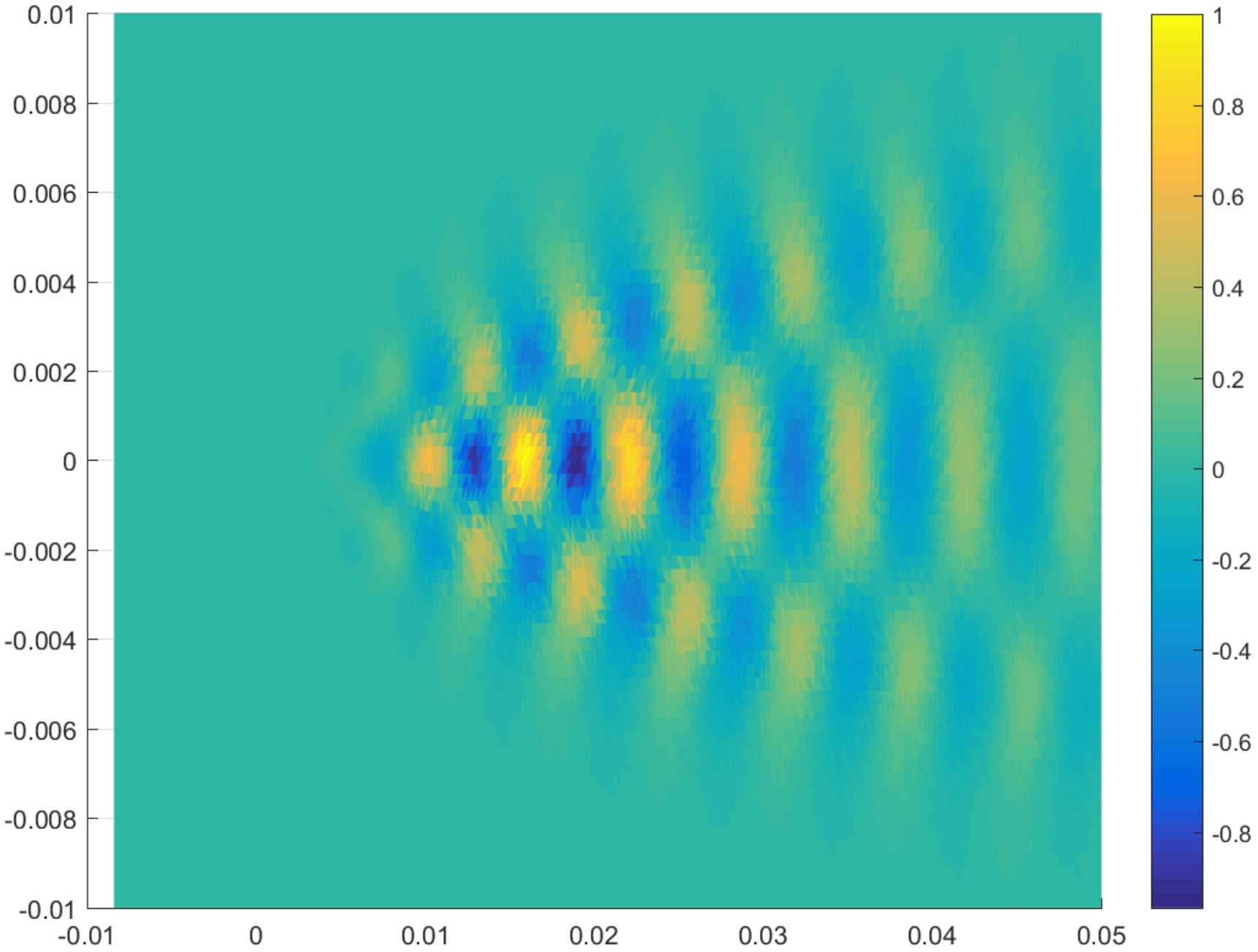}}
 \subfigure[$2 f_0$]{\includegraphics{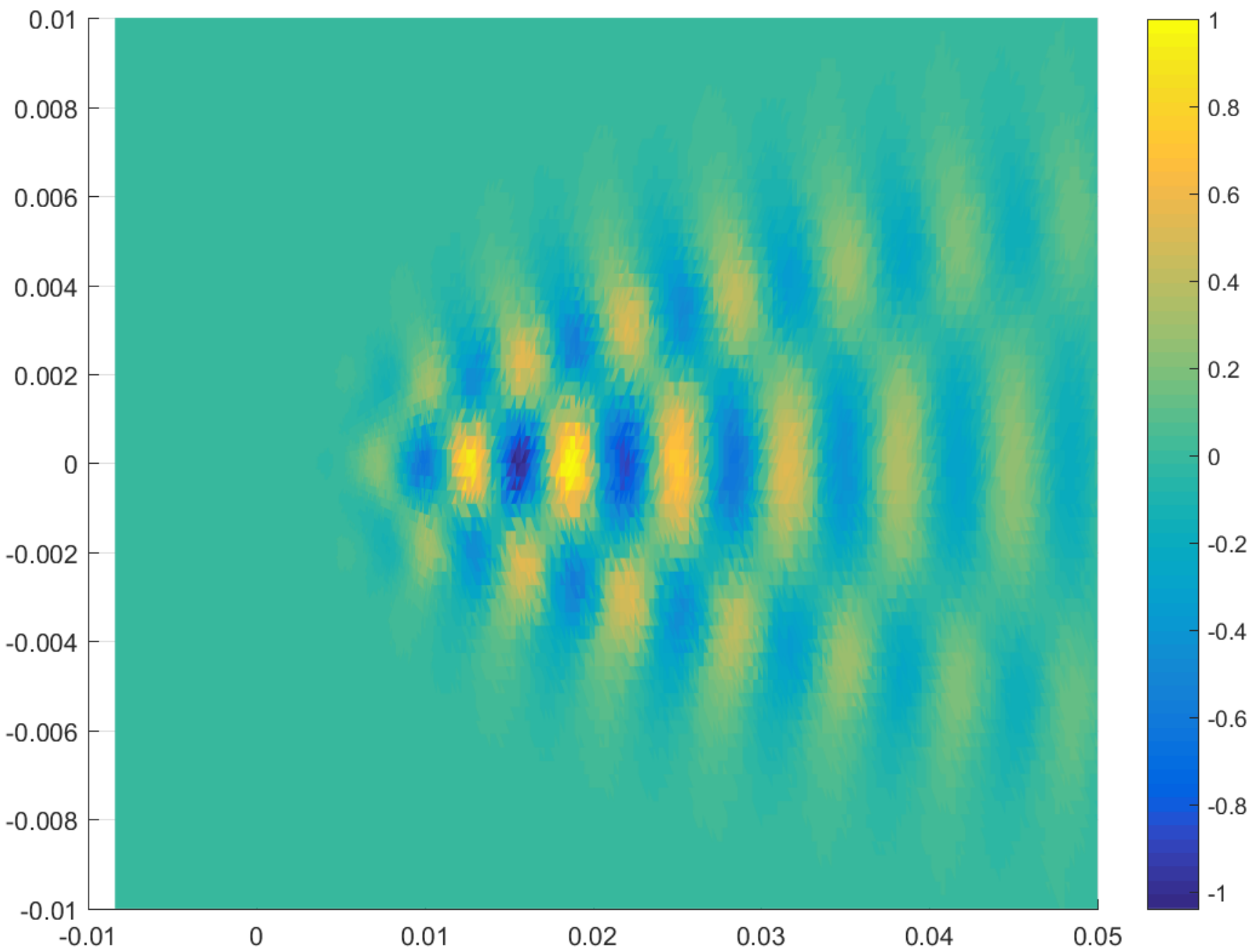}}
 \subfigure[$2f_0 + f$]{\includegraphics{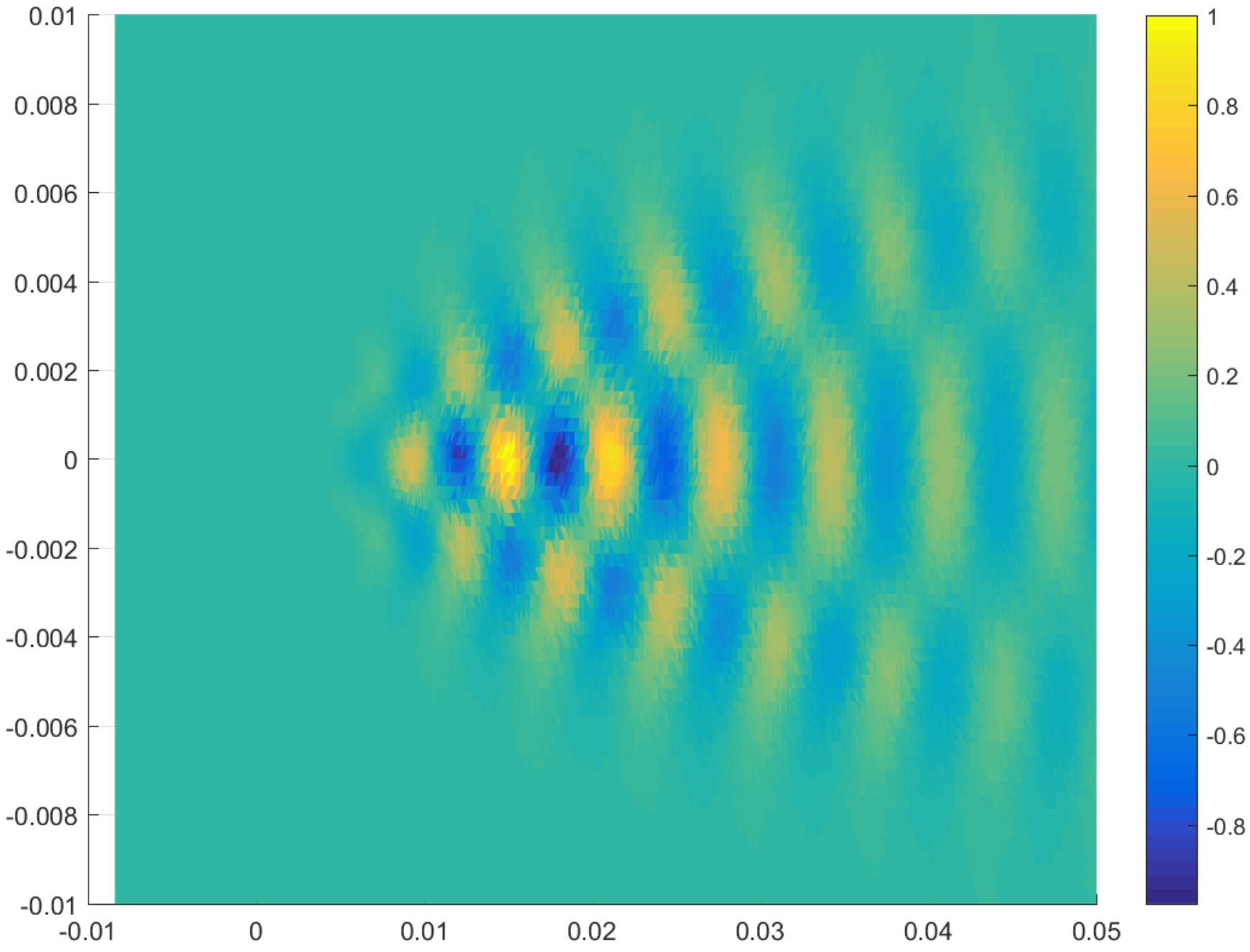}}
  \end{subfigmatrix}
 \caption{Koopman modes (real parts) corresponding to streamwise velocity for the oscillating cylinder ($Re_q = 1$).}
 \label{cyl_koop_modes_osc_small_1}
\end{figure*}

Figure~\ref{cyl_koop_modes_osc_small_1} shows Koopman modes for $Re_q = 1$. The modes correspond to $f_0 - f$, $f_0$, $f_0 + f$, $2 f_0 - f$, $2 f_0$, and $2 f_0 + f$; these are prevalent frequencies in Fig.~\ref{cyl_osc_spectra}(a).  The modes shown in Figs.~\ref{cyl_koop_modes_osc_small_1}(a) - \ref{cyl_koop_modes_osc_small_1}(c) exhibit similar coherent asymmetric spatial structure as the primary shedding mode shown in Fig.~\ref{cyl_koop_modes}(b), while Figs.~\ref{cyl_koop_modes_osc_small_1}(d) - \ref{cyl_koop_modes_osc_small_1}(f) are similar to Fig.~\ref{cyl_koop_modes}(c). Therefore, while the spectral content is beginning to broaden even for low amplitude forcing, the mode shapes of the unforced system are still underlying features of the forced system.

\begin{figure*}
 \begin{subfigmatrix}{3}
 \subfigure[$Re_q = 10$; 20 Hz]{\includegraphics{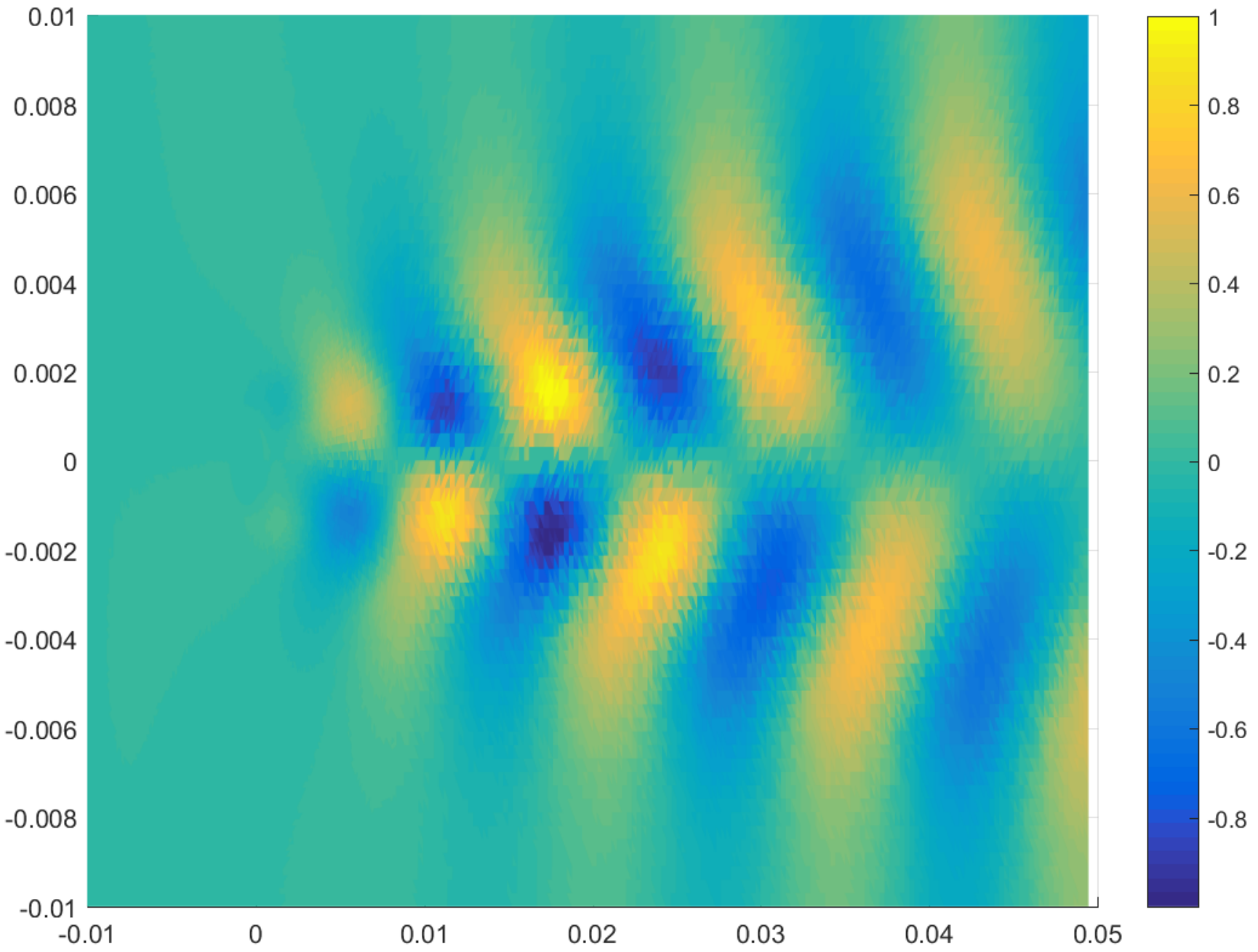}}
 \subfigure[$Re_q = 10$; 33 Hz]{\includegraphics{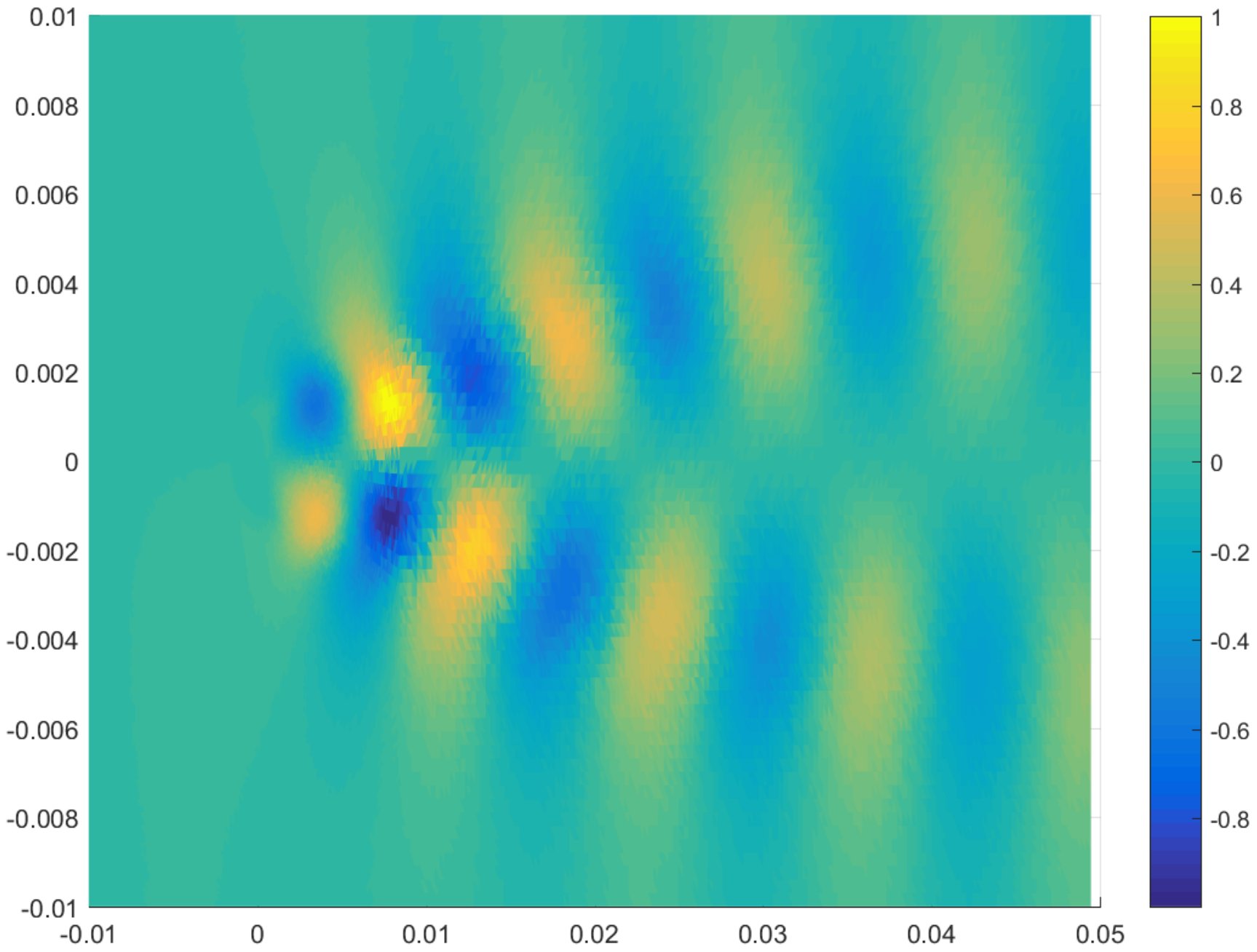}}
 \subfigure[$Re_q = 10$; 34 Hz]{\includegraphics{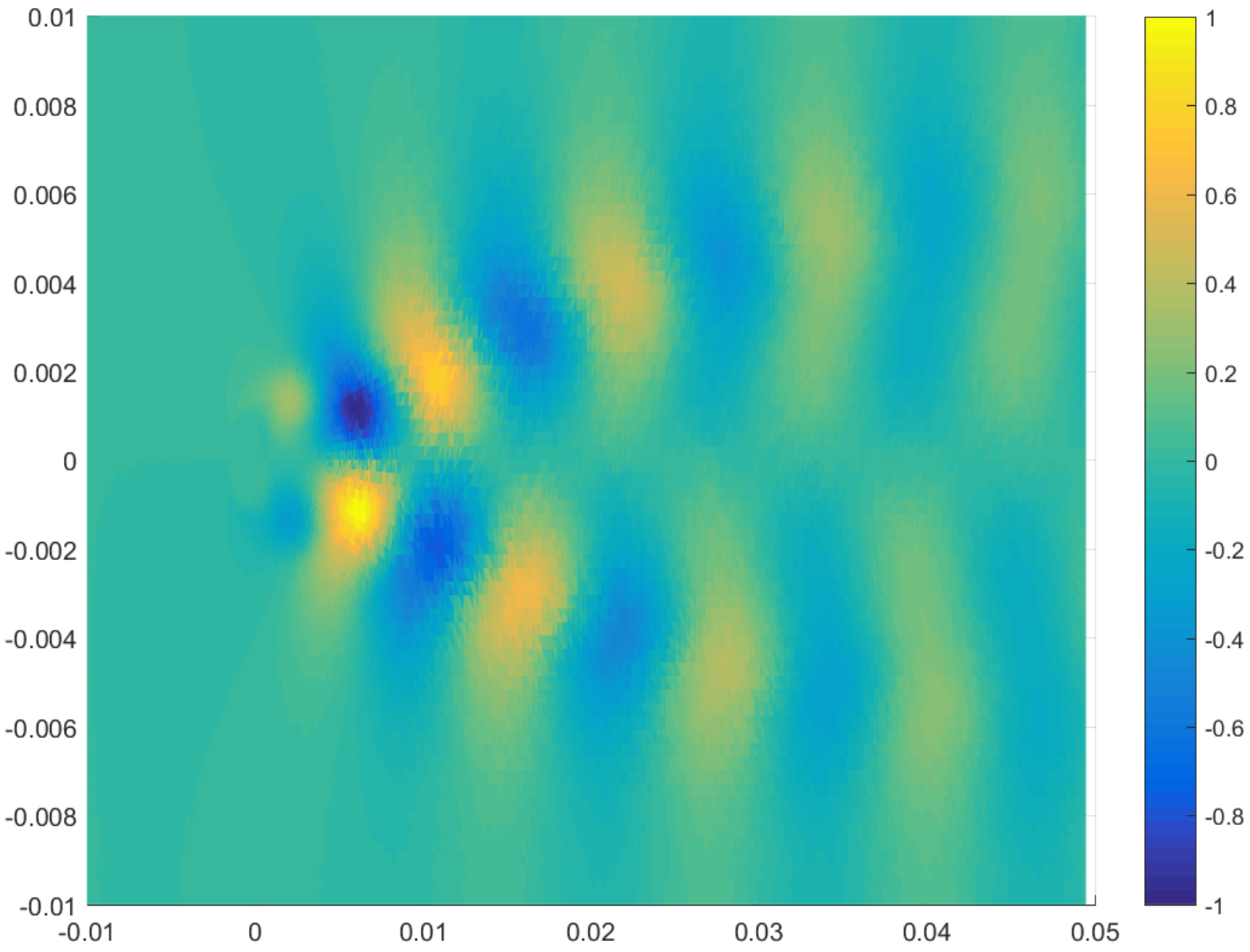}}
  \subfigure[$Re_q = 20$; 15 Hz]{\includegraphics{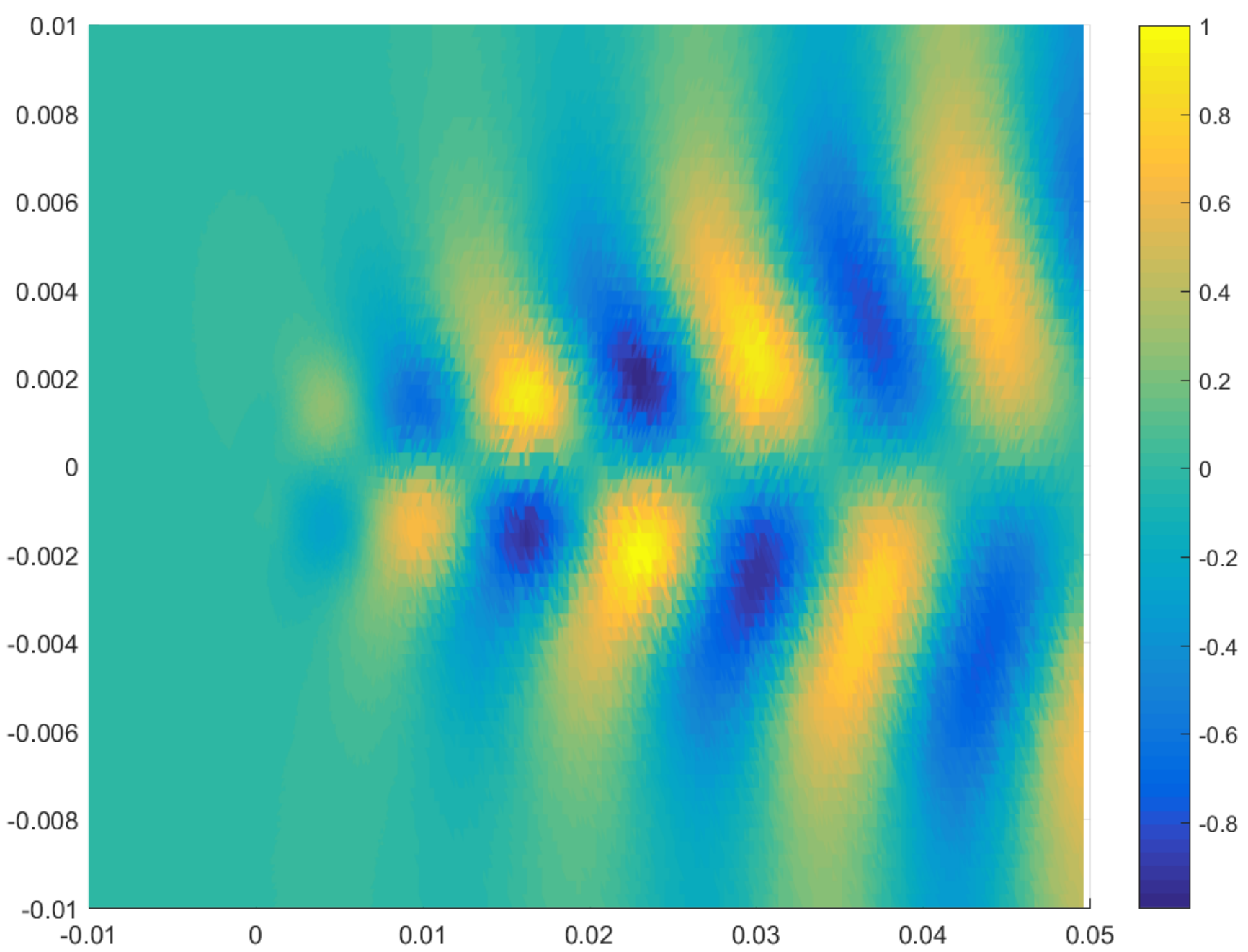}}
    \subfigure[$Re_q = 20$; 38 Hz]{\includegraphics{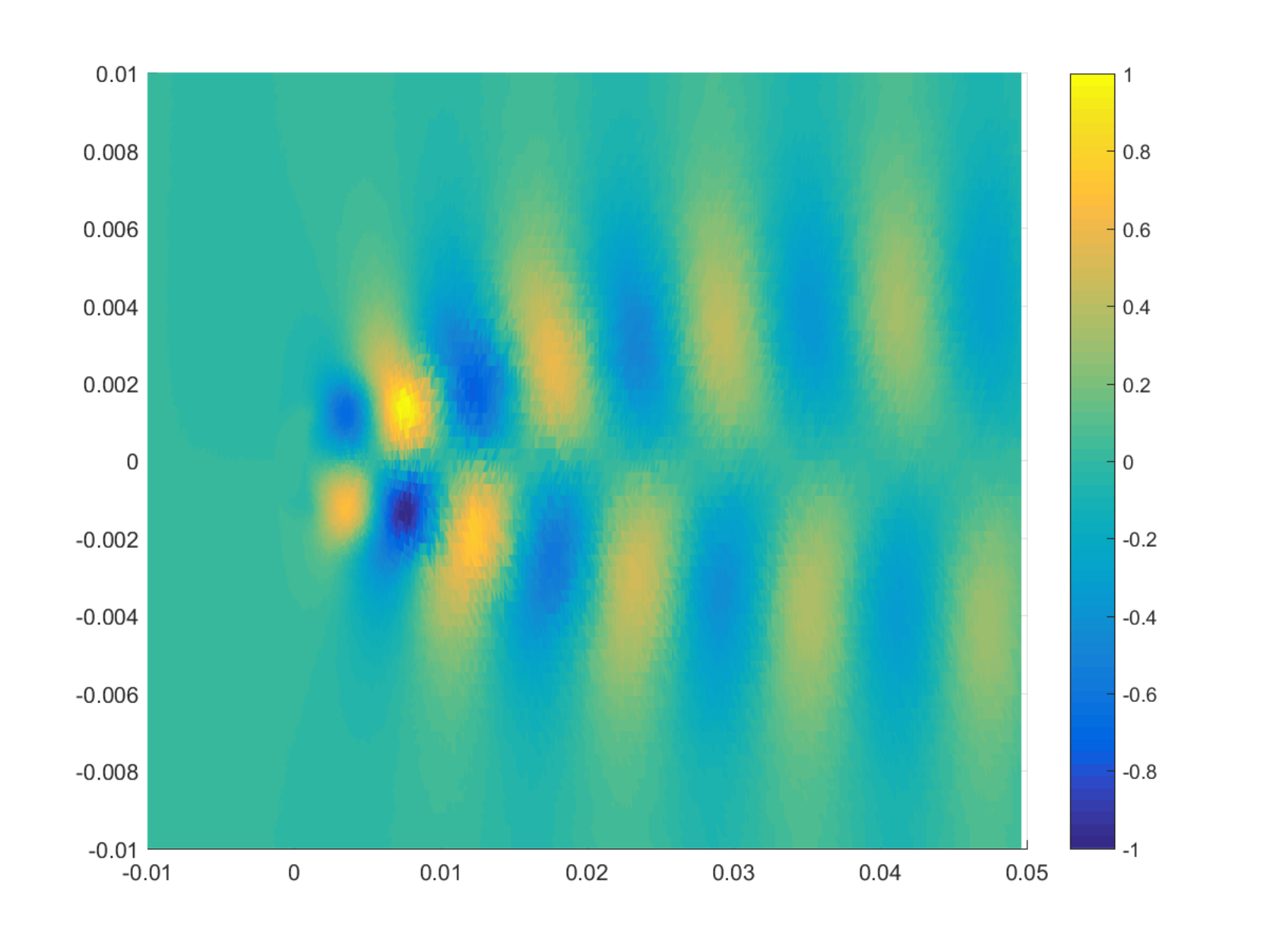}}
        \subfigure[$Re_q = 20$; 42 Hz]{\includegraphics{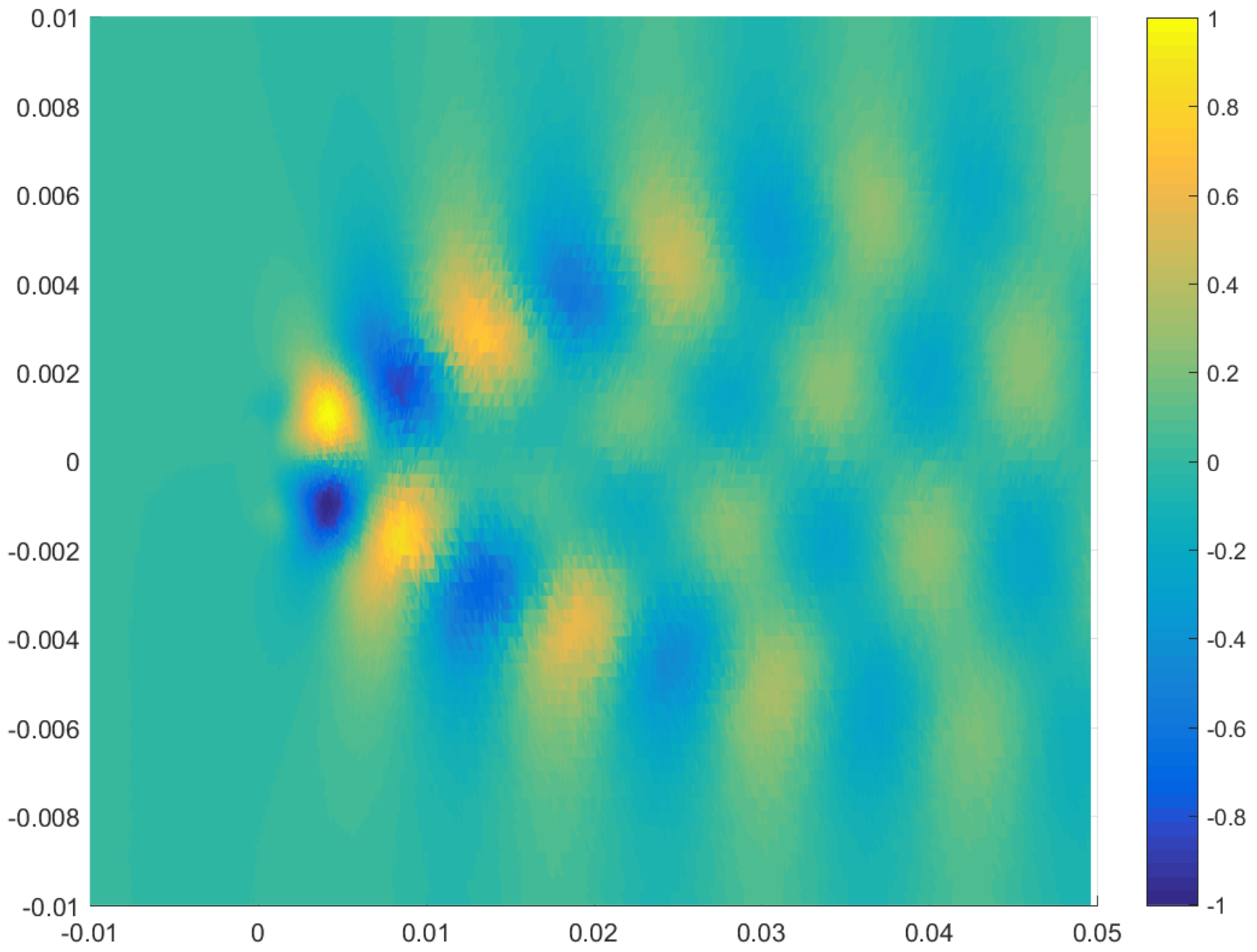}}
  \end{subfigmatrix}
 \caption{Real parts of representative Koopman modes  corresponding to streamwise velocity for various oscillation amplitudes and frequencies.}
 \label{cyl_koop_modes_various}
\end{figure*}

Interestingly, similar behavior is observed for the Koopman modes even as oscillation amplitude is increased and the spectral content broadens. The prevalence of the primary antisymmetric shedding mode shape across multiple amplitudes and frequencies is illustrated in Fig.~\ref{cyl_koop_modes_various}. In Fig.~\ref{cyl_koop_modes_various}, the center column corresponds to the frequency of maximum amplitude from Figs.~\ref{cyl_osc_spectra}(b) and (c), while the left and right columns approximately align with the minimum and maximum frequencies respectively at which the primary shedding modal spatial structure could be clearly discerned.  As the oscillation amplitude is increased, the shedding mode appears across a wider range of frequencies. For instance, the shedding mode appears throughout $20-34$ Hz for $Re_q = 10$, and $15-42$ Hz for $Re_q = 20$. Note that only the primary shedding mode is shown in Fig.~\ref{cyl_koop_modes_various}; occurrences of the second modal structure similar to Fig.~\ref{cyl_koop_modes}(c) were found at higher frequencies but are not shown. These results also support the selection of the unforced shedding Koopman mode as an appropriate basis for reduced-order modeling, as this bifurcation mode clearly remains a dynamically relevant feature of the forced system.  The theoretical underpinnings of Koopman operators provide the underlying theoretical justification for using unforced Koopman modes, even for modeling forced Hopf bifurcation dynamics. This is because the action of the unforced operator $L$ is still prevalent in the forced dynamics in (\ref{Koopevo_excited}). 

Figures \ref{cyl_osc_spectra}-\ref{cyl_koop_modes_various} indicate the potential for the wake shedding dynamics to progress from narrowly banded quasi-periodicity to spectrally broad  dynamics, while still maintaining some of the spatial coherency associated with the unforced bifurcation case. The fundamental mechanism for the spectral broadening, along with with the seemingly contradictory prevalence of the modal/spatial coherency, are explained next using normal form mathematical models from Section~\ref{nf_section}.

\subsection{Normal Forms}

Before proceeding to provide theoretical underpinnings for the numerical Koopman mode results using simplified normal form mathematical models, the reduced-order normal form approximations presented in Section \ref{cmnf_section} are verified by comparing with full-order CFD solutions. To demonstrate the concepts associated with dynamics-of-observables, we select the integrated transverse force coefficient on the cylinder, $c_y$, as the observable of interest since it is a convenient, and commonly used, representation of cylinder wake dynamics, e.g. \citep{osc_cyl_jfm2001, cyl_chaos},
\be
\g(t, \z_0) = c_y(t) .
\ee
Using the methodology outlined by \citet{sreenivasan_cyl}, explicit functions for the normal form coefficients, $\lambda_1$ and $\beta$, expressed in terms of the bifurcation parameter $\mu = (Re - Re_c)\nu / D^2$ are estimated from the stationary cylinder CFD solutions. The following expressions were used:
\be
\lambda_1 = 0.17 \mu + i \left( 38.6 \nu/D^2 + 0.7 \mu \right), 
\ee
\begin{equation}
\Re{[\beta]} = -0.2 \mu / (c_{y, max})^2; \Im[{\beta}] = -3 \Re{[\beta]}.
\end{equation}

\begin{figure}
 \begin{subfigmatrix}{1}
 \subfigure[$Re_q = 1$]{\includegraphics{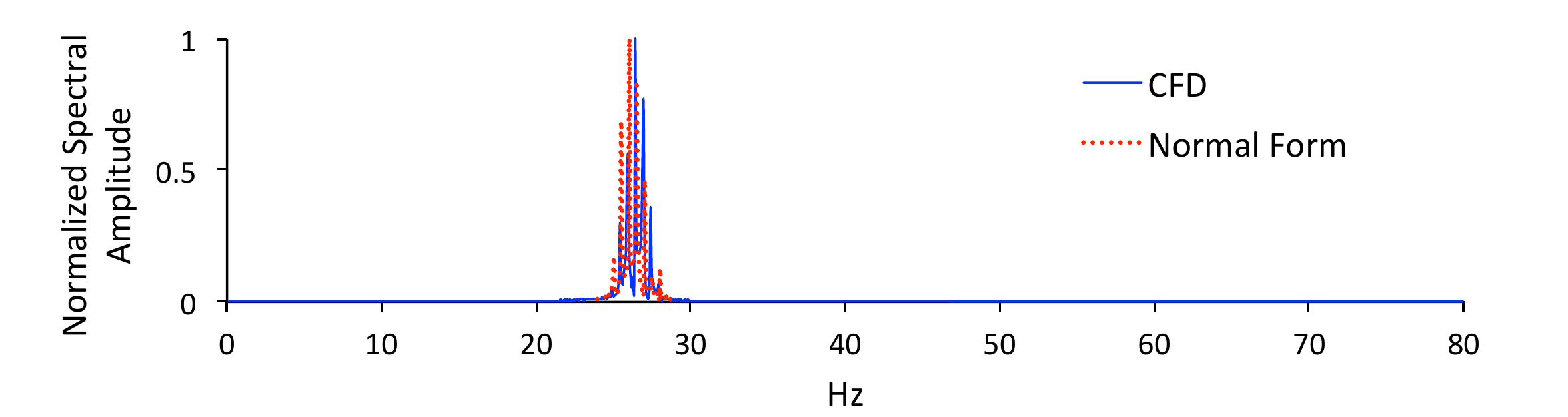}}
\subfigure[$Re_q = 10$]{\includegraphics{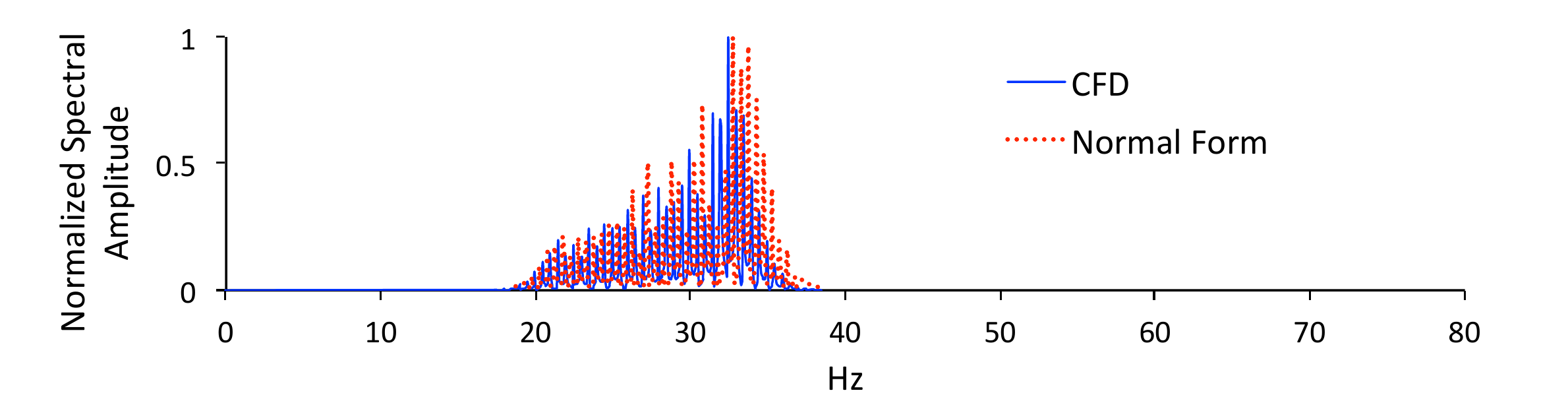}}
\subfigure[$Re_q = 20$]{\includegraphics{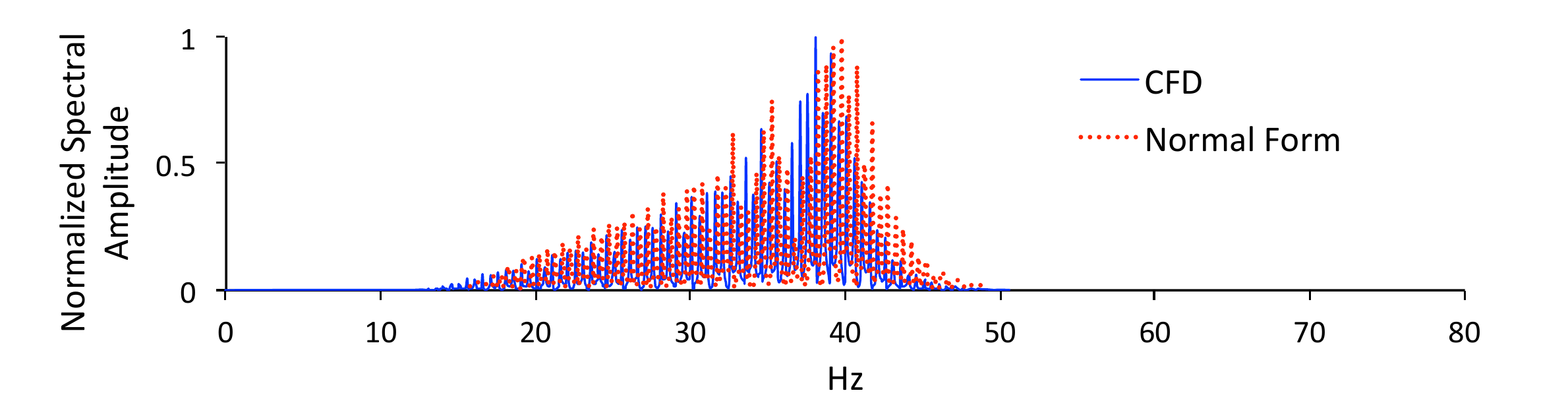}}
  \end{subfigmatrix}
 \caption{Transverse force spectra for $f = 0.5$ Hz oscillating cylinder cases at various oscillation amplitudes; normal form solutions based on (\ref{nf_1}).}
 \label{cfd_v_nf_spectra_wf2}
\end{figure}

\begin{figure}
 \begin{subfigmatrix}{2}
  \subfigure[CFD, $Re_q = 1$]{\includegraphics{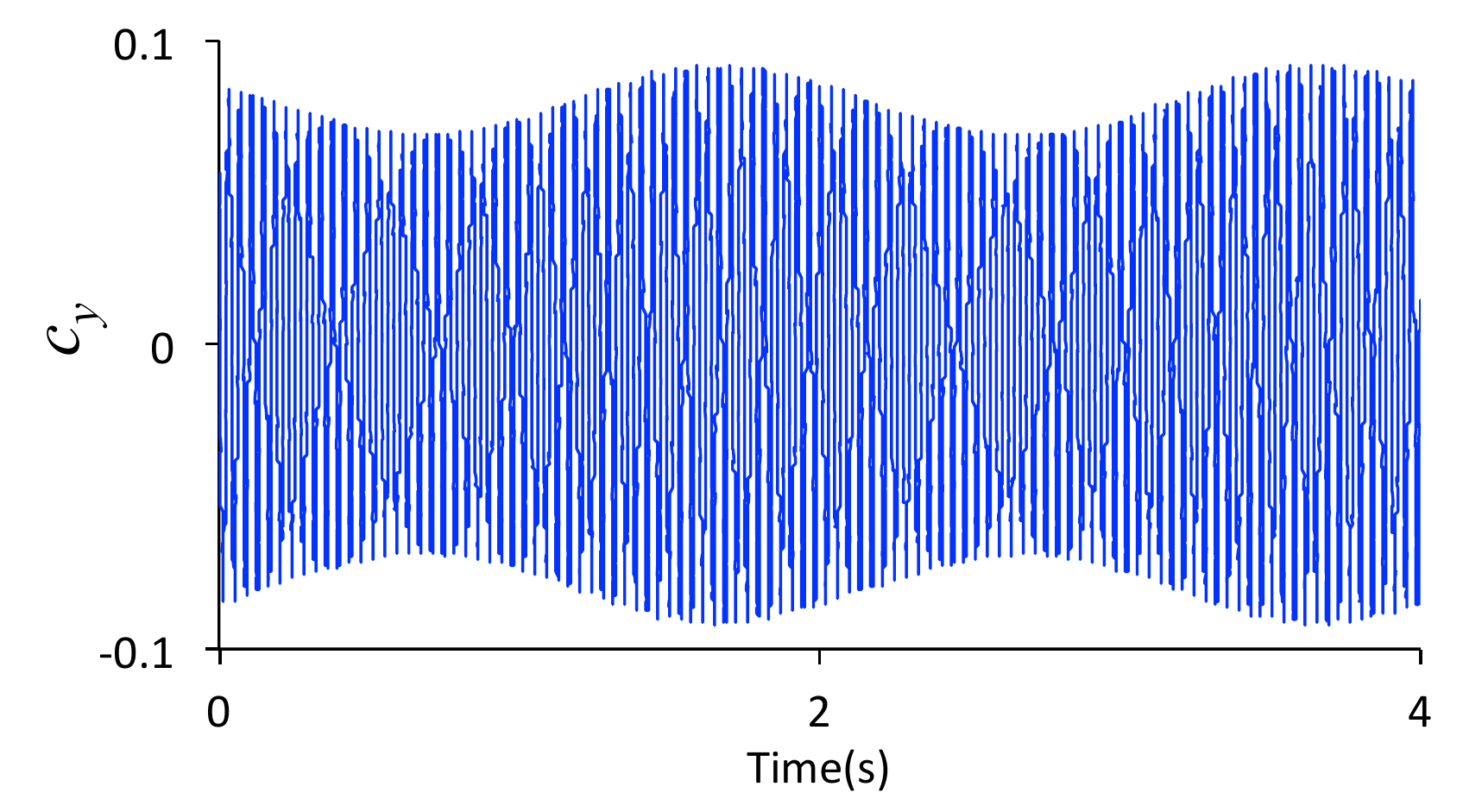}}
  \subfigure[Normal Form, $Re_q = 1$]{\includegraphics{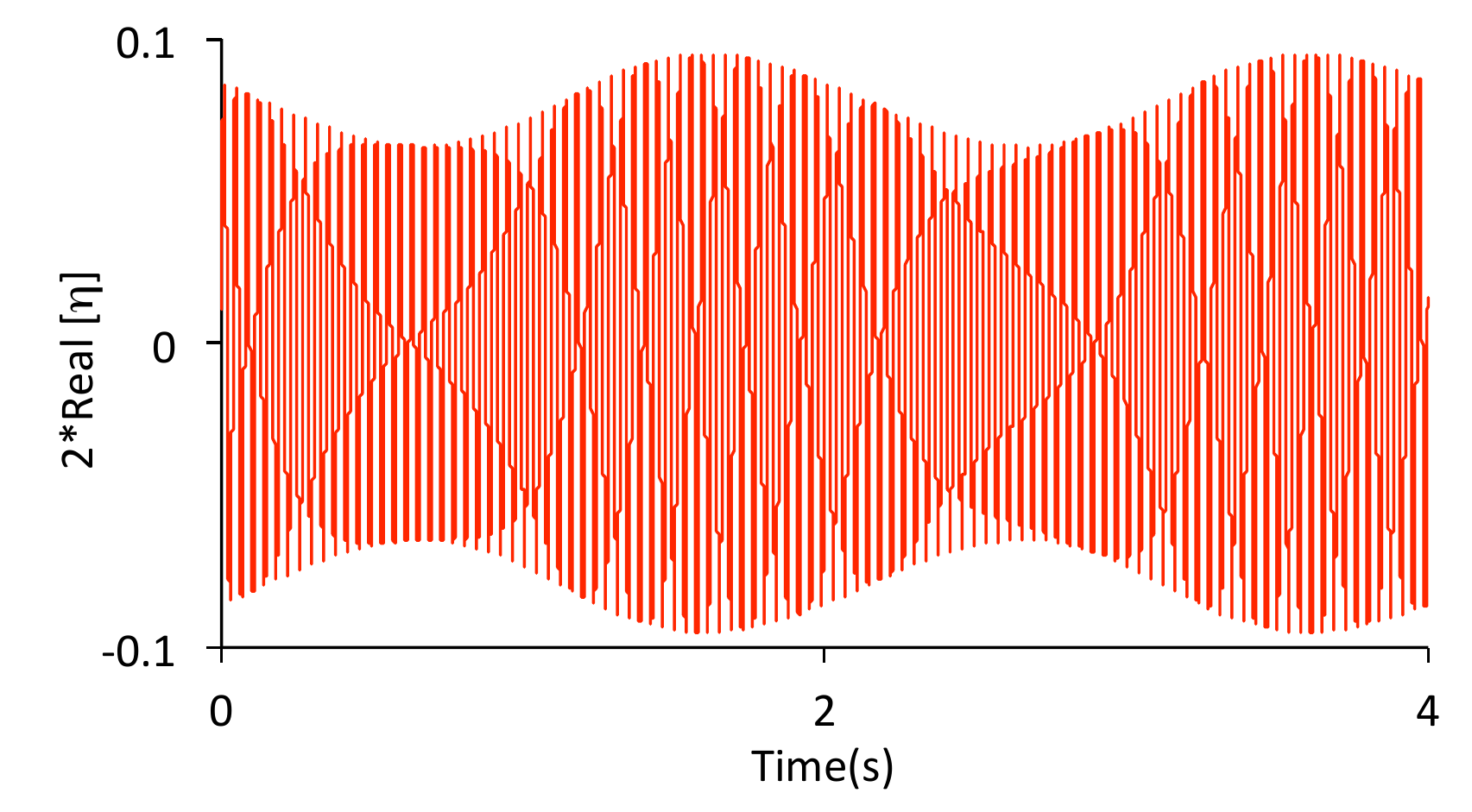}}
   \subfigure[CFD, $Re_q = 10$]{\includegraphics{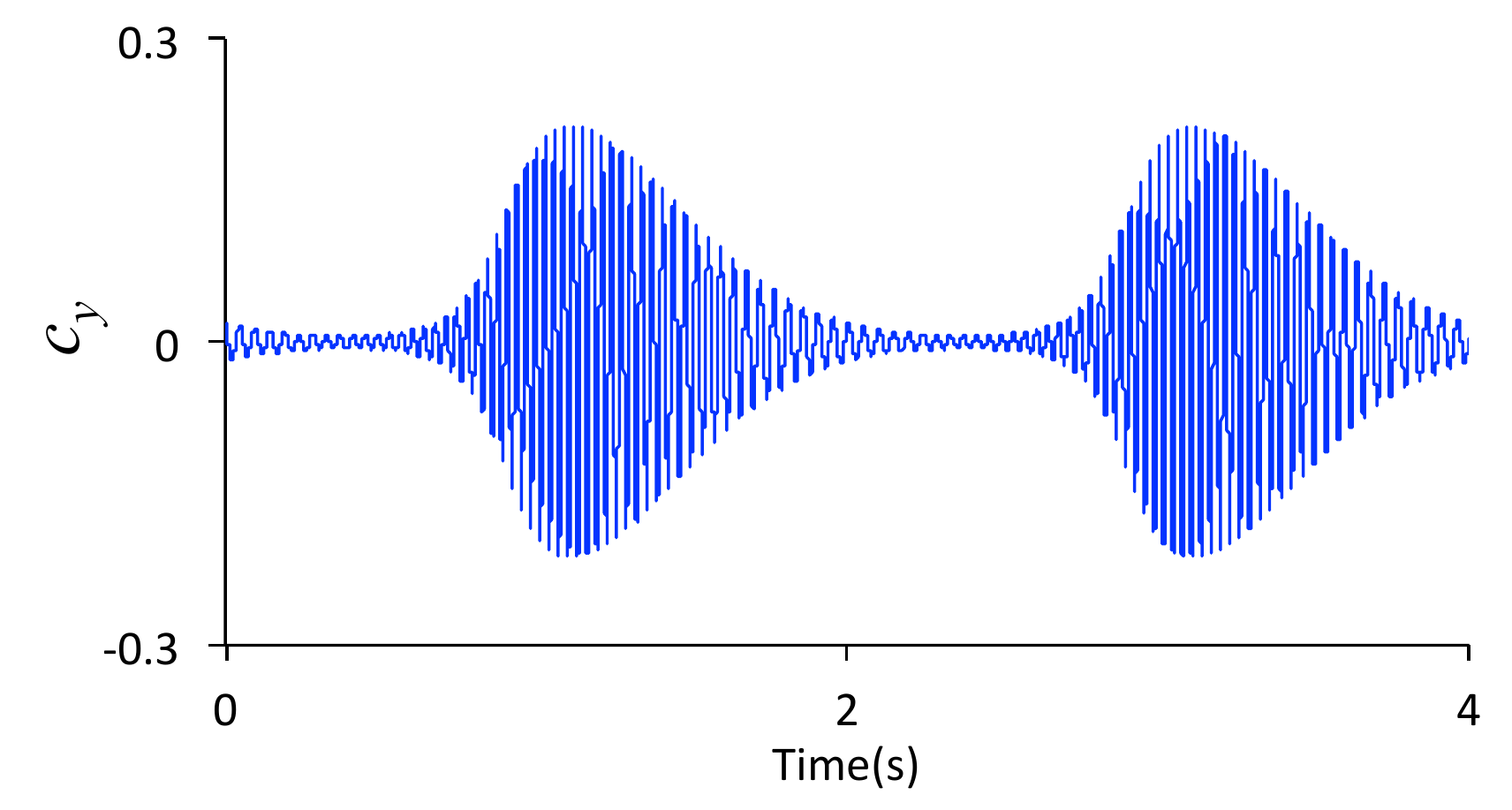}}
  \subfigure[Normal Form, $Re_q = 10$]{\includegraphics{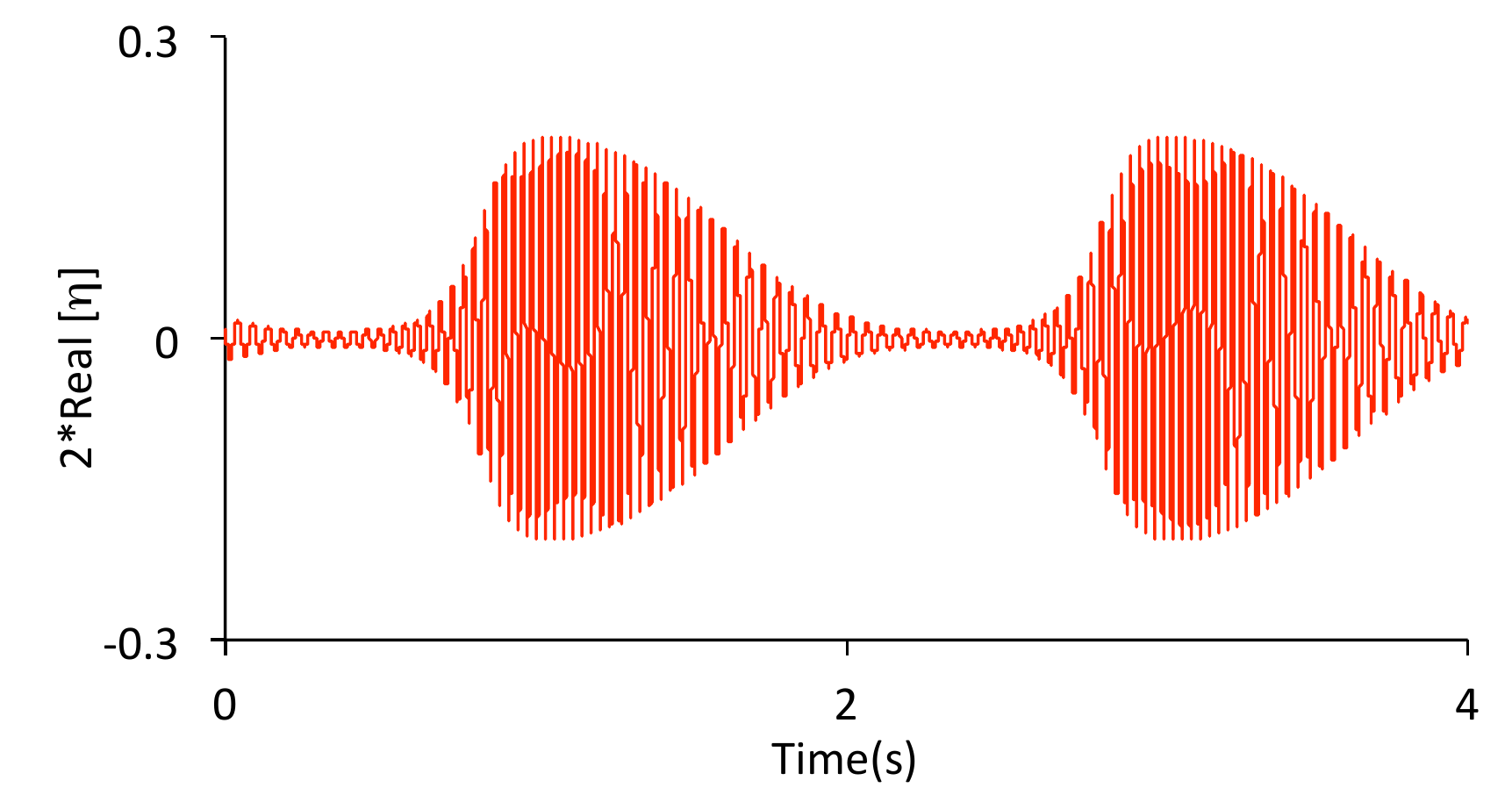}}
    \subfigure[CFD, $Re_q = 20$]{\includegraphics{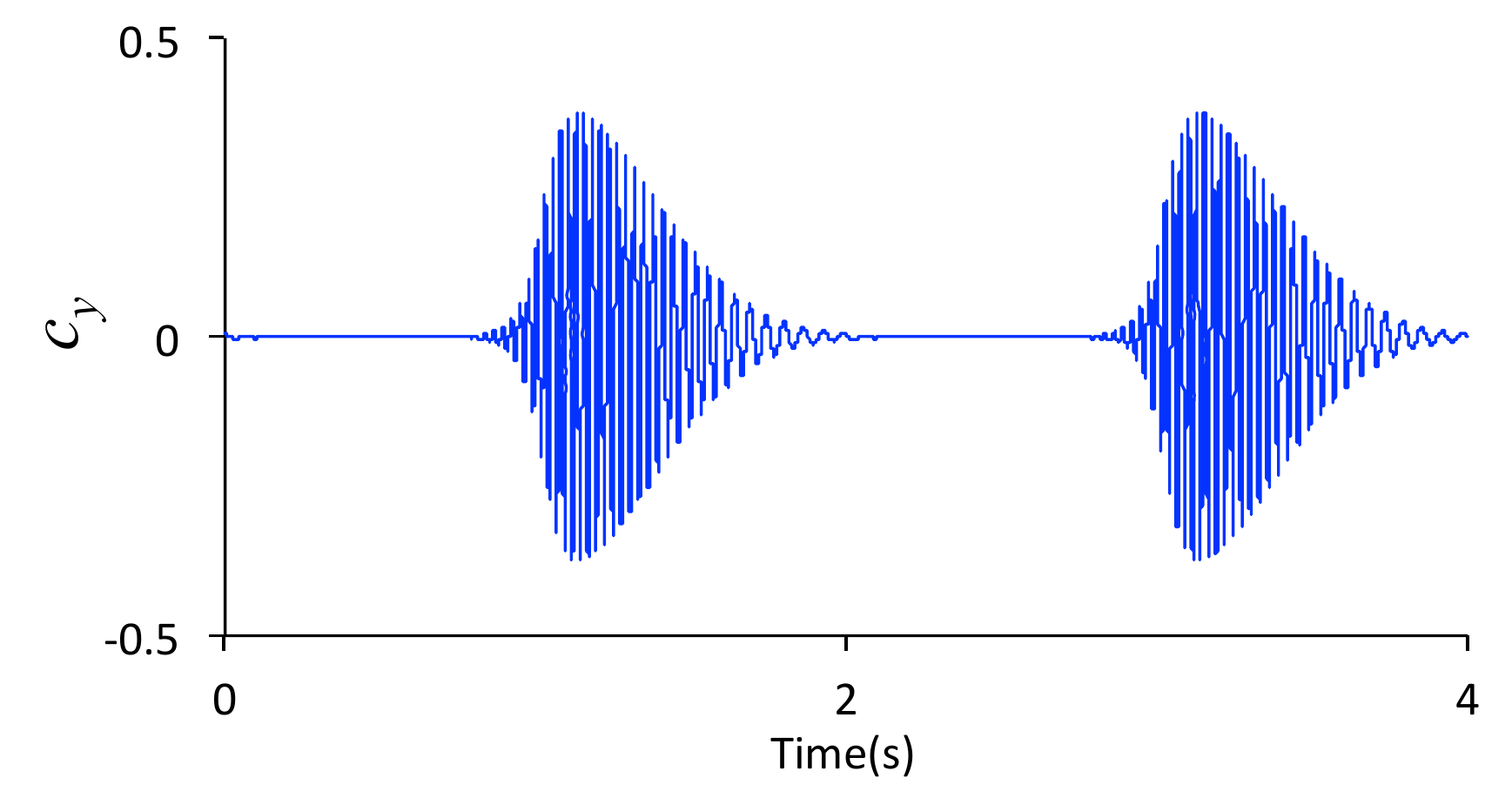}}
  \subfigure[Normal Form, $Re_q = 20$]{\includegraphics{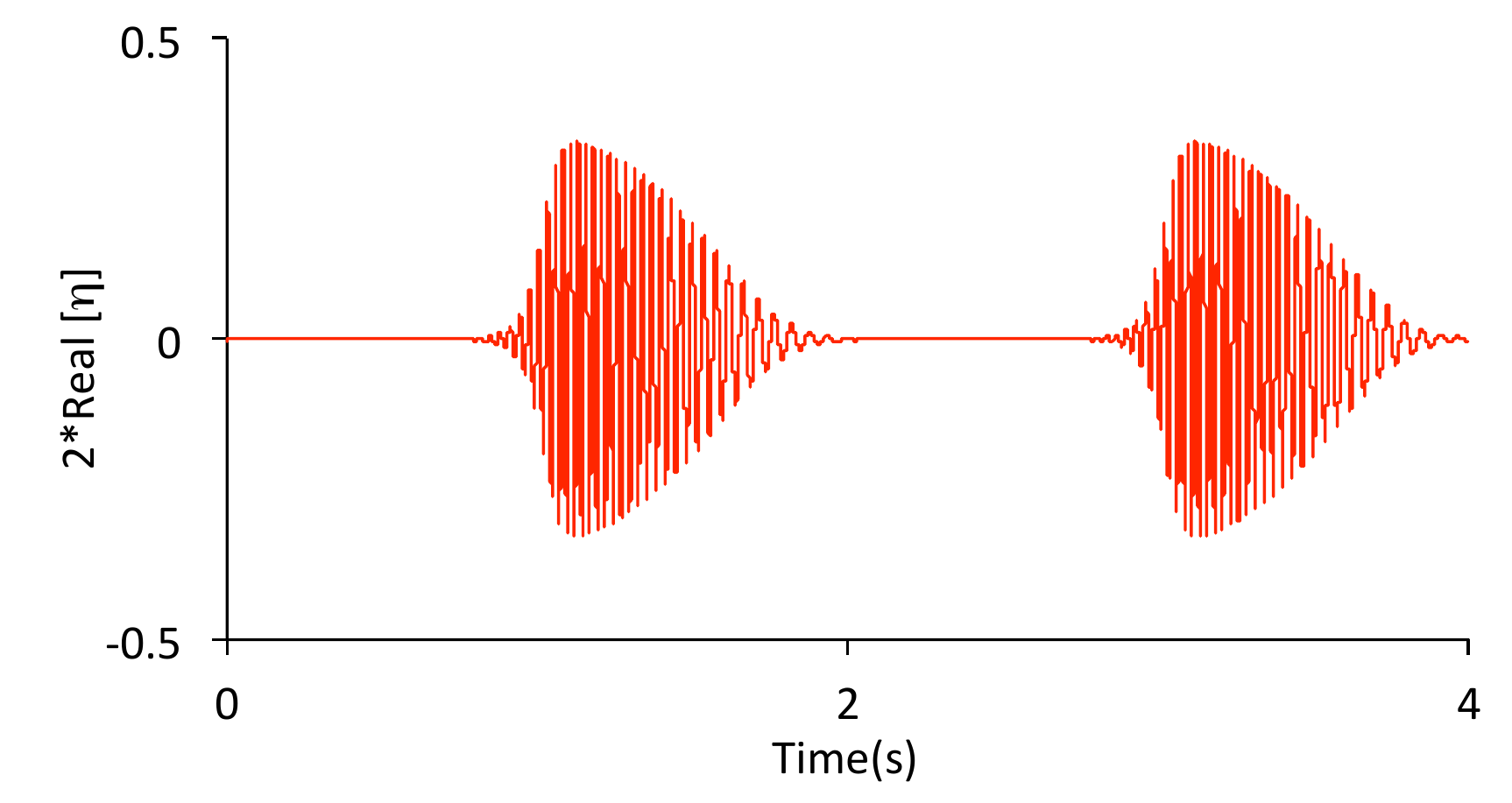}}
    \end{subfigmatrix}
 \caption{Transverse force coefficient time signals for $f = 0.5$ Hz oscillating cylinder cases at various oscillation amplitudes; normal form solutions based on (\ref{nf_1}).}
 \label{cfd_v_normal_form_signal_wf2}
\end{figure}

Spectral and time-signal comparisons between normal form theory and CFD results are presented in Figs. \ref{cfd_v_nf_spectra_wf2} and \ref{cfd_v_normal_form_signal_wf2} for $f << f_0$. The normal form (\ref{nf_1}) was used for the comparisons due to the separation in scale between the forcing and natural frequencies. For simplicity, it was assumed that $Q_I = 0$ since $\omega_0 + \beta_I r^2 >> Q_I$ for the range of parameters that were considered. The leading order normal form approximations accurately capture the spectral broadening exhibited by the CFD solutions as oscillation amplitude is increased. These results indicate that the forcing effects acting on only the primary bifurcation mode are responsible for the similar behavior observed in Fig. \ref{cyl_osc_spectra} for the streamwise velocity fields. 

The transverse force time-signal comparisons corresponding to Fig. \ref{cfd_v_nf_spectra_wf2} are provided in Fig. \ref{cfd_v_normal_form_signal_wf2}. The similarities between full-order and approximate time-signals  further illustrate the excellent agreement between the normal forms based on just one mode. The accuracy of the normal form approximations derived from center manifold reductions support the finding that using the unforced bifurcation mode as the basis can lead to an accurate first-order approximation of a forced Hopf bifurcation system. So while the broad spectral content associated with the CFD would indicate a complicated high-dimensional system, the `complexity' can be simplified quite nicely into a low-order representation when viewed from the correct perspective; in this case, the appropriate perspective is a basis spanned by the unforced Koopman bifurcation mode.

\begin{figure}
\centering
    \includegraphics[width=1\textwidth]{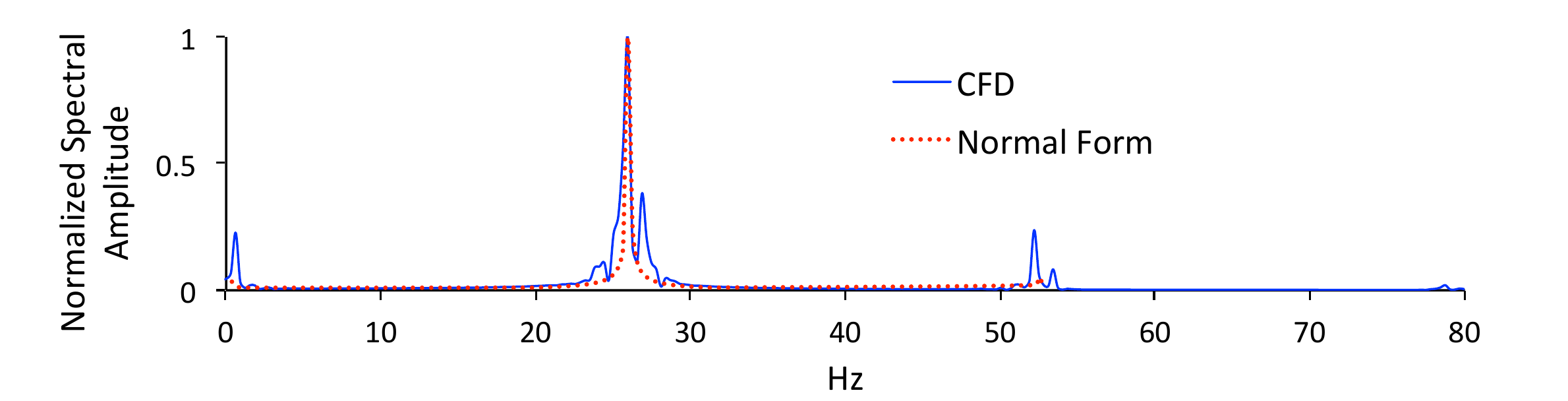}
  \caption{Transverse force spectra for $f = f_0=26.4$ Hz, $Re_q = 10$ oscillating cylinder case; normal form solutions based on (\ref{nf_2}).}
\label{cfd_v_nf2}
\end{figure}

To demonstrate that $f << f_0$ is a necessary condition for spectral broadening due to simple harmonic forcing of a bifurcation mode, CFD and normal form results for a case when $f=f_0$ are contrasted with the $f << f_0$ case in Fig.~\ref{cfd_v_nf2}. As discussed in Section \ref{nf_section}, the proper full normal form when $f$ is not much less than $f_0$ is given by (\ref{nf_2}) rather than (\ref{semi_nf}) or (\ref{nf_1}). The leading order normal form differential equation in (\ref{nf_2}), which is equivalent to the unforced bifurcation system, accurately approximates the full system dynamics as shown in Fig. \ref{cfd_v_nf2}; note the difference compared to the separation of scale case in Fig. \ref{cfd_v_nf_spectra_wf2}(b). In addition to the primary frequency at $f_0$, the CFD solution also exhibits spectral content at $2 f_0$ and 0. Whereas in the $f << f_0$ case the $\eta \zeta$ and $\eta \bar{\zeta}$ terms are retained in the normal form differential equation, when $f$ is not much less than $f_0$ these terms can be eliminated from the differential equation and thus do not affect the $\epsilon^0$ order approximation of $\gamma_1$ in \ref{zeta}. Instead, they are recovered in the higher order $\epsilon^1$ nonlinear transformation terms associated with the function $f_1$, where $\eta \zeta$ corresponds to the $2 f_0$ spectral content and $\eta \bar{\zeta}$ produces the 0 Hz content observed in the CFD. As such, the leading order effects of the forcing will enter as additive terms to the expression for $\gamma_1$ -- rather than terms in the ODE governing $\eta$ -- and thus will always appear as narrowly banded discrete spectral content. Note that the more general semi-normal form (\ref{semi_nf}) would capture the spectral content associated with $f_0$, $2 f_0$, and 0 in the solution of $\eta$. However, the results in Fig. \ref{cfd_v_nf2} using (\ref{nf_2}) demonstrate that it is not necessary to retain the leading order forcing effects in the differential equation. By proceeding with the full normal form simplifications using assumptions on $f$, we are able to show explicitly that $f << f_0$ is a necessary condition for leading order forcing effects to appear in the differential equation, and thus other scenarios for $f$ are precluded from inducing spectral broadening or chaos as a leading order phenomena spanned by the wake shedding mode.

\begin{figure}
 \begin{subfigmatrix}{2}
  \subfigure[CFD, $Re_q=10$ ]{\includegraphics{figs/cfd_signal_Re1_20_05Hz_b.pdf}}
   \subfigure[Normal form (\ref{nf_additive}), $Re_q = 10$]{\includegraphics{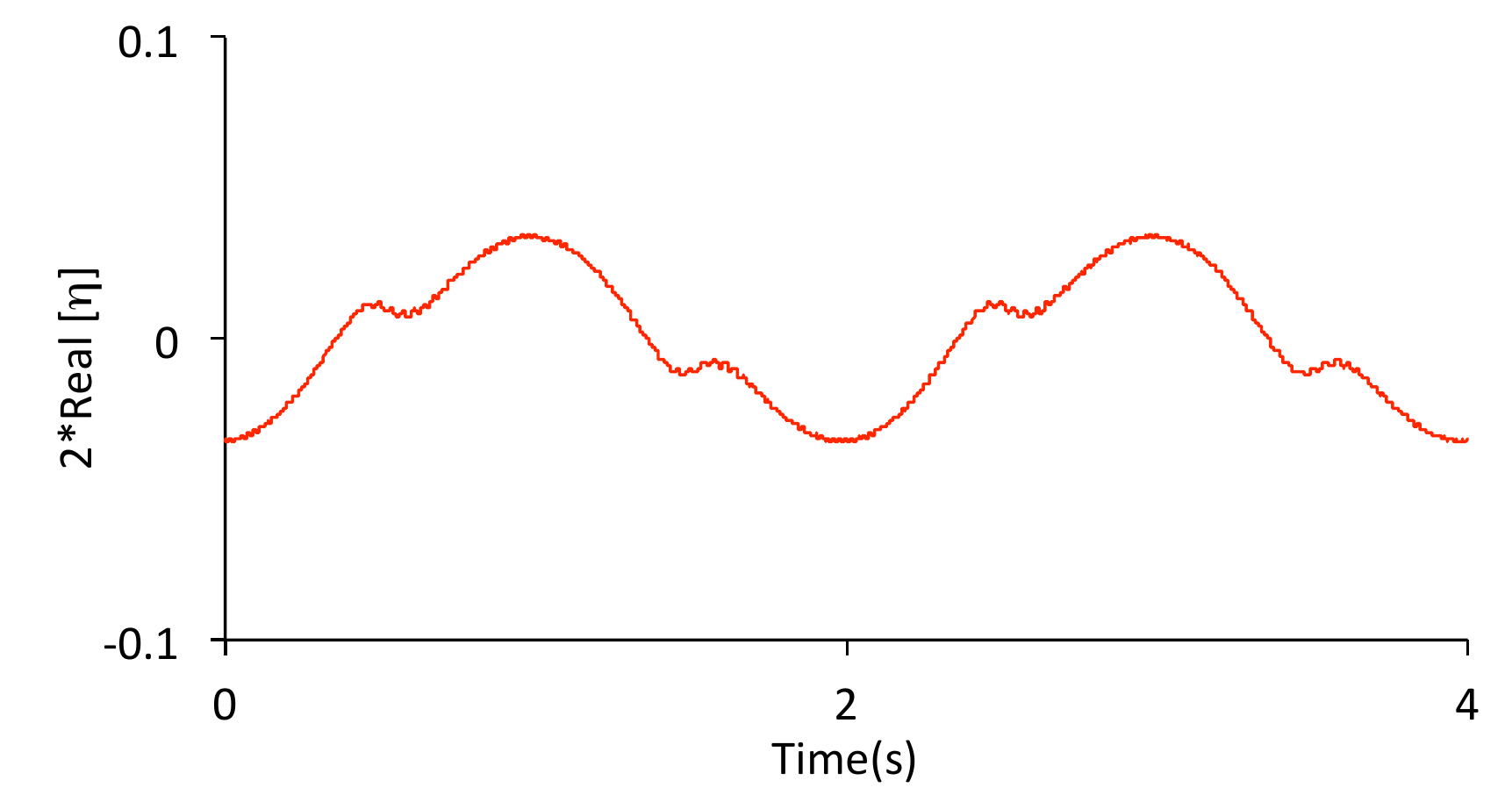}}
    \subfigure[Spectra]{\includegraphics[width=1\textwidth]{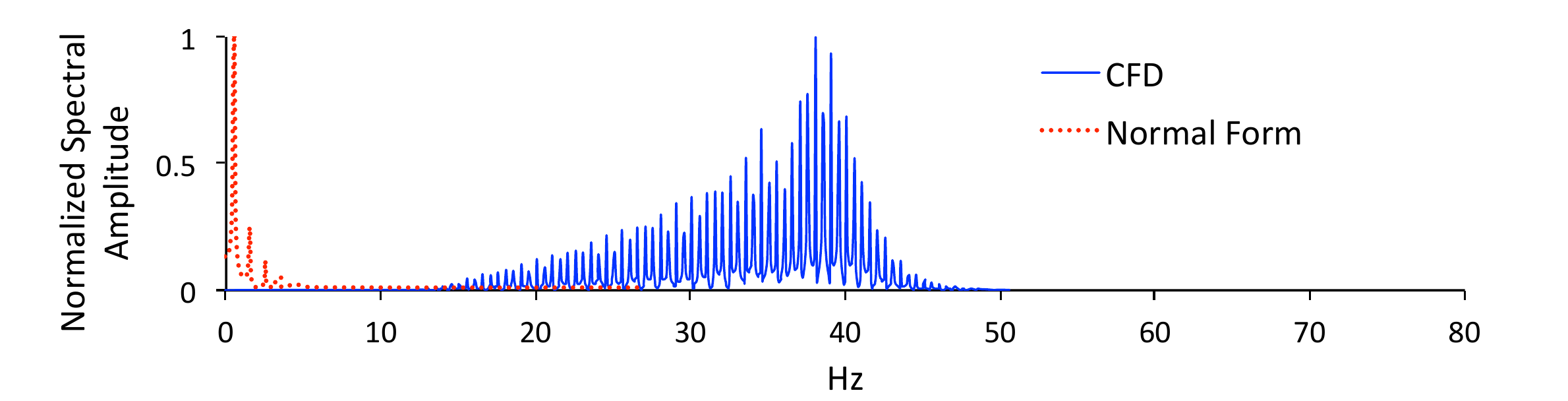}}
  \end{subfigmatrix}
 \caption{Transverse force coefficients for (a) CFD and (b) normal form with additive forcing term for a $f = 0.5$ Hz oscillating cylinder case; (c) corresponding spectral content.}
 \label{cfd_v_nf_spectra_wf2_2}
\end{figure}

Finally, we use normal form theory to demonstrate the importance of the dynamics-of-observables formalism  that leads to a bi-linear structure of forcing/excitation terms when $\g$ is a nonlinear function of $\z$; see (\ref{Koopevo_excited}). Without undertaking the formal development in Section \ref{cmnf_section}, one may be tempted to treat the effects of $\u$ by adding a simple harmonic forcing term to the unforced normal form equation, i.e.
\be
\label{nf_additive}
\dot{\eta} = \lambda_1 (\mu) \eta + \beta(\mu) \eta^2 \bar{\eta} - i Q (\zeta - \bar{\zeta}),
\ee
rather than the correct bi-linear parametric excitation appearing in (\ref{nf_1}). Similar analogies have been suggested to describe oscillating cylinder wake dynamics, e.g. \citet{forced_cyl_1}. However, as shown in this study, such an approach is only valid when $\g$ is a linear function of $\z$. For the more general dynamics-of-observables case, (\ref{nf_additive}) is incorrect. In essence, the additive forcing is appropriate when considering the system states $\z$ as the observable of interest, while the parametric system is appropriate for the dynamics of nonlinear functions of $\z$, such as force coefficients. To illustrate this, we reconsider the oscillating cylinder case for $f = 0.5$ Hz and $Re_q=10$. Except, we now compare with the incorrect additive normal from (\ref{nf_additive}), instead of the correct parametrically excited normal form whose solutions are shown in Figs. \ref{cfd_v_nf_spectra_wf2}(b) and \ref{cfd_v_normal_form_signal_wf2}(d). From Fig. \ref{cfd_v_nf_spectra_wf2_2}, it is clear that adding the forcing term to the unforced normal form equation in an ad hoc manner will lead to incorrect results when considering  dynamics of a generic observable. This finding has significant implications for development of control strategies based on reduced-order models of observables since the proper way to introduce control inputs will be through bi-linear terms. The dynamics-of-observables approach can be viewed as a generalization of the additive normal form approaches, e.g. \citep{vance_forced_bifurcation, Tsarouhas_1, Tsarouhas_2, gabale_nf}, since the correct normal form structure would be recovered if the observable of interest is a linear function of the underlying state $\z$. It is also worth noting that while parametrically excited Hopf bifurcation studies such as \citet{param_Hopf} include a bi-linear term, resonant phenomena (e.g. $f ~\sim 2 f_0$) have typically been studied. In contrast, the dynamics of (\ref{nf_1}) for $f << f_0$ have, to the best of our knowledge, not been presented previously.

\subsection{Quasi-Periodic Intermittency}
\label{chaos_vs_qp}
\begin{figure*}
 \begin{subfigmatrix}{2}
 \subfigure[$Re_q = 1$]{\includegraphics{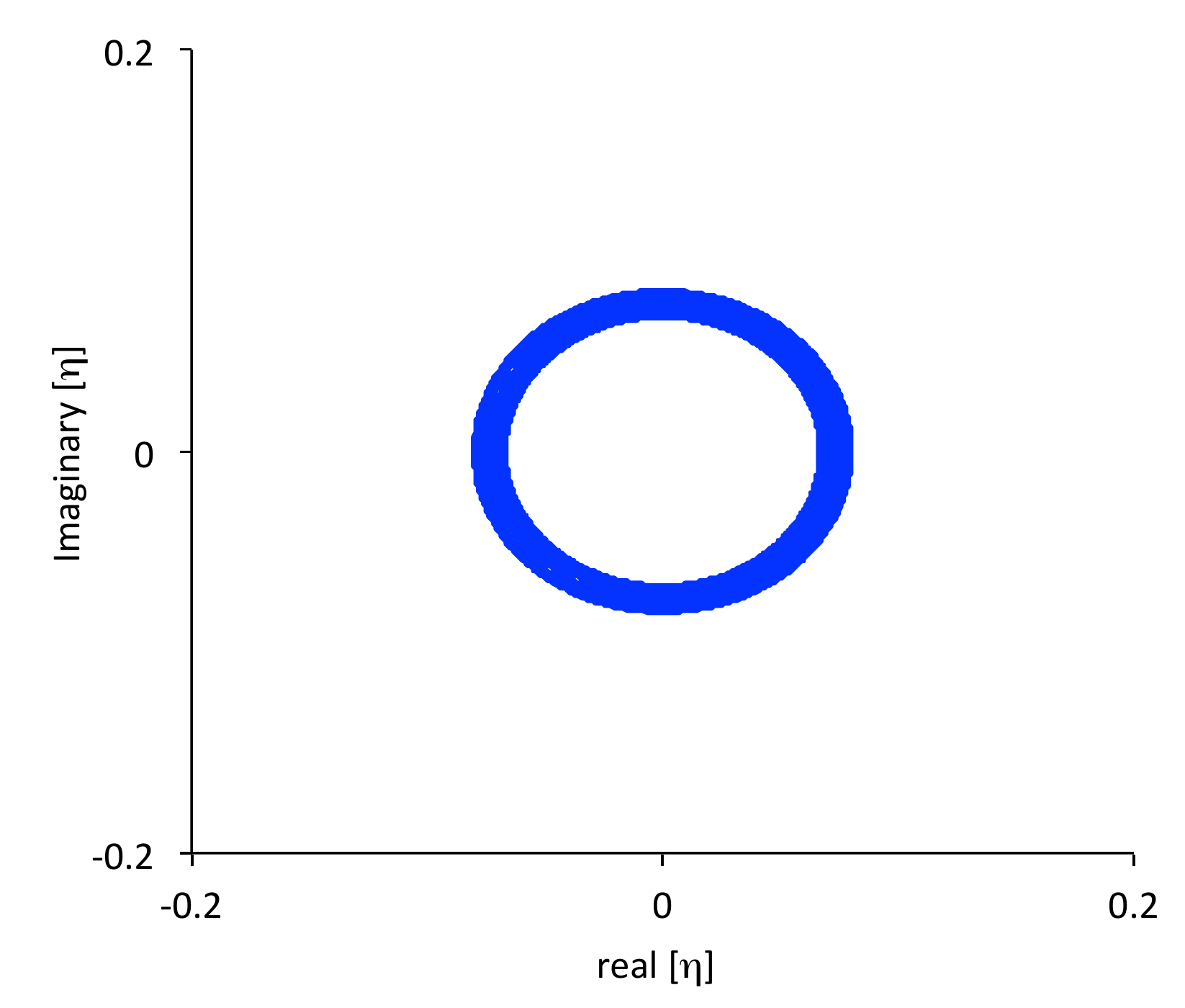}}
  \subfigure[$Re_q = 20$]{\includegraphics{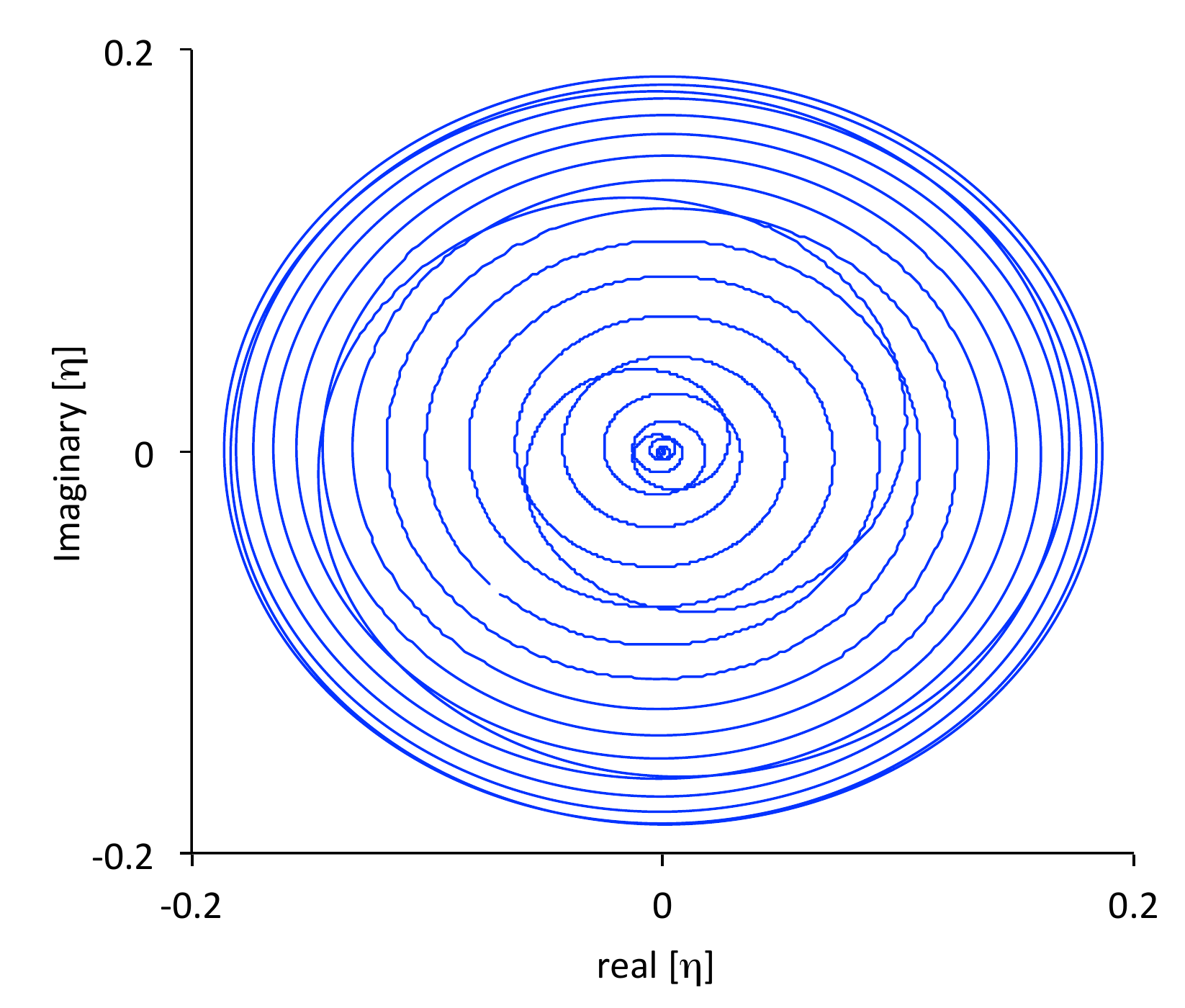}}
  \end{subfigmatrix}
 \caption{Normal form state space trajectories for one period corresponding to the forcing frequency $f$.}
 \label{trajectories_vs_forcing}
\end{figure*}

Having verified the two-dimensional reduced order normal form equations with respect to the full-order CFD solutions, we now proceed with determining the underlying nature of the spectral broadening that can manifest in the wake dynamics when $f << f_0$; i.e. are the dynamics quasi-periodic, chaotic, or something else? First, it is useful to visualize the attractor in state space. From Fig.~\ref{cfd_v_normal_form_signal_wf2}, it appears that the trajectories pass near $r=|\eta| = 0$ as the oscillatory forcing amplitude $Re_q$ is increased. This is confirmed in Fig.~\ref{trajectories_vs_forcing}.

\begin{figure*}
 \begin{subfigmatrix}{2}
 \subfigure[Comparison of trajectories after perturbation]{\includegraphics{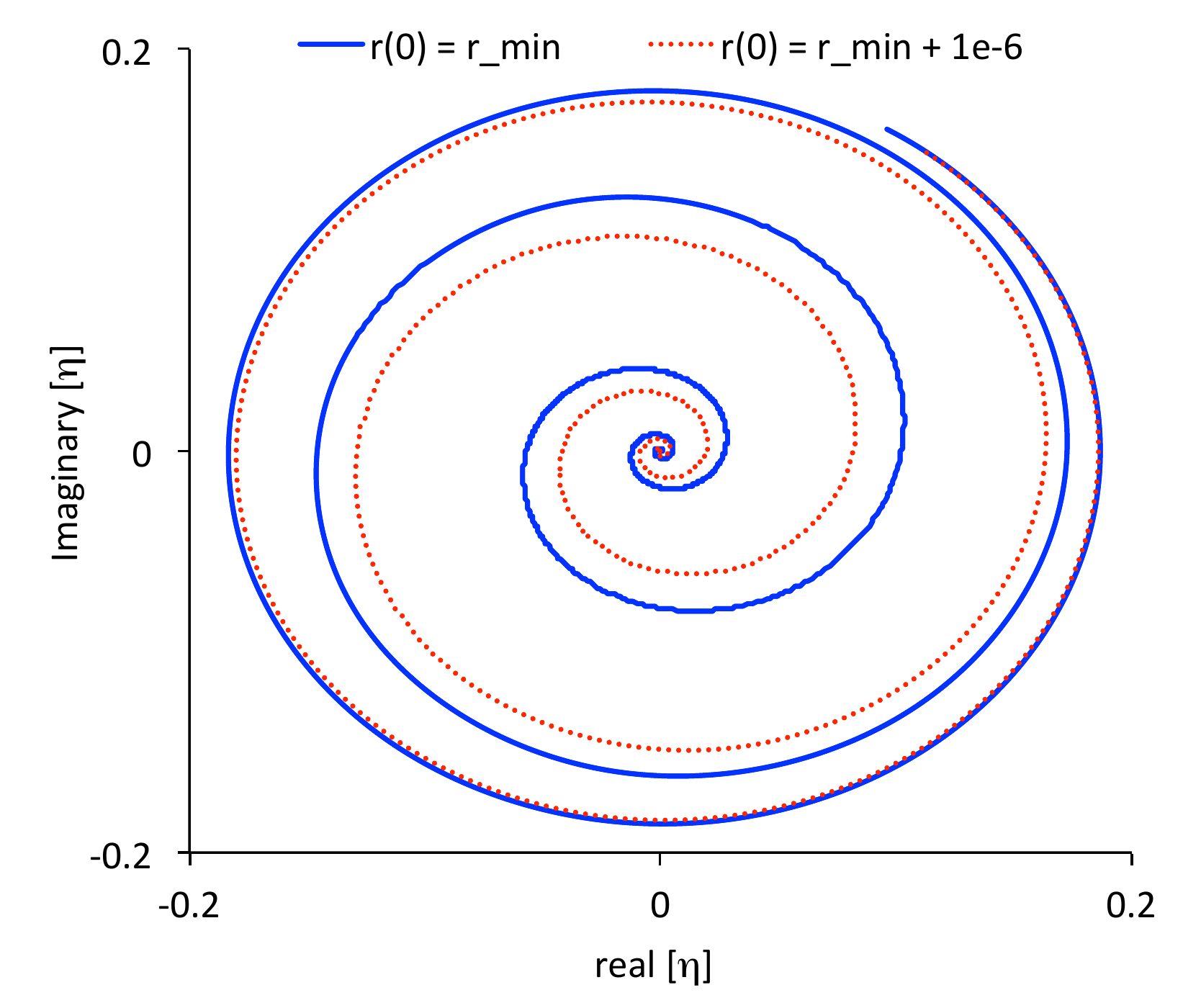}}
  \subfigure[Finite time Lyapunov exponents]{\includegraphics{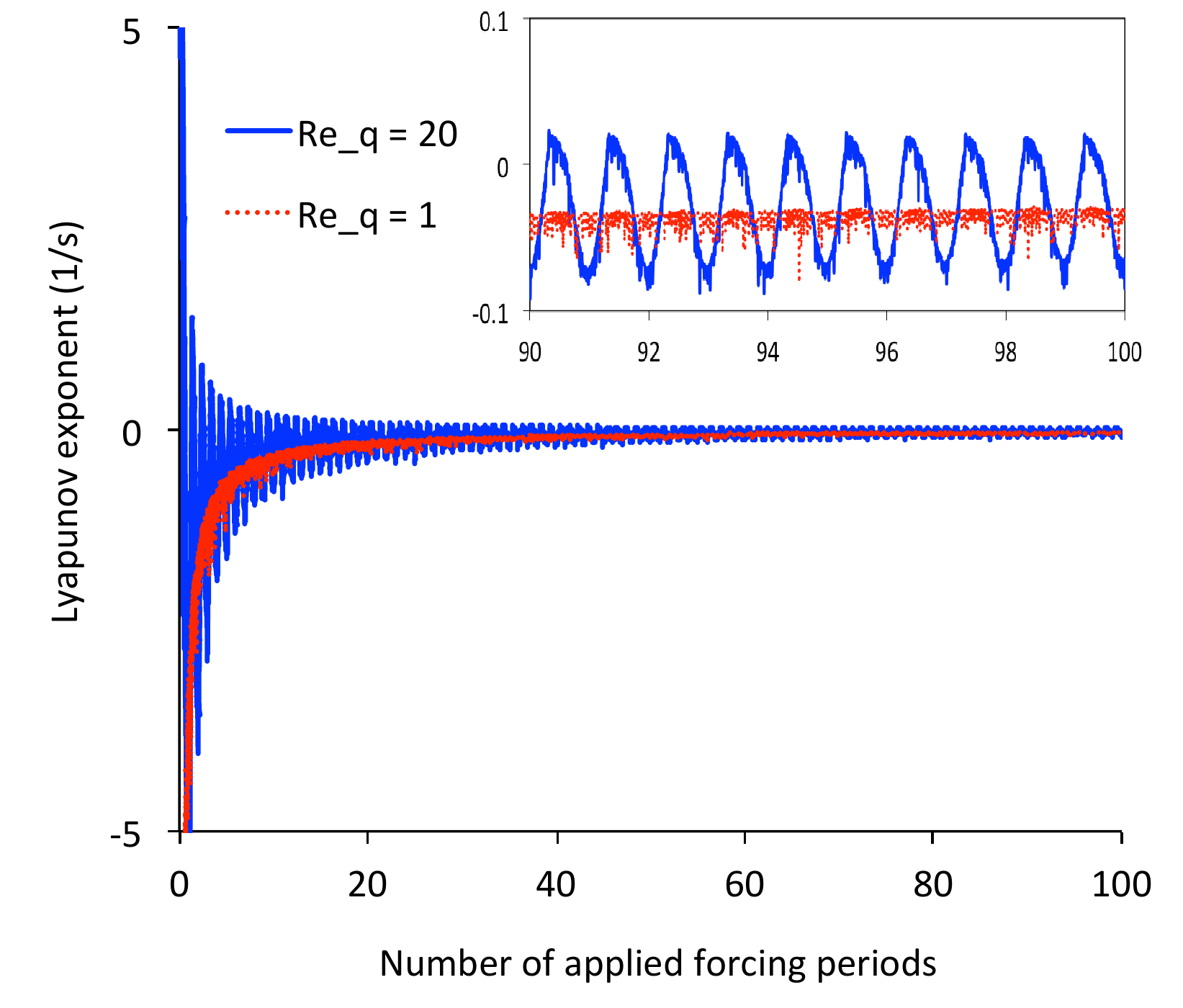}}
  \end{subfigmatrix}
 \caption{(a) Comparison of state space trajectories beginning from perturbed initial conditions on the attractor in the regime where $\beta_R r^3 << \left(\sigma + 2 Q_R \sin \omega_f t \right) r$ for $Re_q = 20$; note that the plot corresponds to 1000 forcing periods after the initial perturbation and only the portions of the trajectories corresponding to $\dot{r} > 0$ are shown for clarity. (b) The development of the finite time Lyapunov exponent for $Re_q = 20$ and $Re_q = 1$ over 100 forcing periods; the inset shows a zoomed in view.}
  \label{chaos_trajectories}
\end{figure*}

How closely the attractor passes near $r=0$ is a critical delineator between classical quasi-periodic dynamics and what we will later define as {\it Quasi-Periodic Intermittency}. This can be shown analytically by considering (\ref{rdot}) which is decoupled from $\theta$. When $r$ is sufficiently small, $\beta_R r^3 << \left(\sigma + 2 Q_R \sin \omega_f t \right) r$, leading to 
\be
\label{rdot_small} 
\dot{r} = \left(\sigma + 2 Q_R \sin \omega_f t \right) r.
\ee
If $\beta_R r^3 << \left(\sigma + 2 Q_R \sin \omega_f t^* \right) r$ at some time  $t=t^*$, then the solution to (\ref{rdot_small}) at $t=t^* + \Delta t$ is
\be
\label{rsol}
r(t^*+ \Delta t) = r(t^*) e^{\lambda_L \Delta t} 
\ee
where 
\be
\label{lyapunov_exp}
\lambda_L = \sigma + 2 Q_R \left(\sin \omega_f t^* + \omega_f \Delta t \cos \omega_f t^* \right)
\ee
and the approximation $\sin \omega_f (t^*+ \Delta t) \approx \sin \omega_f t^* + \omega_f \Delta t \cos \omega_f t^*$ has been used. Now we consider the evolution of two trajectories: the first $r_1(t^* + \Delta t)$ which is the solution beginning from $r(t^*)$, and the second $r_2 (t^* + \Delta t)$ beginning from a perturbed initial condition $r(t^*)+\delta_0$. After substituting these initial conditions into (\ref{rsol}), it is easy to show that the separation between the two trajectories, $\delta_r(t^* + \Delta t) \equiv r_2(t^* + \Delta t) - r_1(t^* + \Delta t)$, is
\be
\label{dr_traj}
\delta_r (t^* + \Delta t) = \delta_0 e^{\lambda_L \Delta t}.
\ee
Furthermore, if $\dot{r} > 0$, then $\lambda_L > 0$. Therefore, the separation between the trajectories exponentially diverges for finite time when $\dot{r} > 0$, until $r$ increases sufficiently for $\beta_R r^3$ to become significant. This behavior is consistent with chaos since the difference between two nearby trajectories on the attractor exponentially diverges. In this context, $\lambda_L$ in (\ref{dr_traj}) can be interpreted as a finite time Lyapunov exponent. Note that the system is only susceptible to exponential divergence of nearby trajectories at points on the attractor corresponding to small $r$ and $\dot{r} > 0$. The exponential divergence of nearby trajectories is illustrated in Fig.~\ref{chaos_trajectories}(a) where the two trajectories diverge as they spiral radially outwards ($\dot{r} > 0$) until reaching a sufficiently large value of $r$. In Fig.~\ref{chaos_trajectories}(a), the deviating trajectories are shown after 1000 forcing periods from the initial instant at which perturbation is introduced. Even though the system was only perturbed at one instant in time, the two trajectories will differ for all time due to the periodic occurrence of exponential divergence. Thus, the system exhibits the sensitive dependence on initial conditions on the attractor typically associated with chaotic systems. We have verified (not shown) that introducing similar perturbations elsewhere on the attractor away from small $r$, or for smaller forcing amplitudes, will not induce exponential divergence and the two nearby trajectories will eventually coalesce and become indistinguishable after sufficient time from the initial perturbation.

The development of the finite time Lyapunov exponent as the number of iterations is increased is shown in Fig.~\ref{chaos_trajectories}(b). Although the system exhibits the classically chaotic feature of sensitive dependance on initial conditions for some points on the attractor, the Lyapunov exponent approaches 0 as $t \rightarrow \infty$. So in contrast to classically chaotic systems, the Lyapunov exponent is not positive over infinite time. Instead, as shown in the Fig.~\ref{chaos_trajectories}(b) inset, the divergence of nearby trajectories for $Re_q=20$ is maintained for all time due to transiently positive finite time Lyapunov exponents as the system periodically passes by $r = 0$. For systems that do not pass near $r = 0$, as shown for the $Re_q =1$ case in Fig.~\ref{chaos_trajectories}(b), the Lyapunov exponent never becomes positive even within finite time windows. 

\begin{figure*}
\centering
 {\includegraphics[width = .75 \textwidth]{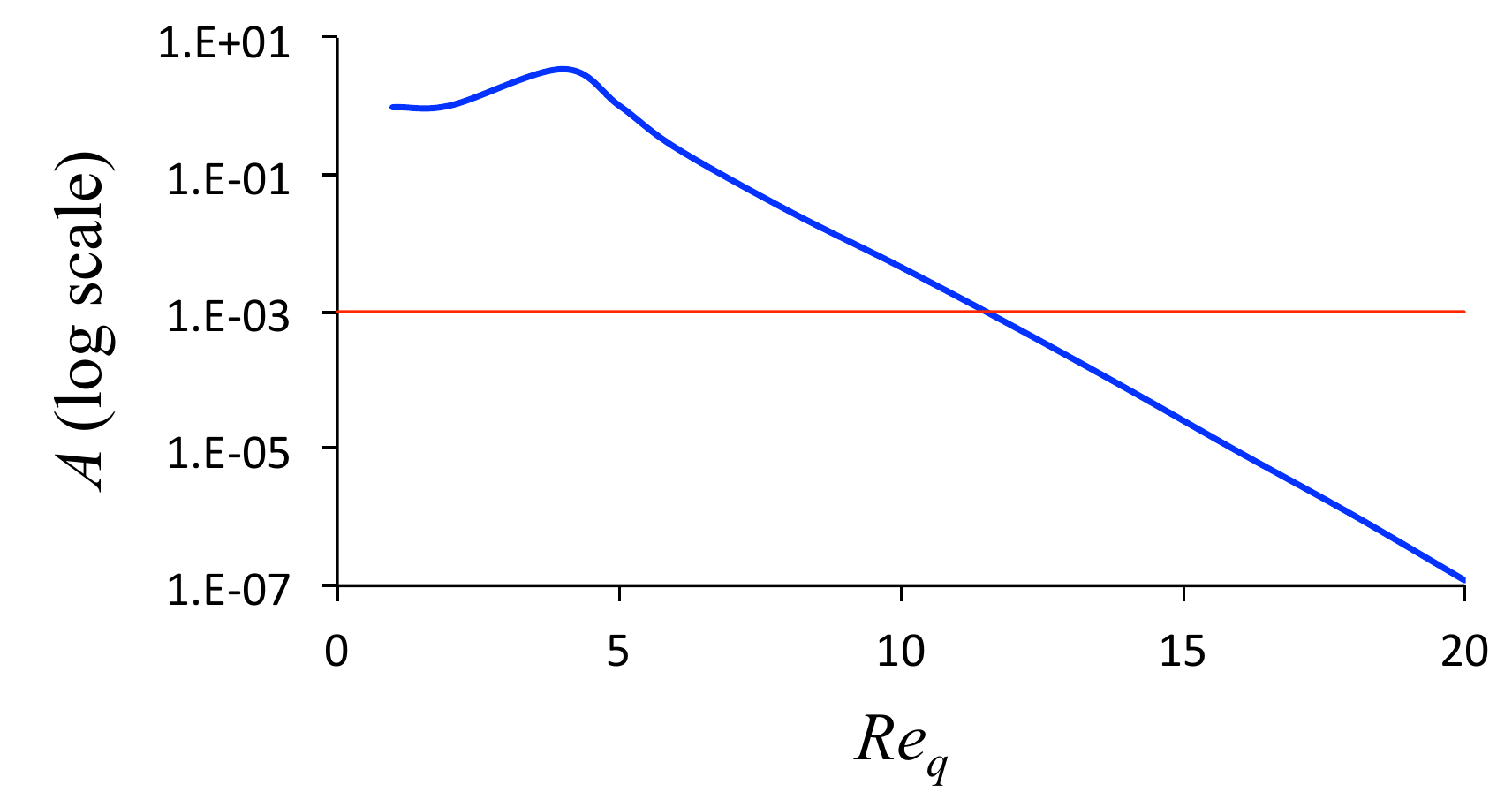}}
 \caption{Variation of the critical parameter delineating the onset of chaos, $A$, with oscillatory Reynolds number amplitude.}
 \label{rmin}
\end{figure*}

The effects of forcing amplitude on the susceptibility to exponential divergence over finite time is shown in Fig.~\ref{rmin}, where the critical parameter for exponential divergence is
\be
A = \frac{\beta_R r_{\mathrm{min}}^2}{\left(\sigma + 2 Q_R \sin \omega_f t_{\mathrm{min}} \right)} 
\ee
in which $r_{\mathrm{min}}$ is the minimum value of $r(t)$ and $t_{\mathrm{min}}$ is the time at which $r_{\mathrm{min}}$ occurs. The horizontal line in Fig.~\ref{rmin} indicates when $A << 1$ and the system is sensitive to perturbation at certain points on the attractor. Note that the value of $10^{-3}$ was chosen arbitrarily as a small value for illustrative purposes. It is clear from Fig.~\ref{rmin} that the oscillatory forcing amplitude must be sufficiently large to drive the system near the fixed point. If $2 Q_R$ is not $\ge \sigma$, then the system will never decay (although it will saturate due to the cubic term). As a result, $ 2 Q_R \ge \sigma$ is a condition on the amplitude that is required for finite time exponential divergence to be possible. Otherwise, such behavior can be ruled out \emph{a priori}. For $Re_q = 20$, $r_{\mathrm{min}} \sim \mathcal{O}(1e^{-8})$. Therefore, for larger oscillatory Reynolds numbers, the dynamics for the transverse force coefficient $c_y$ would have finite-time exponential divergence under realistic experimental conditions since very small perturbations (or lack of measurement precision for small numbers) are realities. 

\begin{figure*}
 \begin{subfigmatrix}{2}
 \subfigure[Normal form spectrum]{\includegraphics[width = .63 \textwidth]{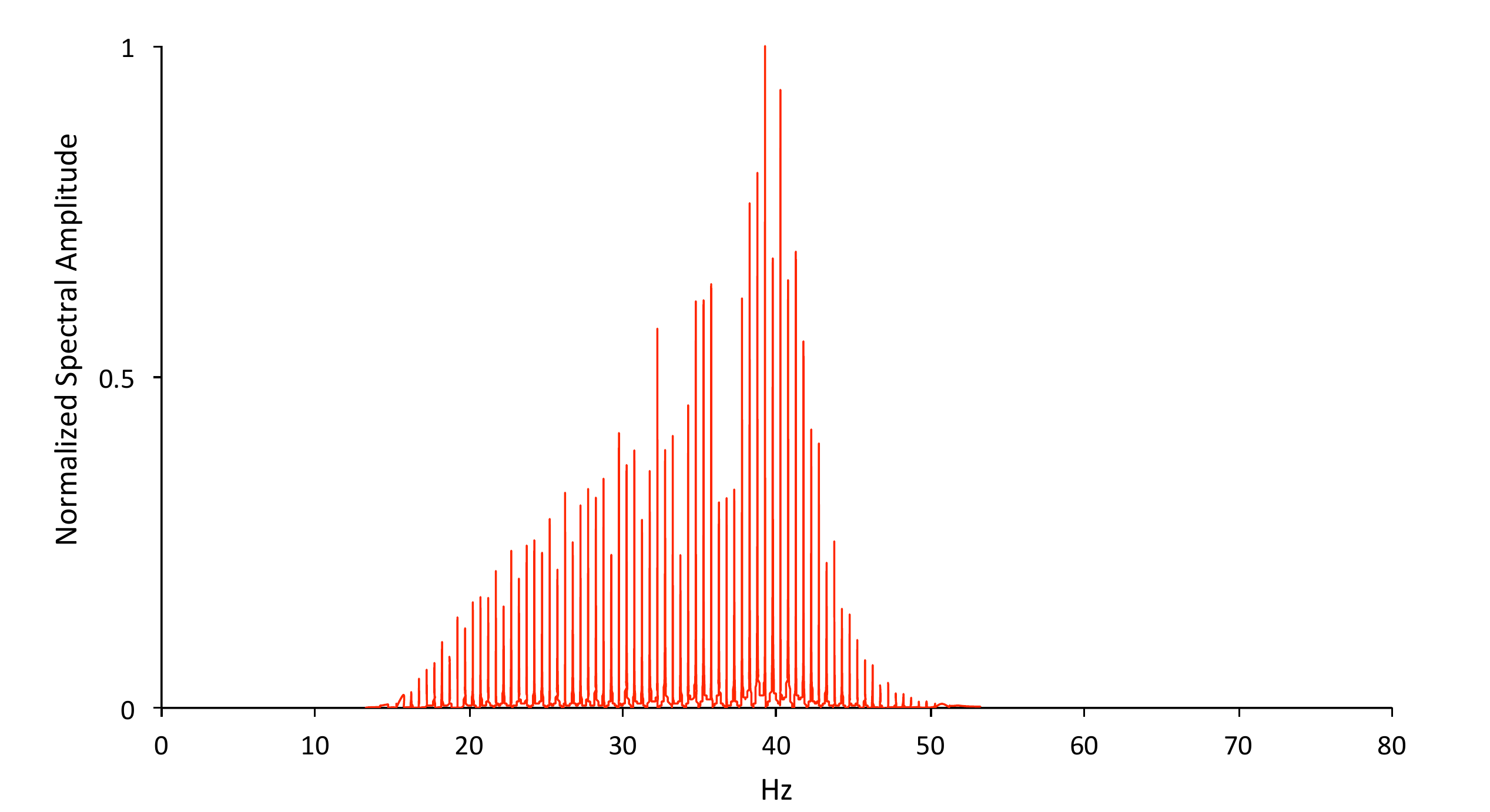}}
  \subfigure[Poincar{\'e} section]{\includegraphics[width = .35 \textwidth]{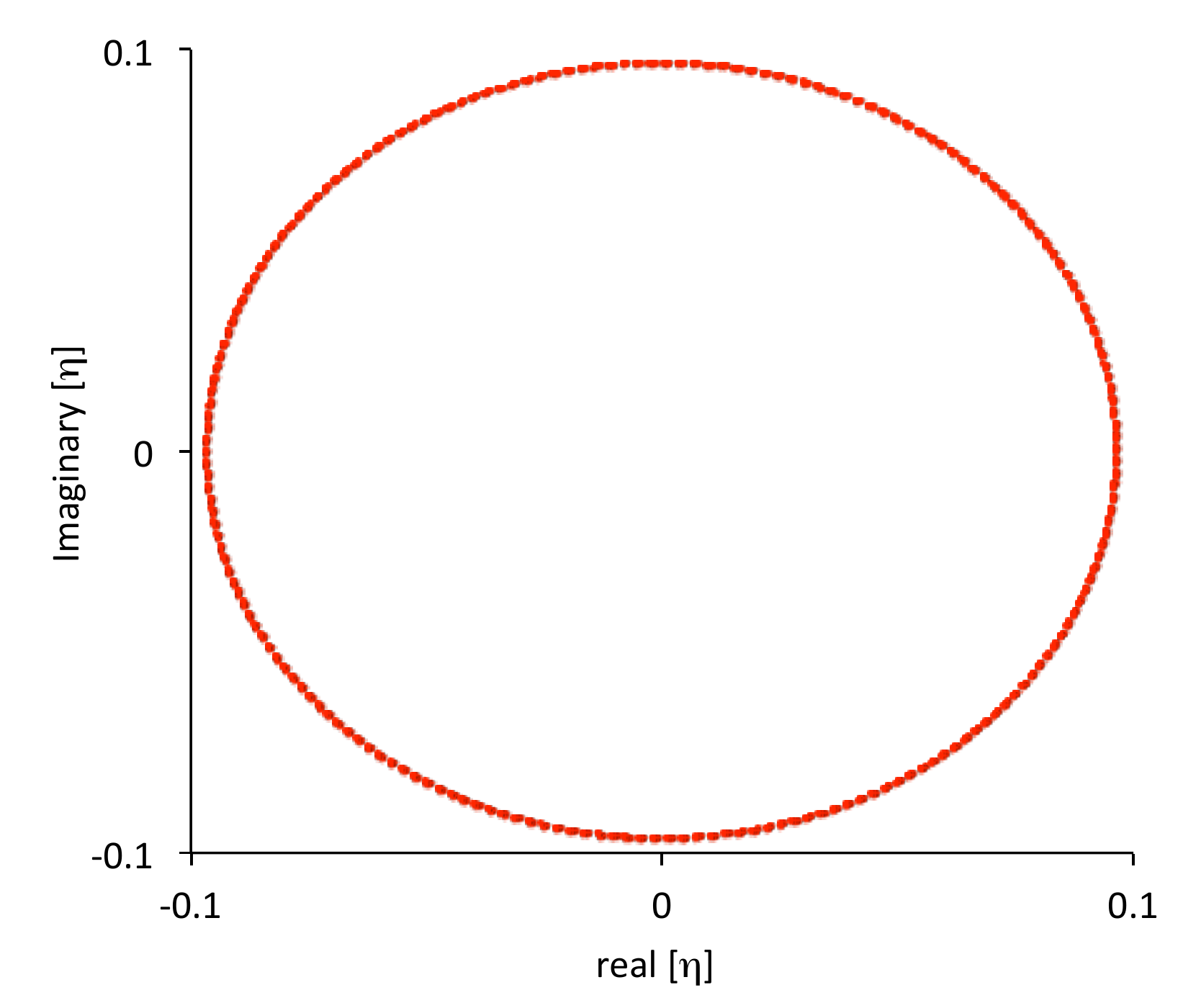}}
  \end{subfigmatrix}
 \caption{Discrete spectrum (a), and Poincar{\'e} section (b) corresponding to the normal form solution for $Re_q=20$; both are computed from 1000 periods of forcing frequency $f = 0.5$ Hz.}
 \label{qp}
\end{figure*}

Interestingly, the system also exhibits quasi-periodic characteristics even after transitioning to exponential divergence for sufficient forcing amplitude. As shown in Fig.~\ref{qp}(a), the normal form solution corresponding to $Re_q=20$ exhibits a discrete spectrum, as opposed to a continuous spectrum conventionally associated with chaos. The corresponding Poincar{\'e} section is shown in Fig.~\ref{qp}(b). The Poincar{\'e} section points are obtained by recording solutions at integer multiples of the forcing period $1/f$. The intersection points fill in a curve in the Poincar{\'e} plane, i.e. a drift ring, which is a clear indication of quasi-periodic dynamics \citep{hilborn_chaos}. Due to the described nature of this physical  phenomenon - featuring finite-time exponential divergence, associated with the bursts shown in Fig. \ref{cfd_v_normal_form_signal_wf2}e, and quasi-periodic dynamics - we call it the Quasi-Periodic Intermittency.   

\begin{figure*}
\centering
 {\includegraphics[width = 1 \textwidth]{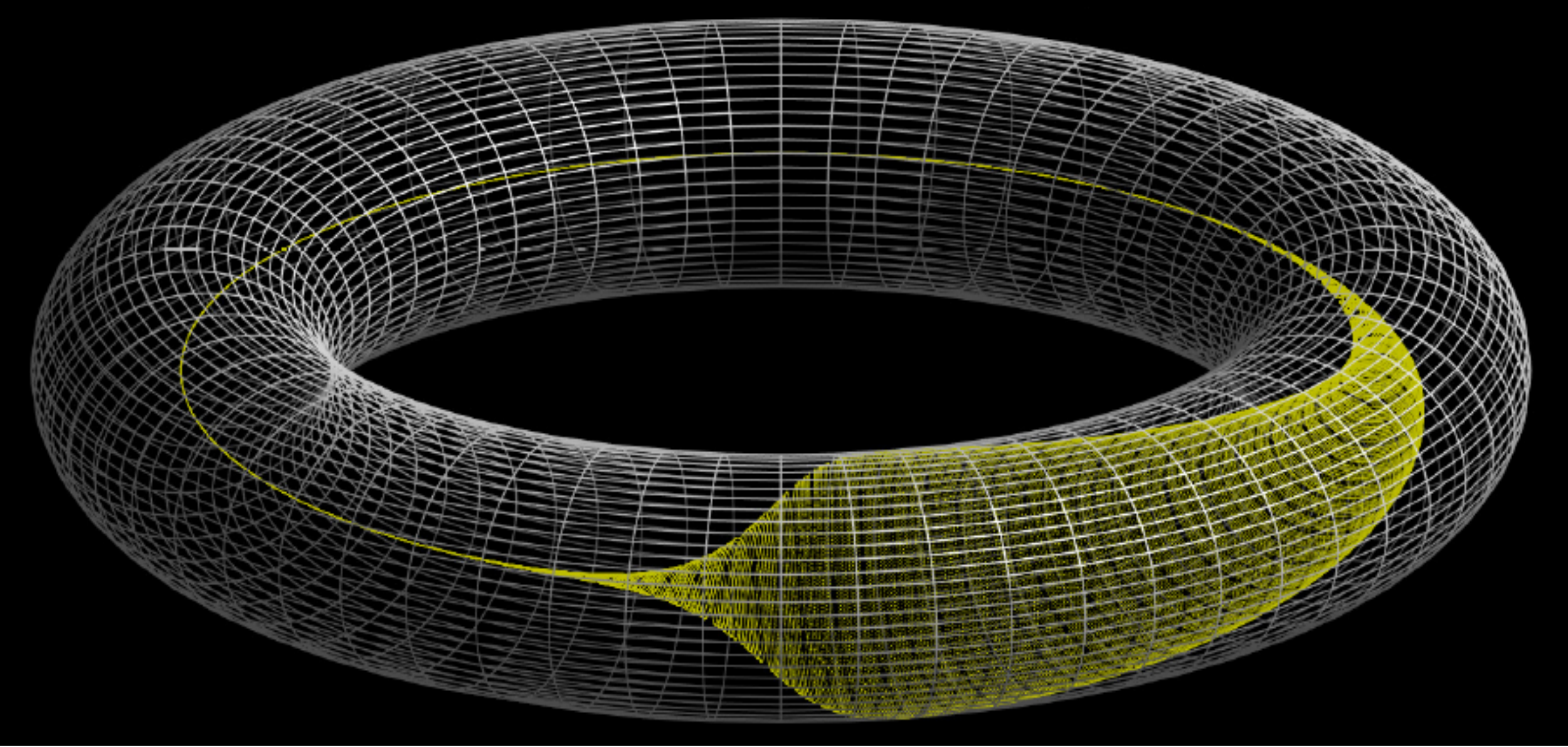}}
 \caption{The Quasi-Periodic Intermittency attractor (yellow) plotted on the torus. The toroidal grid (grey) is shown for reference. Rotation around the large diameter of the torus corresponds to the radial growth/decay that occurs over the slow forcing frequency, while rotation around the smaller diameter (i.e. phase velocity as the attractor spins around cross sections of the torus) correspond to the faster natural frequency of the system. The figure corresponds to $Re_q = 20$ and 24 periods of motion corresponding to the forcing frequency $f$.}
 \label{QPI}
\end{figure*}

Further insight can be gained by viewing a three-dimensional representation of the Quasi-Periodic Intermittency attractor obtained by plotting in toroidal coordinates, which is a common approach to visuaizing quasi-periodic and multi-scale behavior -- see Section 4.7 of \citep{hilborn_chaos} for example. In Fig. \ref{QPI}, the slower forcing scale (i.e. $f$) is represented by rotation around the larger diameter of the torus. It can be observed that the radial growth/decay (i.e. $\dot{r}$) corresponds to rotation around the torus, which is consistent with the fact that the forcing term manifests as a parametric excitation of the bifurcation parameter. Similarly, rotation around the cross-section (i.e. $\dot{\theta}$) corresponds to the faster natural frequency $f_0$. Note that if one were to unfold the torus in Fig.~\ref{QPI} into a cylinder, the cross-sectional view of the cylinder would resemble Fig.~\ref{trajectories_vs_forcing}(b).

\section{Conclusions}

Parametrically excited Hopf bifurcation flows were studied using Koopman Mode Analysis (KMA) of the canonical oscillating cylinder system when there is a separation of scale between the forcing and natural frequencies, i.e. $f << f_0$. Koopman decompositions indicate transition from narrowly distributed discrete spectra for the stationary cylinder case, to more broadly distributed discrete spectra as the oscillation amplitude is increased, even though only a single forcing frequency is applied. Furthermore, the Koopman mode shapes for the forced case indicate that the unforced shedding mode is still a dynamically relevant feature of the forced system. The use of unforced Hopf bifurcation modes as a projection basis for forced bifurcation normal form-based  reduced-order modeling was also justified in the context of Koopman operator theory as the unforced Koopman operator was shown to be a prevalent feature of the forced system. In contrast, this may not be appropriate for reduced-order modeling control studies based on other approaches, such as proper-orthogonal decomposition, which cannot be theoretically interpreted as spanning the action of a linear operator associated with the unforced system that remains a part of the equation governing the evolution of observables under forcing. 

When considering a dynamics-of-observables perspective, it was shown that the effect of forcing (or a control input) appears as a bi-linear excitation term if the observable is a nonlinear function of the underlying state vector. To verify this finding, model order reduction was conducted by projecting onto the unforced Koopman shedding mode, i.e. the center manifold. This system was then further simplified using normal form theory. It was demonstrated that the two-dimensional bi-linear normal form approximation accurately captures the leading order effect of the prescribed forcing on the shedding dynamics of the full-order high-dimensional CFD solutions. These findings will inform future development of control based reduced-order modeling studies using Koopman theory, and are generally applicable to forced Hopf bifurcation systems.

Normal form mathematical models were used to explicitly establish $f << f_0$ as a necessary condition for the applied forcing to induce spectral broadening in the shedding mode dynamics. It was shown that the normal form equations for forced cylinder wake shedding dynamics are a physical realization of a phenomenon that we call Quasi-Periodic Intermittency. The phenomenon exhibits  finite-time exponential divergence of nearby trajectories, while maintaining a discrete spectrum as in quasi-periodic systems. Given the prevalence of Hopf bifurcation systems throughout physics, the interaction between disparate scales described by the Quasi-Periodic Intermittency theory may underpin a variety of multi-scale dynamics phenomena. 


\section*{Acknowledgements}

This work was supported by a U.S. Army Research Laboratory Director's Research Initiative award for B. Glaz. Support for I. Mezi{\'c}, M. Fonoberova, and S. Loire under U.S. Army Research Office project W911NF-14-C-0102, U.S. Army Research Office project W911NF-16-1-0312 (Program Manager Dr. Samuel Stanton) , and U.S. Air Force Office of Scientific Research project FA9550-12-1-0230 (Program Manager Dr. Fariba Fahroo and Dr. Frederick Leve) are gratefully acknowledged.

\bibliographystyle{jfm}
\bibliography{jfm_bglaz}

\end{document}